\documentclass[12pt]{article}
\usepackage[latin9]{inputenc}
\usepackage{geometry}
\usepackage{float}
\usepackage{mathtools}
\usepackage{amsmath}
\usepackage{amsthm}
\usepackage{amssymb}
\usepackage{graphicx}
\usepackage{setspace}
\usepackage[authoryear]{natbib}
\makeatletter

\providecommand{\tabularnewline}{\\}

\usepackage{latexsym}
\usepackage{amsthm}\usepackage{amsfonts}\usepackage{graphicx}\usepackage{epsfig}
\usepackage{bm}
\newcommand*{\patchAmsMathEnvironmentForLineno}[1]{%
      \expandafter\let\csname old#1\expandafter\endcsname\csname #1\endcsname
      \expandafter\let\csname oldend#1\expandafter\endcsname\csname end#1\endcsname
      \renewenvironment{#1}%
         {\linenomath\csname old#1\endcsname}%
         {\csname oldend#1\endcsname\endlinenomath}}%
    \newcommand*{\patchBothAmsMathEnvironmentsForLineno}[1]{%
      \patchAmsMathEnvironmentForLineno{#1}%
      \patchAmsMathEnvironmentForLineno{#1*}}%
    \AtBeginDocument{%
    \patchBothAmsMathEnvironmentsForLineno{equation}%
    \patchBothAmsMathEnvironmentsForLineno{align}%
    \patchBothAmsMathEnvironmentsForLineno{flalign}%
    \patchBothAmsMathEnvironmentsForLineno{alignat}%
    \patchBothAmsMathEnvironmentsForLineno{gather}%
    \patchBothAmsMathEnvironmentsForLineno{multline}%
    }
\usepackage[mathlines,displaymath]{lineno}

\include{def}
\def\dispmuskip{\thinmuskip= 3mu plus 0mu minus 2mu \medmuskip=  4mu plus 2mu minus 2mu \thickmuskip=5mu plus 5mu minus 2mu}
\def\textmuskip{\thinmuskip= 0mu                    \medmuskip=  1mu plus 1mu minus 1mu \thickmuskip=2mu plus 3mu minus 1mu}
\def\beq{\dispmuskip\begin{equation}}    \def\eeq{\end{equation}\textmuskip}
\def\beqn{\dispmuskip\begin{displaymath}}\def\eeqn{\end{displaymath}\textmuskip}
\def\bea{\dispmuskip\begin{eqnarray}}    \def\eea{\end{eqnarray}\textmuskip}
\def\bean{\dispmuskip\begin{eqnarray*}}  \def\eean{\end{eqnarray*}\textmuskip}

\def\R{{\cal R}}

\def\Eq{Eq.}
\def\dlq{\lq\lq}
\def\drq{\rq\rq}
\def\FSV{factor stochastic volatility}

\usepackage[mathlines,displaymath]{lineno}
\usepackage{authblk}
\usepackage{comment}
\usepackage{bbold}
\makeatother

\begin{document}
\title{Variational Approximation of Factor Stochastic
Volatility Models}
\author[1,3]{David Gunawan}
\author[2,3]{Robert Kohn}
\author[4,5]{David Nott}
\affil[1]{School of Mathematics and Applied Statistics, University of Wollongong}
\affil[2]{School of Economics, UNSW Business School, University of New South Wales}
\affil[3]{Australian Center of Excellence for Mathematical and Statistical Frontiers}
\affil[4]{Department of Statistics and Applied Probability, National University of Singapore}
\affil[5]{Institute of Operations Research and Analytics, National University of Singapore}
\renewcommand\Authands{ and }
\maketitle

\begin{abstract}
Estimation and prediction in high dimensional multivariate
factor stochastic volatility models is an important and
active research area because such models allow a parsimonious representation
of multivariate stochastic volatility.
Bayesian inference for factor stochastic volatility models is usually done by
Markov chain Monte Carlo methods, often by particle Markov chain Monte Carlo,
which are usually slow for high dimensional or long time series because of the large number of parameters and latent states involved.
Our article makes two contributions. The first is to propose
fast and accurate variational Bayes methods to approximate the posterior distribution
of the states and parameters in factor stochastic volatility models. The second contribution
is to extend this batch methodology to develop fast sequential variational updates for prediction as new observations arrive. The methods are applied to simulated and real datasets and shown to produce good approximate inference and prediction compared to the latest particle Markov chain Monte Carlo approaches, but are much faster.
\end{abstract}

%


\section{Introduction \label{sec:Introduction}}
Statistical inference and prediction in high dimensional time series models that incorporate  stochastic volatility (SV) is an important and active research area because of its applications in financial econometrics and financial decision making.
One of the main challenges in estimating such models is that
the number of parameters and latent variables involved often increases quadratically with dimension, while the length of each latent variable equals the number of time periods. These problems are lessened by using a factor stochastic volatility (FSV) model.
See, e.g. \citet{pitt:factor} and \citet{Chib2006} who use  independent and  low dimensional latent factors to approximate a fully multivariate SV (MSV) model to achieve an attractive tradeoff between model complexity and model flexibility.

Although FSV models are more  tractable computationally
than high dimensional multivariate SV (MSV) models, they are still challenging to estimate because they are relatively high dimensional in the number of parameters and latent state variables; the likelihood of any FSV model is also intractable as it is a high dimensional integral over the latent state variables. Current state of the art approaches use Markov chain Monte Carlo (MCMC)   \citep[e.g.,][]{Chib2006, Kastner:2017} were developed over a number of years and are very powerful. However, they are less flexible and accurate than recent particle MCMC based approaches  by \citet{mendes:2020}, \citet{Li:scharth} and \citet{Gunawan2020} as they  use the approach  proposed in \citet{Kim:1998} (based on assuming normal errors) to approximate the distribution of the innovations in the log outcomes (usually the log of the squared returns) by a mixture of normals; see \citet{mendes:2020} for a comparison of the accuracy and flexibility of the \cite{Chib2006} approach compared to the particle MCMC approach.
As  well as estimating the FSV model, practitioners are often interested in updating estimates of the parameters and latent volatilities as new observations arrive, and  obtaining one-step or multiple-steps ahead predictive distributions of future volatilities and observations.
However, it is usually computationally very expensive  to estimate a \FSV{} by MCMC or particle MCMC. Sequential updating by MCMC or particle MCMC is even more expensive as it is necessary to carry out a new MCMC simulation as  new observations become available, although such updating is made simpler because good starting values for the MCMC are available from the previous MCMC run. Sequential Monte Carlo methods can also be used instead for estimation and sequential updating, but these can also be expensive.

Variational Bayes (VB) has become increasingly prominent
as a method for conducting parameter inference in a wide range of
challenging statistical models with  computationally difficult posteriors
\citep{Ormerod:2010,Tan2018,Ong2018,Blei2017}.
VB expresses the estimation of the posterior distribution as an optimisation problem
by  approximating the posterior density by a simpler distribution whose parameters are unknown,
e.g., by a multivariate Gaussian with unknown mean and covariance matrix.
VB methods usually produce posterior inference with much less computational cost compared
to the exact methods such as MCMC or particle MCMC, although they are approximate.


Our article makes two contributions. First, it
proposes fast and accurate variational approximations of the
posterior distribution of the latent volatilities and parameters of
a multivariate factor stochastic volatility
model. It does so by carefully studying which dependencies it is important to retain in the
variational posterior to obtain good approximations; and which can be omitted,
thus speeding up the computation. This results in approximate Bayesian inference that is very close to that obtained by the much slower particle methods.

The second contribution is to develop a sequential version of the batch approach in the first contribution which updates the variational approximation as new observations arrive. The sequential version can be used
for one step and multiple steps forecasting as new observations arrive. It is also useful for
time series cross-validation for model selection as well as assessing model fit.
Sections \ref{subsec:Simulation-Study}
and \ref{subsec:Application-to-US StockReturnsData} show that our variational
approach produces forecasts that are very close to those obtained
from the exact particle MCMC methods, even when the corresponding
variational posterior variance is sometimes underestimated.

The importance of the second  contribution  is that
the variational updates and forecasts are produced with much less
computational cost compared to the time taken to produce the corresponding
updates and exact predictions using  particle MCMC.

Although our article focuses on a particular, but quite general, factor  stochastic volatility model,
we believe that the methodology developed can be applied to a wide range of such models. See
Section~\ref{sec:Conclusions} for further remarks.


The current VB literature focuses on a Gaussian VB
for approximating the posterior density. The variational
parameters to be optimised are the mean vector and the covariance
matrix. It is challenging to perform Gaussian variational approximation
with an unrestricted covariance matrix for high dimensional parameters
because the number of elements in the covariance matrix of the variational
density increases quadratically with number of model parameters. It is then very important to find an effective and efficient
parameterisation of the covariance structure of the variational density
when the number of latent states and parameters are large as in the
multivariate factor SV model.

%

Various suggestions in the literature exist for parsimoniously parameterising
the  covariance matrix in the Gaussian variational approximation (GVA).
\citet{Titsias2014} use a Cholesky factorisation of the full
covariance matrix, but they do not allow for parameter sparsity;
the exceptions are diagonal approximations which lose any ability to capture
posterior dependence in the variational approximation. \citet{Kucukelbir2017}
also consider an unrestricted and diagonal covariance matrix and use
similar gradient estimates to \citet{Titsias2014}, but use  automatic differentiation.
\citet{Tan2018}  parameterise the precision matrix in terms of a sparse Cholesky
factor that reflects the conditional
independence structure in the model. This approach  is motivated by the result that a zero on the off-diagonal in the precision matrix means that the corresponding two variables are conditionally independent
\citep{ Tan2018}.
The computations can then be done efficiently through fast sparse matrix
operations. Recently, \citet{Ong2018} consider a factor covariance
structure as a parsimonious representation of the covariance matrix
in the Gaussian variational approximation. They also combine stochastic
gradient ascent optimisation with the \lq \lq reparameterisation
trick\rq \citep{Kingma2014}. \citet{Opper:2009},
\citet{Challis2013}, and \citet{Salimans:2013} consider other parameterisations of the covariance matrix in the Gaussian
variational approximation. \citet{Quiroz2018}
combine the factor parameterisation and sparse precision Cholesky
factors for capturing dynamic dependence structure in high-dimensional
state space model. They apply their method to approximate the posterior
density of the multivariate stochastic volatility model via a Wishart
process. The computational cost of their variational approach increases
with the length of the time series and their variational approach
is currently limited to low dimensional and short time series. Our proposed variational
approach is scalable in terms of number of parameters, latent states,
and observations.

\citet{Tan2019} and \citet{Smith2019} propose more flexible variational
approximations than GVA.
\citet{Tan2019} extend the  approach in \citet{Tan2018} by defining the posterior
distribution as a product of two Gaussian densities. The first is the posterior for the global parameters;
the second is the posterior of the local parameters, conditional on the global parameters.
The method respects a conditional independence structure similar to that of \citet{Tan2018}, but allows the
joint posterior to be non-Gaussian because the conditional mean of the local parameters can be
nonlinear in the global parameters.  Similarly to \citet{Tan2018}, they
also exploit the conditional independence structure in the model and improve
the approximation by using the importance weights approach \citep{Burda2016}.
\citet{Smith2019}
propose an approximation that is based on implicit copula models
for original parameters with a Gaussian or skew-normal copula function
and flexible parametric marginals. They consider the Yeo-Johnson \citep{Yeo2000}
and G\&H families \citep{Tukey1977}  transformations and use the sparse
factor structures proposed by \citet{Ong2018} as the covariance matrix
of the Gaussian or skew-normal densities. Our variational method for
approximating posterior distribution of the multivariate factor SV
model makes use of both conditional independence structures of the
model and the factor covariance structure.
In particular, we make use of the non-Gaussian sparse Cholesky factor
parameterisation proposed by \citet{Tan2019} and implicit Gaussian
copula with Yeo-Johnson transformation and sparse factor covariance
structure proposed by \citet{Smith2019} as the main building blocks
for variational approximations. We use an efficient stochastic
gradient ascent optimisation method based on the so called \lq \lq reparameterisation
trick\rq \rq  for obtaining unbiased estimation of the gradients of the
variational objective function \citep{Salimans:2013,Kingma2014,Rezende2014,Ranganath:2014,Titsias2014};
in practice, the reparameterisation trick  greatly helps
to reduce the variance of gradient estimates. Section \ref{sec:Summary-of-VB approach}
describes these methods further.

%
%
%
%

In related work, \citet{Tomasetti2019} propose a sequential updating variational method,
and apply it to a simple univariate autoregressive model with a tractable likelihood.
\citet{Frazier2019} explore the use of Approximate
Bayesian Computation (ABC) to generate forecasts. The ABC method is a very different approach to approximate Bayesian inference, and is usually only effective in low dimensions.
They prove, under certain regularity conditions, that ABC  produces forecasts that are asymptotically equivalent
to those obtained from the exact Bayesian method with much less computational
cost. However, they only apply their method to a univariate state
space model.

The rest of the article is organised as follows. Section~\ref{sec:Description-of-Factor SV model}
introduces the multivariate factor stochastic volatility model.
Section~\ref{subsec:Approximate-Bayesian-Inference} briefly summarizes VB. Sections \ref{subsec:Sparse-Cholesky-factorTan2019}
and \ref{subsec:Approximation-with-factor Smith2019} review the variational
methods proposed by \citet{Tan2019} and \citet{Smith2019}. Section
\ref{subsec:VB-approach-for factor SV model} discusses our variational
approximation for approximating the posterior of the multivariate
factor SV model. Section \ref{sec:Sequential-Updating-Variational Approximation} discusses the sequential variational algorithm for updating the
posterior density. Section~\ref{subsec:VariationalForecasting} discusses the
variational forecasting method. Section \ref{sec:Simulation-Study-and real data application}
presents results from a simulated dataset and a US stock returns
dataset.
Section~\ref{sec:Conclusions} concludes. The paper has an online supplement which contains some further empirical results.

\section{The factor SV Model \label{sec:Description-of-Factor SV model}}

 Suppose that $P_{t}$ is a $S\times1$
vector of daily stock prices and define $y_{t}\coloneqq\log P_{t}-\log P_{t-1}$
as the vector of stock returns of the stocks. The factor model is
\begin{equation}
y_{t}=\beta f_{t}+\epsilon_{t},\:\left(t=1,...,T\right);\label{eq:factor model}
\end{equation}
$f_{t}$ is a $K\times1$ vector of latent factors (with $K\lll S$),
$\beta$ is a $S\times K$ factor loading matrix of the unknown parameters.
The latent factors $f_{k,t}$, $k=1,...,K$ are assumed independent
with $f_{t}\sim N\left(0,D_{t}\right)$. Section~\ref{sec:factorloadingmat} of the supplement discusses parameterisation and identification issues for the factor loading matrix $\beta$ and the latent factors $f_{t}$. The time-varying variance
matrix $D_{t}$ is a diagonal matrix with $k$th diagonal element
$\exp\left(\tau_{f,k}h_{f,k,t}\right)$. Each $h_{f,k,t}$ is assumed
to follow an independent autoregressive process
\begin{equation}
h_{f,k,1}\sim N\left(0,\frac{1}{1-\phi_{f,k}^{2}}\right),\;h_{f,k,t}=\phi_{f,k}h_{f,k,t-1}+\eta_{f,k,t}\;\eta_{f,k,t}\sim N\left(0,1\right),k=1,...,K.\label{eq:statetransitionfactor}
\end{equation}
The idiosyncratic errors are modeled  as $\epsilon_{t}\sim N\left(0,V_{t}\right)$;
the time-varying variance matrix $V_{t}$ is diagonal with $s$th
diagonal elements $\exp\left(\tau_{\epsilon,s}h_{\epsilon,s,t}+\kappa_{\epsilon,s}\right)$.
Each $h_{\epsilon,s,t}$ follows an independent autoregressive
process
\begin{equation}
h_{\epsilon,s,1}\sim N\left(0,\frac{1}{1-\phi_{\epsilon s}^{2}}\right),\;h_{\epsilon st}=\phi_{\epsilon,s}h_{\epsilon,s,t-1}+\eta_{\epsilon,s,t}\;\eta_{\epsilon,s,t}\sim N\left(0,1\right),s=1,...,S.\label{eq:statetransitionidiosyncratic-1}
\end{equation}
Thus,
\[
y_{t}|\Sigma_{t}\sim N_{S}\left(0,\Sigma_{t}\right),
\quad
\text{with}
\quad
\Sigma_{t}=\beta D_{t}\beta^{\top}+V_{t}.
\]

To simplify the optimisation, the constrained parameters are mapped to the real space $\R$
by letting
$\alpha_{\epsilon,s}=\log\left(\exp\left(\tau_{\epsilon,s}\right)-1\right)$,
$\psi_{\epsilon,s}=\log\left(\phi_{\epsilon,s}/(1-\phi_{\epsilon,s})\right)$
for $s=1,...,S$, $\alpha_{f,k}=\log\left(\exp\left(\tau_{f,k}\right)-1\right)$,
$\psi_{f,k}=\log\left(\phi_{f,k}/(1-\phi_{f,k})\right)$ for
$k=1,...,K$, and log-transform the diagonal elements of the
factor loading matrix $\beta$, $\delta_{k}=\log\left(\beta_{k,k}\right),$
for $k=1,...,K$. The transformations for $\alpha_{f,k}$ and $\alpha_{\epsilon,s}$ are suggested by \citet{Tan2019} who find they are better than $\alpha_{f,k}=\textrm{log}(\tau_{f,k})$ for $k=1,...,K$ and $\alpha_{\epsilon,s}=\textrm{log}(\tau_{\epsilon,s})$ for $s=1,...,S$, the latter  often giving convergence problems.

We follow \citet{Kim:1998} and choose the prior
for persistence parameter $\phi_{\epsilon,s}$ for $s=1,...,S$ and
$\phi_{f,k}$ for $k=1,...,K$ as $\left(\phi+1\right)/2\sim Beta\left(a_{0},b_{0}\right)$,
with $a_{0}=20$ and $b_{0}=1.5$. The prior for each of $\tau_{\epsilon,s}$
and $\tau_{f,k}$ is half Cauchy, i.e. $p\left(\tau\right)\propto I\left(\tau>0\right)/(1+\tau^{2})$
and the prior for $p\left(\kappa_{\epsilon,s}\right)\sim N(0,10)$. For
every unrestricted element of the factor loadings matrix $\beta$, we follow \citet{Kastner:2017} and
choose independent Gaussian distribution $N\left(0,1\right)$. The collection of unknown parameters in the FSV model is $\theta=\left(\left\{ \kappa_{\epsilon,s},\alpha_{\epsilon,s},\psi_{\epsilon,s}\right\} _{s=1}^{S},\left\{ \alpha_{f,k},\psi_{f,k}\right\} _{k=1}^{K},\beta\right)$; the collection of states $x_{1:T}=\left(\left\{ h_{\epsilon,s,1:T}\right\} _{s=1}^{S},\left\{ h_{f,k,1:T}\right\} _{k=1}^{K},\left\{ f_{k,1:T}\right\} _{k=1}^{K}\right)$.

The posterior distribution of the multivariate factor SV model is
\begin{multline}
p\left(\theta,x_{1:T}|y\right)=\left\{ \prod_{s=1}^{S}\prod_{t=1}^{T}p\left(y_{s,t}|\beta_{s},f_{t},h_{\epsilon,s,t},\kappa_{\epsilon,s},\alpha_{\epsilon,s},\psi_{\epsilon,s}\right)p\left\{ h_{\epsilon,s,t}|h_{\epsilon,s,t-1},\kappa_{\epsilon,s},\alpha_{\epsilon,s},\psi_{\epsilon,s}\right\} \right\} \\
\left\{ \prod_{k=1}^{K}\prod_{t=1}^{T}p\left(f_{k,t}|h_{f,k,t},\alpha_{f,k},\psi_{f,k}\right)p\left(h_{f,k,t}|h_{f,k,t-1},\alpha_{f,k},\psi_{f,k}\right)\right\} \\
\left\{ \prod_{s=1}^{S}p\left(\kappa_{\epsilon,s}\right)p\left(\phi_{\epsilon,s}\right)\bigg |\frac{\partial\phi_{\epsilon,s}}{\partial\psi_{\epsilon,s}}\bigg | p\left(\tau_{\epsilon,s}\right)\bigg |\frac{\partial\tau_{\epsilon,s}}{\partial\alpha_{\epsilon,s}}\bigg |\right\} .\\
\left\{ \prod_{k=1}^{K}p\left(\phi_{f,k}\right)\bigg | \frac{\partial\phi_{f,k}}{\partial\psi_{f,k}}\bigg | p\left(\tau_{f,k}\right)\bigg | \frac{\partial\tau_{f,k}}{\partial\alpha_{f,k}}\bigg | \right\} p\left(\beta\right)\prod_{k=1}^{K}\bigg | \frac{\partial\beta_{k,k}}{\partial\delta_{k}}\bigg | .\label{eq:posteriorfactormodel}
\end{multline}
\section{Variational Inference \label{sec:Summary-of-VB approach}}

\subsection{Variational  Bayes Inference \label{subsec:Approximate-Bayesian-Inference}}

Let $\theta$ and $x_{1:T}$ be the vector of parameters and latent
states in the model, respectively. Let $y_{1:T}=\left(y_{1},...,y_{T}\right)^{\top}$
be the observations and consider Bayesian inference for $\theta$ and
$x_{1:T}$ with a prior density $p\left(x_{1:T}|\theta\right)p\left(\theta\right)$.
Denote the density of $y_{1:T}$ conditional on $\theta$ and $x_{1:T}$ by $p\left(y_{1:T}|\theta,x_{1:T}\right)$;
the posterior density is $p\left(\theta,x_{1:T}|y_{1:T}\right)\propto p\left(y_{1:T}|\theta,x_{1:T}\right)p\left(x_{1:T}|\theta\right)p\left(\theta\right)$.
Denote $h\left(\theta,x_{1:T}\right):=p\left(y_{1:T}|\theta,x_{1:T}\right)p\left(x_{1:T}|\theta\right)p\left(\theta\right)$.
We consider the variational density $q_{\lambda}\left(\theta,x_{1:T}\right)$,
indexed by the variational parameter $\lambda$, to approximate $p\left(\theta,x_{1:T}|y_{1:T}\right)$.
The variational Bayes (VB) approach approximates the posterior distribution
of $\theta$ and $x_{1:T}$ by  minimising over $\lambda $ the Kullback-Leibler
(KL) divergence between $q_{\lambda}\left(\theta,x_{1:T}\right)$
and $p\left(\theta,x_{1:T}|y_{1:T}\right)$, i.e.,
\[
\textrm{KL}\left(\lambda\right)=\textrm{KL}\left(q_{\lambda}\left(\theta,x_{1:T}\right)||p\left(\theta,x_{1:T}|y_{1:T}\right)
\right)=\int\log\frac{q_{\lambda}\left(\theta,x_{1:T}\right)}{p\left(\theta,x_{1:T}|y_{1:T}\right)}q_{\lambda}\left(\theta,x_{1:T}\right)d\theta dx_{1:T}.
\]

%
%
%
%
%
%

Minimizing the KL divergence between $q_{\lambda}\left(\theta,x_{1:T}\right)$
and $p\left(\theta,x_{1:T}|y_{1:T}\right)$ is  equivalent
to  maximising a variational lower bound (ELBO) on the log marginal
likelihood $\log p\left(y_{1:T}\right)$ (where $p\left(y_{1:T}\right)=\int p\left(y_{1:T}|\theta,x_{1:T}\right)p\left(x_{1:T}|\theta\right)p\left(\theta\right)d\theta dx_{1:T}$) \citep{Blei2017}.


The ELBO
\begin{equation}
\mathcal{L}\left(\lambda\right)=\int\log\frac{h\left(\theta,x_{1:T}\right)}{q_{\lambda}\left(\theta,x_{1:T}\right)}q_{\lambda}\left(\theta,x_{1:T}\right)d\theta dx_{1:T}
\end{equation}
can be used as a tool for model selection, for example, choosing the number of latent factors in the multivariate factor SV models.
In non-conjugate models, the variational lower bound $\mathcal{L}\left(\lambda\right)$
may not have a closed form solution. When it cannot be evaluated in
closed form, stochastic gradient methods are usually used \citep{Hoffman:2013,Kingma2014,Nott:2012,Paisley:2012,Salimans:2013,Titsias2014,Rezende2014}
to maximise $\mathcal{L}\left(\lambda\right)$.
An initial value $\lambda^{\left(0\right)}$ is updated according
to the iterative scheme
\begin{equation}
\lambda^{\left(j+1\right)}=\lambda^{\left(j\right)}+a_{j}\circ
\widehat{\nabla_{\lambda}\mathcal{L}\left(\lambda^{\left(j\right)}\right)},\label{eq:stochastic gradient update}
\end{equation}
where $\circ$ denotes the Hadamard (element by element) product of
two vectors. Updating \Eq{}\eqref{eq:stochastic gradient update}
is continued until a stopping criterion is satisfied. The $\{a_{j}$,
$j\geq 1\}$ is a sequence of vector valued learning rates, $\nabla_{\lambda}\mathcal{L}\left(\lambda\right)$
is the gradient vector of $\mathcal{L}\left(\lambda\right)$ with
respect to $\lambda$, and $\widehat{\nabla_{\lambda}\mathcal{L}\left(\lambda\right)}$
denotes an unbiased estimate of $\nabla_{\lambda}\mathcal{L}\left(\lambda\right)$.
The learning rate sequence is chosen to satisfy the Robbins-Monro
conditions $\sum_{j}a_{j}=\infty$ and $\sum_{j}a_{j}^{2}<\infty$
\citep{Robbins1951}, which ensures that the iterates $\lambda^{\left(j\right)}$
converge to a local optimum as $j\rightarrow\infty$ under suitable
regularity conditions \citep{Bottou2010}. We consider adaptive learning
rates based on the ADAM approach \citep{Kingma2014a} in our examples.



Reducing the variance of the gradient estimates $\widehat{\nabla_{\lambda}\mathcal{L}\left(\lambda\right)}$
is important to ensure the stability and fast convergence of the algorithm.
Our article uses gradient estimates based on the so-called reparameterisation
trick \citep{Kingma2014,Rezende2014}. To apply this approach, we represent
samples from $q_{\lambda}\left(\theta,x_{1:T}\right)$ as $\left(\theta,x_{1:T}\right)=u\left(\eta,\lambda\right)$,
where $\eta$ is a random vector with a fixed density $p_{\eta}\left(\eta\right)$
that does not depend on the variational parameters. In the case of
a Gaussian variational distribution parameterized in terms of a mean
vector $\mu$ and the Cholesky factor $C$ of its covariance matrix,
we can write $\left(\theta,x_{1:T}\right)=\mu+C\eta$, where $\eta\sim N\left(0,I\right)$.


Then,
\begin{eqnarray*}
\mathcal{L}\left(\lambda\right) & = & E_{q}\left(\log h\left(\theta,x_{1:T}\right)-\log q_{\lambda}\left(\theta,x_{1:T}\right)\right)\\
 & = & E_{p_{\eta}}\left(\log h\left(u\left(\eta,\lambda\right)\right)-\log q_{\lambda}\left(u\left(\eta,\lambda\right)\right)\right).
\end{eqnarray*}
Differentiating under the integral sign,
\begin{eqnarray*}
\nabla_{\lambda}\mathcal{L}\left(\lambda\right) & = & E_{p_{\eta}}\left(\nabla_{\lambda}\log h\left(u\left(\eta,\lambda\right)\right)-\nabla_{\lambda}\log q_{\lambda}\left(u\left(\eta,\lambda\right)\right)\right)\\
 & = & E_{p_{\eta}}\left(\nabla_{\lambda}u\left(\eta,\lambda\right)\left\{ \nabla_{\theta,x_{1:T}}\log h\left(\theta,x_{1:T}\right)-\nabla_{\theta,x_{1:T}}\log q_{\lambda}\left(\theta,x_{1:T}\right)\right\} \right);
\end{eqnarray*}
this is an expectation with respect to $p_{\eta}$ that can be estimated
unbiasedly if it is possible to sample from $p_{\eta}$. Note that the gradient estimates
obtained by the reparameterisation trick use gradient information from
the model; it has been shown empirically that it  greatly reduces
the variance compared to alternative approaches.

The rest of this section is organised
as follows. Section \ref{subsec:Sparse-Cholesky-factorTan2019} discusses
the non-Gaussian sparse Cholesky factor parameterisation of the precision
matrix proposed by \citet{Tan2019}. Section \ref{subsec:Approximation-with-factor Smith2019}
discusses the implicit Gaussian copula variational approximation with a factor
structure for the covariance matrix proposed by \citet{Smith2019}.
Section \ref{subsec:VB-approach-for factor SV model} discusses our
proposed variational approximation for the multivariate factor stochastic
volatility model.

%
%

\subsection{Conditionally Structured  Gaussian Variational Approximation (CSGVA)
\label{subsec:Sparse-Cholesky-factorTan2019}}

When the vector of parameters and latent variables is high dimensional,
taking the variational covariance matrix $\widetilde{\Sigma}$ in
the Gaussian variational approximation to be dense  is computationally
expensive and impractical. An alternative is to assume that the variational
covariance matrix $\widetilde{\Sigma}$ is diagonal, but this loses
any ability to model the dependence structure of the target posterior density.
\citet{Tan2018} consider an approach which parameterises
the precision matrix $\Omega=\widetilde{\Sigma}^{-1}$ in terms of
its Cholesky factor, $\Omega=CC^{\top}$, and impose on it a sparsity structure
that reflects the conditional independence structure in the model.
\citet{Tan2018} note that sparsity is important for reducing the
number of variational parameters that need to be optimised and it
allows the Gaussian variational approximation to be extended to very
high-dimensional settings.

For an $R\times R$
matrix $A$, let $\textrm{vec}\left(A\right)$ be the vector of length
$R^{2}$ obtained by stacking the columns of $A$ under each other
from left to right and $\textrm{vech}\left(A\right)$ be the vector of length
$R\left(R+1\right)/2$ obtained from $\textrm{vec}\left(A\right)$
by removing all the superdiagonal elements of the matrix $A$. Tan, Bhaskaran, and Nott (TBN) extend their previous
approach to allow non-Gaussian variational approximation.
Suppose that $\xi$ is a vector of length $G$ and the $\zeta$ is a vector of length $H$. They consider the variational approximation $q_{\lambda}^{TBN}\left(\xi,\zeta\right)$
of the posterior distribution $p\left(\xi,\zeta|y_{1:T}\right)$
of the form
\begin{equation}
q_{\lambda}^{TBN}\left(\xi,\zeta\right)=q_{\lambda}^{TBN}\left(\xi\right)q_{\lambda}^{TBN}\left(\zeta|\xi\right),\label{eq:TanVA}
\end{equation}
where $q_{\lambda}^{TBN}\left(\xi\right)=N\left(\mu_{G},\Omega_{G}^{-1}\right)$,
$q_{\lambda}^{TBN}\left(\zeta|\xi\right)=N\left(\mu_{L},\Omega_{L}^{-1}\right)$. The $\mu_{G}$ and $\mu_{L}$ are mean vectors with length $G$ and $H$, respectively.
The $\Omega_{G}$ and $\Omega_{L}$ are the precision (inverse covariance)
matrices of variational densities $q_{\lambda}^{TBN}\left(\theta\right)$
and $q_{\lambda}^{TBN}\left(x_{1:T}|\theta\right)$ of orders $G$ and $H$, respectively.
Let $C_{G}C_{G}^{\top}$ and $C_{L}C_{L}^{\top}$ be the unique cholesky
factorisations of $\Omega_{G}$ and $\Omega_{L}$, respectively, where
$C_{G}$ and $C_{L}$ are lower triangular matrices with positive
diagonal entries. We now explain the idea that imposing sparsity in
$C_{G}$ and $C_{L}$ reflects the conditional independence relationship
in the precision matrix $\Omega_{G}$ and $\Omega_{L}$, respectively.
For example, for a Gaussian distribution, $\Omega_{L,i,j}=0$, corresponds
to variables $i$ and $j$ being conditionally independent given the
rest. If $C_{L}$ is a lower triangular matrix, Proposition~1
of \citet{Rothman2010} states that if $C_{L}$ is row banded then
$\Omega_{L}$ has the same row-banded structure.

To allow for unconstrained optimisation of the variational parameters,
they define $C_{G}^{*}$ and $C_{L}^{*}$ to also be lower triangular
matrices such that $C_{G,ii}^{*}=\log\left(C_{G,ii}\right)$ and $C_{G,ij}^{*}=C_{G,ij}$
if $i\neq j$. The $C_{L}^{*}$ are defined similarly. In their
parameterisation, the $\mu_{L}$ and $\textrm{vech}\left(C_{L}^{*}\right)$ are
linear functions of $\xi$:
\begin{equation}
\mu_{L}=d+C_{L}^{-\top}D\left(\mu_{G}-\xi\right),\qquad \textrm{vech}\left(C_{L}^{*}\right)=f^{*}+F\xi;\label{eq:muL}
\end{equation}
where, $d$ is a vector of length $H$, $D$ is a $H\times G$ matrix,
$f^{*}$ is a vector of length $H\left(H+1\right)/2$, $F$ is a $H\left(H+1\right)/2\times G$
matrix, and $G$ is the length of vector parameter $\xi$. The
parameterisation of $q_{\lambda}^{TBN}\left(\xi,\zeta\right)$
in \Eq~\eqref{eq:TanVA} is Gaussian if and only if $F=0$ in
\Eq~\eqref{eq:muL}. The set of variational parameters to be optimised
is denoted as
\begin{equation}
\lambda=\left\{ \mu_{G}^{\top},v\left(C_{G}^{*}\right)^{\top},d^{\top},\textrm{vec}\left(D\right)^{\top},f^{*\top},\textrm{vec}\left(F\right)^{\top}\right\} ^{\top}.
\end{equation}
The closed form reparameterisation gradients in \citet{Tan2019} can
be used directly.

%
%
%


\subsection{Implicit Copula Variational Approximation through transformation
\label{subsec:Approximation-with-factor Smith2019}}

Another way to parameterise the dependence structure parsimoniously
in a variational approximation is to use a sparse factor parameterisation
for the variational covariance matrix $\widetilde{\Sigma}$. This
factor parameterisation is very useful when the prior and the density of $y_{1:T}$ conditional on
$\theta$ and $x_{1:T}$ do not have a special structure.
 The variational
covariance matrix is parameterised in terms of a low-dimensional latent
factor structure. The number of variational parameters to be optimised
is reduced when the number of latent factors is much less than the
number of parameters in the model. \citet{Ong2018} assume that the
Gaussian variational approximation with sparse factor covariance structure
$q_{\lambda}^{ONS}\left(\theta,x_{1:T}\right)=N\left(\mu,BB^{\top}+D^{2}\right)$,
where $\mu$ is the $R\times 1$ mean vector, $B$ is a $R\times p$ full rank
matrix $p\lll R$ and $D$ is a diagonal matrix with diagonal elements
$d=\left(d_{1},...,d_{R}\right)$, where $R$ is the total number
of parameters and latent states. \citet{Smith2019} extend their previous
approach to go beyond Gaussian variational approximation using the implicit
copula method, but still use the sparse factor parameterisation for
the variational covariance matrix. They propose an implicit copula based
on Gaussian and skew-Normal copulas with Yeo-Johnson \citep{Yeo2000}
and G\&H families \citep{Tukey1977} transformations for the marginals. Our article
uses a Gaussian copula with a factor structure for the variational covariance
matrix with Yeo-Johnson (YJ) transformations for the marginals.

Let $t_{\gamma}$ be a Yeo-Johnson one to one transformation onto
the real line with parameter vector $\gamma$. Then,
each parameter and latent state is transformed  as $\xi_{\theta_{g}}=t_{\gamma_{\theta_{g}}}\left(\theta_{g}\right)$, for $g=1,...,G$,
and $\xi_{x_{t}}=t_{\gamma_{x_{t}}}\left(x_{t}\right)$ for $t=1,...,T$. Let $R$ be the total number of parameters and state variables. We use a multivariate
normal distribution function with mean $\mu_{\xi}$ and covariance
matrix $\Sigma_{\xi}$, $F\left(\xi;\mu_{\xi},\Sigma_{\xi}\right)=\Phi_{R}\left(\xi;\mu_{\xi},\Sigma_{\xi}\right)$,
where $\xi=\left(\xi^{\top}_{\theta_{1:G}},\xi^{\top}_{x_{1:T}}\right)^{\top}$ and $\mu_{\xi}$ is the $R \times 1$ mean vector. We use a factor structure \citep{Ong2018} for the covariance matrix
$\Sigma_{\xi}=B_{\xi}B_{\xi}^{\top}+D_{\xi}^{2}$, where $B_{\xi}$ is
an $R\times p$ full rank matrix $\left(p\ll R\right)$, with the upper
triangle  of $B_{\xi}$ set to zero for identifiability;
$D_{\xi}$ is a diagonal matrix with diagonal elements $d_{\xi}=\left(d_{\xi,1},...,d_{\xi,R}\right)$.
If $p\left(\xi;\mu_{\xi},\Sigma_{\xi}\right)=(\partial^{R}/\partial\xi_{1}...\partial\xi_{R})F\left(\xi;\mu_{\xi},\Sigma_{\xi}\right)$
is the multivariate normal density with mean $\mu_{\xi}$ and covariance
matrix $\Sigma_{\xi}$, then the density of the variational approximation
is
\begin{equation*}
q_{\lambda}^{SRN}\left(\theta\right)=p\left(\xi;\mu_{\xi},\Sigma_{\xi}\right)
\prod_{i=1}^{R}t_{\gamma_{i}}^{'}\left(\theta_{i}\right);
\end{equation*}
the variational parameters are $\lambda=\left(\gamma_{1}^{\top},...,\gamma_{R}^{\top},\mu_{\xi}^{\top},\textrm{vech}\left(B_{\xi}\right)^{\top},d_{\xi}^{\top}\right)$. We generate $\xi\sim N\left(\mu_{\xi},B_{\xi}B_{\xi}^{T}+D_{\xi}^{2}\right)$ and then
$\theta_{g}=t_{\gamma_{\theta_{g}}}^{-1}\left(\xi_{\theta_{g}}\right)$ for $g=1,...,G$ and
$x_{t}=t_{\gamma_{x_{t}}}^{-1}\left(\xi_{x_{t}}\right)$ for $t=1,...,T$. Constrained parameters are transformed to the real line.
\citet{Smith2019} give details of the Yeo-Johnson transformation, its inverse, and derivatives with respect
to the model parameters and the closed form reparameterisation gradients.
The inverse of the dense $R\times R$
matrix $\left(B_{\xi}B_{\xi}^{\top}+D_{\xi}^{2}\right)$ is required for the gradient estimate;
it is efficiently computed using the Woodbury formula
\[
\left(B_{\xi}B_{\xi}^{\top}+D_{\xi}^{2}\right)^{-1}=D_{\xi}^{-2}-
D_{\xi}^{-2}B_{\xi}\left(I+B_{\xi}^{\top}D_{\xi}^{-2}B_{\xi}\right)^{-1}B_{\xi}^{\top}D_{\xi}^{-2}.
\]

\subsection{Variational Approximation for multivariate factor SV model \label{subsec:VB-approach-for factor SV model}}

This section discusses our variational approach for approximating
the posterior of the multivariate factor SV model described in Section
\ref{sec:Description-of-Factor SV model}, which aims to select
variational approximations that balance the accuracy and the computational
cost. We use the non-Gaussian sparse Cholesky factor parameterisation
of the precision matrix (CSGVA) defined in Section \ref{subsec:Sparse-Cholesky-factorTan2019}
and the Gaussian copula factor parameterisation of the variational
covariance matrix defined in Section \ref{subsec:Approximation-with-factor Smith2019}
as the main building blocks to approximate the posterior of the multivariate
factor SV model. The CSGVA is used to approximate the idiosyncratic and factor log-volatilities because of their AR1 structures in Eqs. \eqref{eq:statetransitionfactor} and \eqref{eq:statetransitionidiosyncratic-1}, respectively. The implicit Gaussian copula variational approximation is used to approximate the latent factors and the factor loading matrix because they do not have a special structure.

%
%

\begin{doublespace}
The parameters and latent states of the idiosyncratic and factor log-volatilities are defined as  $\theta_{G,\epsilon,s} \coloneqq\left(\kappa_{\epsilon,s},\alpha_{\epsilon,s},\psi_{\epsilon,s}\right)$ with length $G_{\epsilon}$;
$h_{\epsilon,s,1:T}$ with length $T$;  $\theta_{G,f,k}\coloneqq\left(\alpha_{f,k},\psi_{f,k}\right)$ with length $G_{f}$;
and $h_{f,k,1:T}$, with length $T$  for $s=1,...,S$ and $k=1,...,K$.
From \Eq{} \eqref{eq:posteriorfactormodel}, $h_{\epsilon,s,t}$
is conditionally independent of all the other states in the posterior
distribution,  given the parameters $\left(\kappa_{\epsilon,s},\alpha_{\epsilon,s},\psi_{\epsilon,s}\right)$
and the neighbouring states $h_{\epsilon,s,t-1}$ and $h_{\epsilon,s,t+1}$
for $s=1,...,S$; $h_{f,k,t}$ is conditionally independent of
all other states in the posterior distribution given, the parameters
$\left(\alpha_{f,k},\psi_{f,k}\right)$ and the neighbouring states
$h_{f,k,t-1}$ and $h_{f,k,t+1}$ for $k=1,...,K$.

It is therefore
reasonable to take advantage of this conditional independence structure
in the variational approximations, by letting $$q_{\lambda_{\epsilon,s}}^{TBN}\left(\theta_{G,\epsilon,s},
h_{\epsilon,s,1:T}\right)=q_{\lambda_{\epsilon,s}}^{TBN}\left(\theta_{G,\epsilon,s}\right)
q_{\lambda_{\epsilon,s}}^{TBN}\left(h_{\epsilon,s,1:T}|\theta_{G,\epsilon,s}\right), s=1,...,S$$
and $$q_{\lambda_{f,k}}^{TBN}\left(\theta_{G,f,k},h_{f,k,1:T}\right)
=q_{\lambda_{{f,k}}}^{TBN}\left(\theta_{G,f,k}\right)q_{\lambda_{{f,k}}}^{TBN}
\left(h_{f,k,1:T}|\theta_{G,f,k}\right), k=1,...,K.$$
The sparsity structure imposed on $\Omega_{L,\epsilon,s}$
and $C_{L,\epsilon,s}$ for $s=1,...,S$  is{\scriptsize{}
\[
\Omega_{L,\epsilon,s}=\left[\begin{array}{ccccc}
\Omega_{L,\epsilon,s,11} & \Omega_{L,\epsilon,s,12} & 0 & \cdots & 0\\
\Omega_{L,\epsilon,s,21} & \Omega_{L,\epsilon,s,22} & \Omega_{L,\epsilon,s,23} & \cdots & 0\\
0 & \Omega_{L,\epsilon,s,32} & \Omega_{L,\epsilon,s,33} & \cdots & 0\\
\vdots & \vdots & \vdots & \ddots & \vdots\\
0 & 0 & 0 & \cdots & \Omega_{L,\epsilon,s,TT}
\end{array}\right],
\]
\[
C_{L,\epsilon,s}=\left[\begin{array}{ccccc}
C_{L,\epsilon,s,11} & 0 & 0 & \cdots & 0\\
C_{L,\epsilon,s,21} & C_{L,\epsilon,s,22} & 0 & \cdots & 0\\
0 & C_{L,\epsilon,s,32} & C_{L,\epsilon,s,33} & \cdots & 0\\
\vdots & \vdots & \vdots & \ddots & \vdots\\
0 & 0 & 0 & \cdots & C_{L,\epsilon,s,TT}
\end{array}\right].
\]
}The number of non-zero elements in $\textrm{vech}\left(C_{L,\epsilon,s}^{*}\right)$
is $2T-1$. If we set $f^{*}_{\epsilon,s,i}=0$ and $F_{\epsilon,s,ij}=0$
for all indices $i$ in $\textrm{vech}\left(C_{L,\epsilon,s}^{*}\right)$ which
are fixed at zero, then the number of variational parameters to be
optimised is reduced from $T\left(T+1\right)/2$ to $2T-1$ for $f^{*}_{\epsilon,s,i}$,
and from $T\left(T+1\right)G_{\epsilon}/2$ to $\left(2T-1\right)G_{\epsilon}$ for $F_{\epsilon,s,i}$.
Similar sparsity structures are imposed on $\Omega_{L,f,k}$ and $C_{L,f,k}$
for $k=1,...,K$. We can use the Gaussian copula with a factor structure
for the factor loading matrix $\beta$ and the latent factors $\left(f_{1:T}\right)$
that do not have any special structures. We now propose three variational
approximations for  the posterior distribution of the
multivariate factor SV model. The first variational approximation is
\[
q_{\lambda}^{I}\left(\theta,x_{1:T}\right):=\prod_{s=1}^{S}q_{\lambda_{\epsilon,s}}^{TBN}\left(\theta_{G,\epsilon,s},h_{\epsilon,s,1:T}\right)\prod_{k=1}^{K}q_{\lambda_{f,k}}^{TBN}\left(\theta_{G,f,k},h_{f,k,1:T}\right)q_{\lambda_{f_{k,1:T}}}^{SRN}\left(f_{k,1:T}\right)q_{\lambda_{\beta}}^{SRN}\left(\beta\right).
\]
Our empirical work uses $p=0$ factors for the variational density
$q_{\lambda_{f_{k,1:T}}}^{SRN}\left(f_{k,1:T}\right)$ and $p=4$ factors
for the variational density $q_{\lambda_{\beta}}^{SRN}\left(\beta\right)$.
The first VB approximation $q_{\lambda}^{I}\left(\theta,x_{1:T}\right)$
ignores the posterior dependence between $(\theta_{G,\epsilon,s},h_{\epsilon,s,1:T})$ and
$(\theta_{G,\epsilon,j},h_{\epsilon,j,1:T})$ for $s\neq j$ and the posterior dependence
of $\theta_{G,\epsilon,s}$ and $h_{\epsilon,s,1:T}$ with other parameters
and latent states $\left(\left\{ \theta_{G,f,k},h_{f,k,1:T}\right\} _{k=1}^{K},\beta,\left\{ f_{k,1:T}\right\} _{k=1}^{K}\right)$
for $s=1,...,S$;  ignores the posterior dependence between $(\theta_{G,f,k},h_{f,k,1:T})$
and $(\theta_{G,f,j},h_{f,j,1:T})$ for $k\neq j$ and the posterior dependence of
$\theta_{G,f,k}$ and $h_{f,k,1:T}$ with other parameters $\left(\left\{ \theta_{G,\epsilon,s},h_{\epsilon,s,1:T}\right\} _{s=1}^{S},\beta,\left\{ f_{k,1:T}\right\} _{k=1}^{K}\right)$;
and ignores the dependence between the latent factors $f_{k,1:T}$
and $f_{j,1:T}$ for $k\neq j$ and the dependence between the latent
factors $f_{1:T}$ and $\beta$ as well as ignores the dependence in the $k$th latent factors over time, $f_{k,t}$, for $t=1,...,T$.
\end{doublespace}

%
%
%
%
%

The second variational approximation
is
\[
q_{\lambda}^{II}\left(\theta,x_{1:T}\right)=\prod_{s=1}^{S}
q_{\lambda_{\epsilon,s}}^{TBN}\left(\theta_{G,\epsilon,s},h_{\epsilon,s,1:T}\right)\prod_{k=1}^{K}q_{\lambda_{f,k}}^{TBN}
\left(\theta_{G,f,k},h_{f,k,1:T}\right)q_{\lambda_{\beta,f_{k,1:T}}}^{SRN}\left(f_{k,1:T},\beta_{.,k}\right),
\]
where $\beta_{.,k}=\left(\beta_{k,k},...,\beta_{S,k}\right)^{\top}$.
The second variational approximation $q_{\lambda}^{II}\left(\theta\right)$
takes into account the dependence in the $k$th latent factor and between $f_{k,1:T}$ and $\beta_{\cdot,k}$,
but it ignores the dependence between $\left(f_{k,1:T},\beta_{\cdot,k}\right)$
and $\left(f_{j,1:T},\beta_{\cdot,j}\right)$ for $k\neq j$.
The empirical work  below uses $p=4$ for the variational density $q_{\lambda_{\beta,f_{k,1:T}}}^{SRN}\left(f_{k,1:T},\beta_{\cdot,k}\right)$.

The posterior distribution of the FSV model can be written as
\begin{equation}
p\left(\theta,x_{1:T}|y\right)=p\left(f_{1:T}|\theta,x_{1:T,-f_{1:T}},y\right)p\left(\theta,x_{1:T,-f_{1:T}}|y\right),\label{eq:posteriortdist-1}
\end{equation}
where $x_{1:T,-f_{1:T}}$ is all the latent variables in the FSV models,
except the latent factors $\left(f_{k,1:T}\right)$ for $k=1,...,K$.
The full conditional distribution $p\left(f_{1:T}|\theta,x_{1:T,-f_{1:T}},y\right)$
is available in closed form as
\[
p\left(f_{k,t}|\theta,x_{1:T,-f_{1:T}},y\right)=N\left(\mu_{f_{k,t}},\Sigma_{f_{k,t}}\right)\;\textrm{for }k=1,...,K\;\textrm{and}\;\textrm{for }t=1,...,T,
\]
where $\mu_{f_{k,t}}=\Sigma_{f_{k,t}}\beta^{\top}\left(V_{t}^{-1}y_{t}\right)$
and $\Sigma_{f_{k,t}}=\left(\beta V_{t}^{-1}\beta^{\top}+D_{t}^{-1}\right)^{-1}$.
It is unnecessary to approximate the posterior distribution of the
latent factors because it is easy to sample from the full conditional
distribution $p\left(f_{1:T}|\theta,x_{1:T,-f_{1:T}},y\right)$. Similar
ideas are explored by \citet{Maya2020}. Thus, the third variational
approximation is
\begin{equation}
q_{\lambda}^{III}\left(\theta,x_{1:T}\right)=p\left(f_{1:T}|\theta,x_{1:T,-f_{1:T}},y\right)q_{\lambda}^{*}\left(\theta,x_{1:T,-f_{1:T}}|y\right),
\end{equation}
where
\begin{equation}
q_{\lambda}^{*}\left(\theta,x_{1:T,-f_{1:T}}\right)=\prod_{s=1}^{S}q_{\lambda_{\epsilon,s}}^{TBN}\left(\theta_{\epsilon,s},h_{\epsilon,s,1:T}\right)\prod_{k=1}^{K}q_{\lambda_{f,k}}^{TBN}\left(\theta_{G,f,k},h_{f,k,1:T}\right)q_{\lambda_{\beta}}^{SRN}\left(\beta\right).
\end{equation}
Section~\ref{sec:Reparameterisation-Gradients-for q3} of the supplement discusses
the reparameterisation gradient for the variational approximation
$q_{\lambda}^{III}\left(\theta,x_{1:T}\right)$. This variational approximation takes into account the dependence between the latent factors $f_{t}$ and the other parameters and latent state variables in the FSV model.

Algorithm 1 below discusses the updates for the variational approximation
$q_{\lambda}^{I}\left(\theta,x_{1:T}\right)$. It is straightforward
to modify it to obtain the updates for the variational
approximations $q_{\lambda}^{II}\left(\theta,x_{1:T}\right)$ and
$q_{\lambda}^{III}\left(\theta,x_{1:T}\right)$; Section~\ref{sec:Reparameterisation-Gradients-for q3} gives these updates. Sections~\ref{sec:Deriving-the-Gradients for FSV Models} and \ref{variational approximation and gradient FSV with t distribution} give the required gradients for the FSV with normal and t-distributions, respectively.

\subsection*{Algorithm 1}
Initialise all the variational parameters $\lambda$. At each iteration
$j$, (1) Generate Monte Carlo samples for all the parameters and
latent states from their variational distributions, (2) Compute the
unbiased estimates of gradient of the lower bound with respect to
each of the variational parameter and update the variational parameters
using Stochastic Gradient methods. The ADAM method to set the learning
rates is given in the Section~\ref{subsec:Learning-Rate-1}.

\subsection*{STEP 1 }
\begin{itemize}
\item \textbf{For} $s=1:S$,
\begin{itemize}
\item Generate Monte Carlo samples for $\theta_{G,\epsilon,s}$ and $h_{\epsilon,s,1:T}$
from the variational distribution $q_{\lambda_{\epsilon,s}}^{TBN}\left(\theta_{G,\epsilon,s},h_{\epsilon,s,1:T}\right)$
\begin{enumerate}
\item Generate $\eta_{\theta_{G,\epsilon,s}}\sim N\left(0,I_{G_{\epsilon}}\right)$
and $\eta_{h_{\epsilon,s,1:T}}\sim N\left(0,I_{T}\right)$, where
$G_{\epsilon}$ is the number of parameters of idiosyncratic log-volatilities
\item Generate $\theta_{G,\epsilon,s}=\mu_{G,\epsilon,s}^{\left(j\right)}+\left(C_{G,\epsilon,s}^{\left(j\right)}\right)^{-\top}\eta_{\theta_{G,\epsilon,s}}$
and $h_{\epsilon,s,1:T}=\mu_{L,\epsilon,s}^{\left(j\right)}+\left(C_{L,\epsilon,s}^{\left(j\right)}\right)^{-\top}\eta_{h_{\epsilon,s,1:T}}$.
\end{enumerate}
\end{itemize}
\item \textbf{For} $k=1:K$,
\begin{itemize}
\item Generate Monte Carlo samples for $\theta_{G,f,k}$ and $h_{f,k,1:T}$
from the variational distribution $q_{\lambda_{f,k}}^{TBN}\left(\theta_{G,f,k},h_{f,k,1:T}\right)$
\begin{enumerate}
\item Generate $\eta_{\theta_{G,f,k}}\sim N\left(0,I_{G_{f}}\right)$ and
$\eta_{h_{f,k,1:T}}\sim N\left(0,I_{T}\right)$, where $G_{f}$ is
the number of parameters of factor log-volatilities
\item Generate $\theta_{G,f,k}=\mu_{G,f,k}^{\left(j\right)}+\left(C_{G,f,k}^{\left(j\right)}\right)^{-\top}\eta_{\theta_{G,f,k}}$
and $h_{f,k,1:T}=\mu_{L,f,k}^{\left(j\right)}+\left(C_{L,f,k}^{\left(j\right)}\right)^{-\top}\eta_{h_{f,k,1:T}}$.
\end{enumerate}
\item Generate Monte Carlo samples for $f_{k:1:T}$ from the variational
distribution $q_{\lambda_{f_{k,1:T}}}^{SRN}\left(f_{k,1:T}\right)$
\begin{enumerate}
\item Generate $\left(\eta_{f_{k}}\right)\sim N\left(0,I_{T}\right)$ and
calculating $\xi_{f_{k}}=\mu_{\xi_{f_{k}}}^{\left(j\right)}+d_{\xi_{f_{k}}}^{\left(j\right)}\circ\eta_{f_{k}}$.
\item Generate $f_{k,t}=t_{\gamma_{f_{k,t}}}^{-1{\left(j\right)}}\left(\xi_{f_{k,t}}\right)$,
for $t=1,...,T$.
\end{enumerate}
\end{itemize}
\item Generate Monte Carlo samples for the factor loading $\beta$ from
the variational distribution
\begin{enumerate}
\item Generate $z_{\beta}\sim N\left(0,I_{p}\right)$ and $\eta_{\beta}\sim N\left(0,I_{R_{\beta}}\right)$
calculating $\xi_{\beta}=\mu_{\xi_{\beta}}^{\left(j\right)}+B_{\xi_{\beta}}^{\left(j\right)}z_{\beta}+d_{\xi_{\beta}}^{\left(j\right)}\circ\eta_{\beta}$,
and $R_{\beta}$ is the total number of parameters in the factor loading
matrix $\beta$.
\item Generate $\beta_{i}=t_{\gamma_{\beta_{i}}}^{-1{\left(j\right)}}\left(\xi_{\beta_{i}}\right)$,
for $i=1,...,R_{\beta}$.
\end{enumerate}
\end{itemize}

\subsection*{STEP 2}
\begin{itemize}
\item \textbf{For} $s=1:S$, Update the variational parameters $\lambda_{\epsilon,s}$
of the variational distributions $q_{\lambda_{\epsilon,s}}^{TBN}\left(\theta_{\epsilon,s},h_{\epsilon,s,1:T}\right)$
\begin{enumerate}
\item Construct unbiased estimates of gradients $\widehat{\nabla_{\lambda_{\epsilon,s}}\mathcal{L}\left(\lambda_{\epsilon,s}\right)}$.
\item Set adaptive learning rates $a_{j,\epsilon,s}$ using ADAM method.
\item Set $\lambda_{\epsilon,s}^{\left(j+1\right)}=\lambda_{\epsilon,s}^{\left(j\right)}+a_{j,\epsilon,s}\widehat{\nabla_{\lambda_{\epsilon,s}}\mathcal{L}\left(\lambda_{\epsilon,s}\right)}^{\left(j\right)}$,

where $\lambda_{\epsilon,s}=\left\{ \mu_{G,\epsilon,s}^{\top},\textrm{vech}\left(C_{G,\epsilon,s}^{*}\right)^{\top},d_{\epsilon,s}^{\top},\textrm{vec}\left(D_{\epsilon,s}\right)^{\top},f_{\epsilon,s}^{*\top},\textrm{vec}\left(F_{\epsilon,s}\right)^{\top}\right\} ^{\top}$
as defined in Section \ref{subsec:Sparse-Cholesky-factorTan2019}.
\end{enumerate}
\item For $k=1:K$,
\begin{itemize}
\item Update the variational parameters $\lambda_{f,k}$ of the variational
distributions $q_{\lambda_{f,k}}^{TBN}\left(\theta_{G,f,k},h_{f,k,1:T}\right)$
\begin{enumerate}
\item Construct unbiased estimates of gradients $\widehat{\nabla_{\lambda_{f,k}}\mathcal{L}\left(\lambda_{f,k}\right)}$.
\item Set adaptive learning rates $a_{j,f,k}$ using ADAM method.
\item Set $\lambda_{f,k}^{\left(j+1\right)}=\lambda_{f,k}^{\left(j\right)}+a_{j,f,k}
    \widehat{\nabla_{\lambda_{f,k}}\mathcal{L}\left(\lambda_{f,k}\right)}^{\left(j\right)}$,
where $\lambda_{f,k}=\left\{ \mu_{G,f,k}^{\top},\textrm{vech}\left(C_{G,f,k}^{*}\right)^{\top},d_{f,k}^{\top},\textrm{vec}\left(D_{f,k}\right)^{\top},f_{f,k}^{*\top},\textrm{vec}\left(F_{f,k}\right)^{\top}\right\} ^{\top}$.
\end{enumerate}
\item Update the variational parameters $\lambda_{f_{k,1:T}}$ of the variational
distributions $q_{\lambda_{f_{k,1:T}}}^{SRN}\left(f_{k,1:T}\right)$
\begin{enumerate}
\item Construct unbiased estimates of gradients $\widehat{\nabla_{\lambda_{f_{k}}}\mathcal{L}\left(\lambda_{f_{k}}\right)}$.
\item Set adaptive learning rates $a_{j,f_{k}}$ using ADAM method.
\item Set $\lambda_{f_{k}}^{\left(j+1\right)}=\lambda_{f_{k}}^{\left(j\right)}+a_{j,f_{k}}\widehat{\nabla_{\lambda_{f_{k}}}\mathcal{L}\left(\lambda_{f_{k}}\right)}^{\left(j\right)}$,
where $\lambda_{f_{k}}=\left(\gamma_{f_{k,1}}^{\top},...,\gamma_{f_{k,T}}^{\top},\mu_{\xi_{f_{k}}}^{\top},d_{\xi_{f_{k}}}^{\top}\right)^{\top}$.
\end{enumerate}
\end{itemize}
\item Update the variational parameters $\lambda_{\beta}$ from the variational
distribution $q_{\lambda_{\beta}}^{SRN}\left(\beta\right)$.
\begin{enumerate}
\item Construct unbiased estimates of gradients $\widehat{\nabla_{\lambda_{\beta}}\mathcal{L}\left(\lambda_{\beta}\right)}$.
\item Set adaptive learning rates $a_{j,\beta}$ using ADAM method.
\item Set $\lambda_{\beta}^{\left(j+1\right)}=\lambda_{\beta}^{\left(j\right)}+a_{j,\beta}\widehat{\nabla_{\lambda_{\beta}}\mathcal{L}\left(\lambda_{\beta}\right)}^{\left(j\right)}$,
where $\lambda_{\beta}=\left(\gamma_{\beta_{1}}^{\top},...,\gamma_{\beta_{R_{\beta}}}^{\top},\mu_{\xi_{\beta}}^{\top},\textrm{vech}(B_{\xi_{\beta}})^{\top},d_{\xi_{\beta}}^{\top}\right)^{\top}$.
\end{enumerate}
\end{itemize}

%

Note that the updates for the variational parameters $\lambda_{\epsilon,s}$, $\lambda_{f,k}$, and  $\lambda_{f_{k,1:T}}$ are easily parallelised for $s=1,...,S$ and $k=1,...,K$.

\subsection{Sequential Variational Approximation \label{sec:Sequential-Updating-Variational Approximation}}
This section extends the batch variational approach  above  to sequential variational approximation of the latent states and parameters as new observations arrive. In this context, both MCMC and particle MCMC can become very expensive and
time consuming as both methods need to repeatedly run the full
MCMC  as new observations arrive. In this setting, it is necessary  to update the joint posterior
of the latent states and parameters sequentially, to take into account the newly arrived data.
In more detail, let $1$, $2$,...,$t$ be a sequence
of time points, $x_{1:t}$ and $\theta$ be the latent states and parameters
in the model, respectively. The goal is to estimate a sequence of posterior
distributions of states and parameters, $p\left(\theta,x_{1}|y_{1}\right)$,
$p\left(\theta,x_{1:{2}}|y_{1:{2}}\right)$, ..., $p\left(\theta,x_{1:T}|y_{1:T}\right)$,
where $T$ is the total length of the time series.
The sequential algorithm, which we call, seq-VA, first estimates the
variational approximation at time $t-1$ as $q_{\lambda^{*}}\left(\theta,x_{1:t-1}\right)$.
That is, we approximate the posterior $p\left(\theta,x_{1:t-1}|y_{1:t-1}\right)$
by $q_{\lambda}\left(\theta,x_{1:t-1}\right)$. Then, after
observing the additional data at time $t$, the seq-VA approximates the posterior $p\left(\theta,x_{1:t}|y_{1:t}\right)$
by $q_{\lambda}\left(\theta,x_{1:t}\right)$. The optimal value of the variational parameters $\lambda$ at time $t-1$ can be used as the initial values of the variational parameters $\lambda$ at time $t$; Section~\ref{sec:Simulation-Study-and real data application} shows that this may substantially reduce the number of iterations required for the algorithm to converge. The seq-VA can start if only part of the data has been observed.

\citet{Tomasetti2019} propose an alternative sequential updating variational method and  apply it to a simple univariate autoregressive model with a tractable likelihood. With the
availability of the posterior at time $t-1$, given by its posterior
density $p\left(\theta,x_{1:t-1}|y_{1:t-1}\right)$, the updated (exact)
posterior at time $t$ is 
\begin{equation}
p\left(\theta,x_{1:t}|y_{1:t}\right)\propto p\left(y_{t}|\theta,x_{t}\right)p\left(x_{t}|x_{t-1},\theta\right)p\left(\theta,x_{1:t-1}|y_{1:t-1}\right).
\label{eq:updatetrueposterior}
\end{equation}
\citeauthor{Tomasetti2019} replace the posterior construction defined in \Eq{} \eqref{eq:updatetrueposterior},
with the available approximation $q_{\lambda}\left(\theta,x_{1:t-1}\right)$,
\begin{equation}
\widehat{p}\left(\theta,x_{1:t}|y_{1:t}\right)\propto p\left(y_{t}|\theta,x_{t}\right)p\left(x_{t}|x_{t-1},\theta\right)q_{\lambda^{*}}\left(\theta,x_{1:t-1}\right).\label{eq:approximateposterior}
\end{equation}
This sequential approach targets the \lq approximate posterior\rq{}
 at time $t$, and not the true posterior.  
  In our experience, this causes a loss of accuracy relative to the sequential update described above, which targets the true posterior.
A potential disadvantage of the proposed new sequential approach is that the likelihoods need to be computed from time $1$ to $t$.

\section{Variational Forecasting \label{subsec:VariationalForecasting}}

Let $Y_{T+1}$ be the  unobserved value of the dependent variable at time
$T+1$. Given the joint posterior distribution of all the parameters
$\theta$ and latent state variables $x_{1:T}$
of the multivariate factor SV model up to time $T$,
the forecast density of $Y_{T+1}=y_{T+1}$ that accounts for the uncertainty about $\theta$ and $x_{1:T}$ is
\begin{equation}
p\left(y_{T+1}|y_{1:T}\right)=\int p\left(y_{T+1}|\theta,x_{T+1}\right)p\left(x_{T+1}|x_{T},\theta\right)p\left(\theta,x_{1:T}|y_{1:T}\right)d\theta dx_{1:T}; \label{eq:predictivedenseqn}
\end{equation}
$p\left(\theta,x_{1:T}|y_{1:T}\right)$ is the exact posterior
density that can be estimated  from MCMC or particle MCMC.
The draws from MCMC or particle MCMC can be used to produce the simulation-consistent
estimate of this predictive density
\begin{equation}
\widehat{p\left(y_{T+1}|y_{1:T}\right)}=\frac{1}{M}\sum_{m=1}^{M}
p\left(y_{T+1}|y_{1:T},\theta^{\left(m\right)},x_{T+1}^{\left(m\right)}\right);\label{eq:preddensRB}
\end{equation}
it is necessary to know the conditional density
$p\left(y_{T+1}|y_{1:T,}\theta^{\left(m\right)},x_{T+1}^{\left(m\right)}\right)$
in closed form to obtain the estimates in
\Eq~\eqref{eq:preddensRB}.  If Eq. \eqref{eq:preddensRB} is
evaluated at the observed value $y_{T+1}=y_{T+1}^{obs}$, it is commonly
referred to as one-step ahead predictive likelihood at time $T+1$,
denoted $PL_{T+1}$. Alternatively, we can obtain $M$ draws
of $y_{T+1}$ from $p\left(y_{T+1}|y_{1:T,}\theta^{\left(m\right)},x_{T+1}^{\left(m\right)}\right)$
that can be used to obtain the kernel density estimate $\widehat{p\left(y_{T+1}|y_{1:T}\right)}$
of $p\left(y_{T+1}|y_{1:T}\right)$.
Assuming that the MCMC has converged, these are two simulation-consistent estimates
of the exact \dlq Bayesian\drq{} predictive density.

The motivation for using the variational approximation in this setting
is that obtaining the exact posterior density $p\left(\theta,x_{1:T}|y_{1:T}\right)$
is expensive for high dimensional multivariate factor SV models. To obtain an approximation of the posterior density $q_{\lambda}\left(\theta,x_{1:T}|y_{1:T}\right)$
is usually substantially faster than MCMC or particle MCMC.
Given that we have the variational approximation of the joint posterior
distribution of the parameters and latent variables, we can define
\begin{equation}
g\left(y_{T+1}|y_{1:T}\right)=\int p\left(y_{T+1}|\theta,x_{T+1}\right)p\left(x_{T+1}|x_{T},\theta\right)q_{\lambda}\left(\theta,x_{1:T}|y_{1:T}\right)d\theta dx_{1:T},\label{eq:predictivedenseqnVB}
\end{equation}
where $q_{\lambda}\left(\theta,x_{1:T}|y_{1:T}\right)$ is the variational
approximation to the joint posterior distribution of the parameters
and latent variables of the multivariate factor SV model. As with
most quantities in Bayesian analysis, computing the approximate predictive
density in \Eq~\eqref{eq:predictivedenseqnVB} can be challenging
because it involves high dimensional integrals which cannot be solved
analytically. However, it can be approximated through Monte Carlo
integration,
\begin{equation}
\widehat{g\left(y_{T+1}|y_{1:T}\right)}\approx\frac{1}{M}\sum_{m=1}^{M}p\left(y_{T+1}|y_{1:T},\theta^{\left(m\right)},x_{T+1}^{\left(m\right)}\right),
\label{eq:predictivedenseqnVB2}
\end{equation}
where $\theta^{\left(m\right)}$ and $x_{1:T}^{\left(m\right)}$ denotes
the $m$th draw from the variational approximation $q_{\lambda}\left(\theta,x_{1:T}|y_{1:T}\right)$
up to time $T$. We can obtain the approximate
Bayesian predictive density $\widehat{g\left(y_{T+1}|y_{1:T}\right)}$
by generating $M$ draws of $y_{T+1}$ from $p\left(y_{T+1}|y_{1:T,}\theta^{\left(m\right)},x_{T+1}^{\left(m\right)}\right)$
that can be used to obtain the kernel density estimate of $g\left(y_{T+1}|y_{1:T}\right)$.
Similarly, it is also straightforward to obtain multiple-step ahead predictive densities
$g\left(y_{T+h}|y_{1:T}\right)$, for $h=1,...,H$. If Eq. \eqref{eq:predictivedenseqnVB2}
is evaluated at the observed value $y_{T+1}=y_{T+1}^{obs}$, we refer
to one-step ahead approximate predictive likelihood at time $T+1$,
denoted $APL_{T+1}$. The APL can be used to compare between competing
models $A$ and $B$ or to choose the number of latent factors in
FSV model. We consider cumulative log approximate predictive likelihood
(CLAPL) between time points $t_{1}$ and $t_{2}$, denoted by $CLAPL=\sum_{t=t_{1}}^{t_{2}}\log APL_{t}$
and choose the best model with the highest CLAPL.

For the FSV, given $M$ draws of $\theta$ and $x_{1:T}$ from the
variational approximation $q_{\lambda}\left(\theta,x_{1:T}|y_{1:T}\right)$
we can obtain the $APL_{T+1}$ by averaging over $M$ densities of
$N_{S}\left(y_{T+1}^{obs}|0,\beta D_{T+1}\beta^{\top}+V_{T+1}\right)$
evaluated at the observed value $y_{T+1}^{obs}$, where $D_{T+1}=\textrm{diag}\left(\exp\left(\tau_{f,1}h_{f,1,T+1}\right),...,\exp\left(\tau_{f,K}h_{f,K,T+1}\right)\right)$
and $V_{T+1}=\textrm{diag}\left(\exp\left(\tau_{\epsilon,1}h_{\epsilon,1,T+1}+\kappa_{\epsilon,1}\right),...,\exp\left(\tau_{\epsilon,S}h_{\epsilon,S,T+1}+\kappa_{\epsilon,S}\right)\right)$.
This requires evaluating the full $S$-variate Gaussian density
evaluation for each draw $m$ and is thus computationally expensive.
The computational cost can be reduced by using the Woodbury matrix
identity $\Sigma_{t}^{-1}=V_{t}^{-1}-V_{t}^{-1}\beta\left(D_{t}^{-1}+\beta^{\top}V_{t}^{-1}\beta\right)^{-1}\beta^{\top}V_{t}^{-1}$
and the matrix determinant lemma $\textrm{det}\left(\Sigma_{t}\right)=\textrm{det}\left(D_{t}^{-1}+\beta^{\top}V_{t}^{-1}\beta\right)\textrm{det}\left(V_{t}\right)\textrm{det}\left(D_{t}\right)$.

\section{Simulation Study and Real Data Application \label{sec:Simulation-Study-and real data application}}


This section applies the variational approximation methods developed
in Section \ref{sec:Summary-of-VB approach} for the multivariate
factor SV model to simulated and real datasets.

\subsection{Simulation Study \label{subsec:Simulation-Study}}

We conducted a simulation study for the multivariate factor SV model
discussed in Section \ref{sec:Description-of-Factor SV model} to
compare the variational approximations  $q_{\lambda}^{I}\left(\theta,x_{1:T}\right)$, $q_{\lambda}^{II}\left(\theta,x_{1:T}\right)$, $q_{\lambda}^{III}\left(\theta,x_{1:T}\right)$ and the mean-field variational approximation $q_{\lambda}^{MF}\left(\theta,x_{1:T}\right)$ \footnote{We use the terms \lq\lq mean-field\rq\rq{} variational approximation to refer to the case where the covariance matrix is diagonal (Gaussian with diagonal covariance matrix; see \citet{Xu2018})}
that  ignores  all posterior dependence structures in the model to the exact particle MCMC method of
\citet{Gunawan2020} in terms of computation time and the accuracy of
posterior and predictive densities approximation. By comparing our variational approximations with mean-field Gaussian variational approximation, we can investigate the importance of taking into account some of the posterior dependence structures in the model.
Posterior distributions estimated using the exact particle MCMC method of \citet{Gunawan2020} with $N=100$ particles are treated
as the ground truth for comparing the accuracy of the posterior
and predictive density approximations. Section \ref{sec:Particle-MCMC-for FSV Models} briefly discusses the particle MCMC method used to estimate the FSV model. All the computations were done using Matlab  on a single desktop computer with 6-CPU cores. The particle MCMC method is run using all 6-CPU cores with the parallelization module activated (\lq\lq parfor\rq\rq{} command), whereas the variational approximation is run without the \lq\lq parfor\rq\rq{}  command activated (using only a single core).

We simulated data with $T=1000$, $S=100$ stock returns, and $K=1$, and $4$ latent factors. The particle MCMC was run for 31000 iterations and discarded the initial
$1000$ iterates as warmup. Fifty thousand $(50000)$ iterations of a stochastic
gradient ascent optimisation algorithm were run, with learning rates chosen
adaptively according to the ADAM approach discussed in Section~\ref{subsec:Learning-Rate-1} of the supplement.
All the variational parameters are initialised randomly.

Table \ref{tab:Simulated-data.-CPU times-1-1} displays the empirical CPU time per iteration for the MCMC and all variational methods. It is clear that the mean-field VA $q_{\lambda}^{MF}$ is the fastest approach, followed by $q_{\lambda}^{I}$, $q_{\lambda}^{III}$, and $q_{\lambda}^{II}$. When $T=10000$, $q_{\lambda}^{I}$ and $q_{\lambda}^{III}$ are $47.07$ and $28.24$ faster than MCMC approach, respectively. The $q_{\lambda}^{II}$ method is only about $5.24$ times faster than MCMC. The large difference in CPU time between $q_{\lambda}^{I}$ and $q_{\lambda}^{II}$ is due to the computational cost of inverting the matrix for the term $\prod_{k=1}^{K}q_{\lambda_{\beta,f_{k,1:T}}}^{SRN}\left(f_{k,1:T},\beta_{.,k}\right)$ using the Woodbury formula. See Sections \ref{subsec:Approximation-with-factor Smith2019} and \ref{subsec:VB-approach-for factor SV model} for further details.

Table \ref{tab:Simulated-data.TotalNumberofParameters} displays the total number of parameters for 1-factor and 4-factors FSV models and also the total number of variational parameters for different variational approximations. We include the variational approximation $q_{\lambda}^{N}$, the naive Gaussian variational approximation with full covariance matrix. The variational approximation $q_{\lambda}^{N}$ has more than 5 billion variational parameters. The $q_{\lambda}^{MF}$\footnote{We sometimes write $q_{\lambda}\left(\theta,x_{1:T}\right)$ as $q_{\lambda}$} has the lowest number of variational parameters, followed by $q_{\lambda}^{III}$, $q_{\lambda}^{I}$, and $q_{\lambda}^{II}$.

\begin{figure}[H]
\caption{Plot of Lower Bound for variational approximations $q_{\lambda}^{I}$, $q_{\lambda}^{II}$, $q_{\lambda}^{III}$, and  $q_{\lambda}^{MF}$ for simulated datasets.
\label{fig:Plot-of-Lower bound simulated dataset}}

\centering{}\includegraphics[width=15cm,height=6cm]{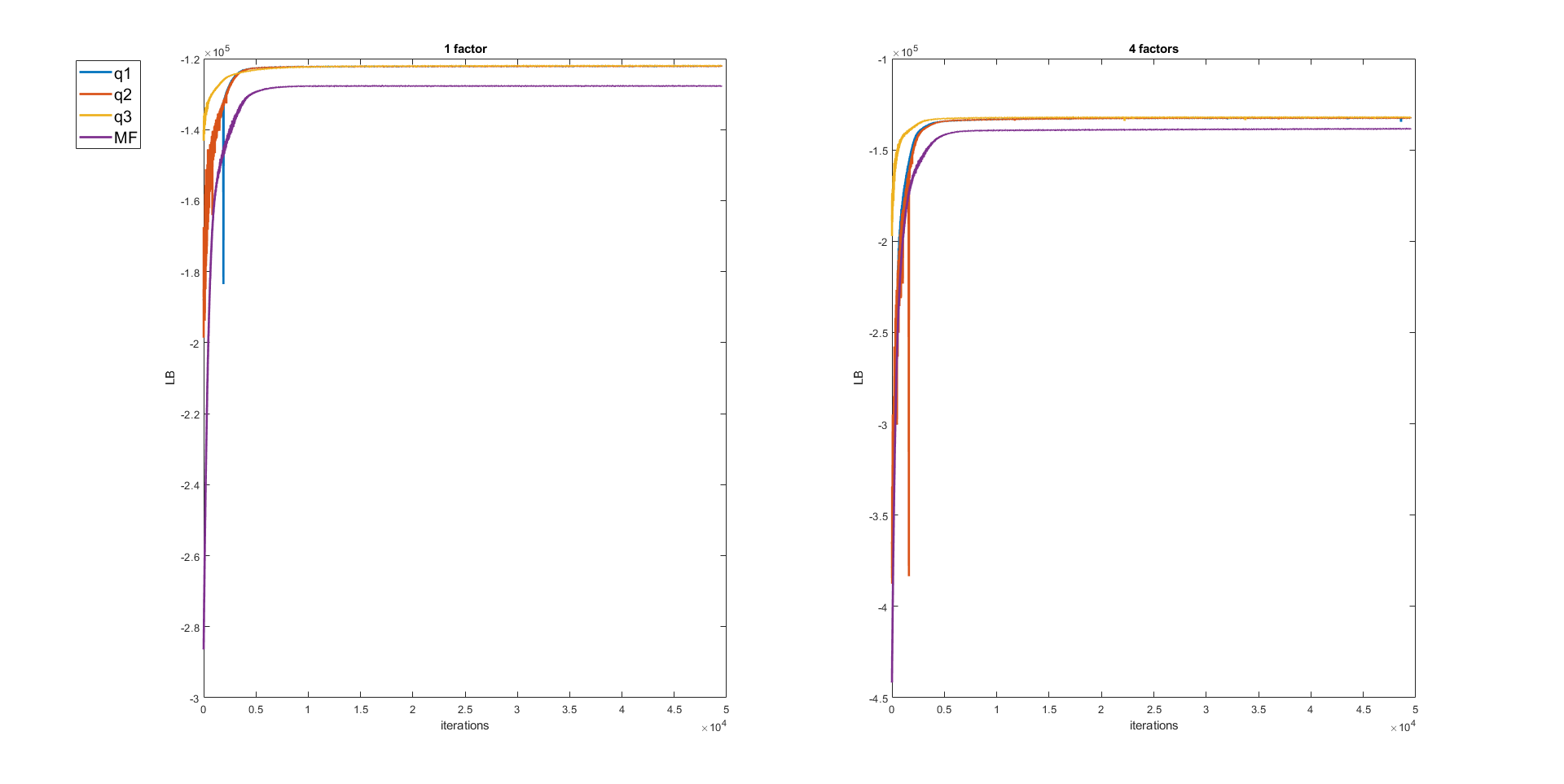}
\end{figure}

Figure~\ref{fig:Plot-of-Lower bound simulated dataset} monitors the
convergence of the variational approximations $q_{\lambda}^{I}$,
$q_{\lambda}^{II}$, $q_{\lambda}^{III}$, and $q_{\lambda}^{MF}$ via the estimated
values of their lower bounds $\widehat{\mathcal{L}^{I}\left(\lambda\right)}$,  $\widehat{\mathcal{L}^{II}\left(\lambda\right)}$,  $\widehat{\mathcal{L}^{III}\left(\lambda\right)}$ and
$\widehat{\mathcal{L}^{MF}\left(\lambda\right)}$ using a single
Monte Carlo sample. The figure shows that the lower bound of all variational approximations increase at the start and then stabilise after around $20000$ iterations; all four
variational approximations converge.

\begin{table}[H]
\caption{The CPU time per iteration on a single core CPU using Matlab for the
variational approximations, $q_{\lambda}^{I}$,
$q_{\lambda}^{II}$, $q_{\lambda}^{III}$,
and $q_{\lambda}^{MF}$ and a six core
CPU for the MCMC for four-factors FSV models with time series of lengths $T=500$, 1000, 2000, 5000, and 10000 \label{tab:Simulated-data.-CPU times-1-1}}

\centering{}%
\begin{tabular}{cccccc}
\hline
$T$ & MCMC & \multicolumn{4}{c}{Variational Approximations}\tabularnewline
\hline
 &  & $q_{\lambda}^{I}$ & $q_{\lambda}^{II}$ & $q_{\lambda}^{III}$ & $q_{\lambda}^{MF}$\tabularnewline
\hline
500 & 1.86 & 0.09 & 0.10 & 0.10 & 0.03\tabularnewline
1000 & 3.73 & 0.13 & 0.20 & 0.17 & 0.04\tabularnewline
2000 & 7.70 & 0.23 & 0.44 & 0.29 & 0.07\tabularnewline
5000 & 21.38 & 0.63 & 2.30 & 0.85 & 0.22\tabularnewline
10000 & 55.07 & 1.17 & 10.51 & 1.95 & 0.39\tabularnewline
\hline
\end{tabular}
\end{table}

\begin{table}[H]
\caption{Total number of model parameters and latent variables and total number
of variational parameters. The $q_{\lambda}^{N}\left(\theta,x_{1:T}\right)$ is the naive Gaussian variational approximation with full covariance matrix $\Sigma$. \label{tab:Simulated-data.TotalNumberofParameters}}

\centering{}%
\begin{tabular}{ccccccc}
\hline
$K$ & Model Parameters  & \multicolumn{5}{c}{Variational Parameters }\tabularnewline
\hline
 & $\left(\theta,x_{1:T}\right)$ & $q_{\lambda}^{N}\left(\theta,x_{1:T}\right)$ & $q_{\lambda}^{I}\left(\theta,x_{1:T}\right)$ & $q_{\lambda}^{II}\left(\theta,x_{1:T}\right)$ & $q_{\lambda}^{III}\left(\theta,x_{1:T}\right)$ & $q_{\lambda}^{MF}\left(\theta,x_{1:T}\right)$\tabularnewline
\hline
1 & 102402 & 5243238405 & 1213196 & 1217202 & 1210196 & 204804\tabularnewline
4 & 108702 & 5908225455 & 1251260 & 1267266 & 1239260 & 217404\tabularnewline
\hline
\end{tabular}
\end{table}

\begin{table}[H]
\caption{Simulated data for FSV models with $K=1$ and $4$ latent factors. Number of iterations $I$ (in thousands) and CPU time per iterations (in seconds) of the MCMC and variational methods.
Average and standard deviation of the lower bound value over the last 5000 steps are also reported.
\label{tab:Simulated-data.-CPU times}}

\centering{}%
\begin{tabular}{cccccc}
\hline
$K$ & Methods & $\widehat{\mathcal{L}\left(\lambda\right)}$ & $I$ & CPU time & Total Time\tabularnewline
\hline
\hline
1 & PMCMC & NA & $30000$ & $3.54$ & $106200$\tabularnewline
 & $q_{\lambda}^{I}$ & $-122130.41 (27.45)$ & $50000$ & $0.11$ & $5500$\tabularnewline
 & $q_{\lambda}^{II}$ & $-122140.72 (28.31)$ & $50000$ & $0.18$ & $9000$\tabularnewline
 & $q_{\lambda}^{III}$ & $-122020.77 (23.99)$ & $50000$ & $0.14$ & $7000$\tabularnewline
 & $q_{\lambda}^{MF}$ & $-127707.31 (37.13)$ & $50000$ & $0.03$ & $1500$\tabularnewline
\hline
4 & PMCMC & NA & $30000$ & $3.73$ & $111900$\tabularnewline
 & $q_{\lambda}^{I}$ & $-132501.73 (48.89)$ & $50000$ & $0.13$ & $6500$\tabularnewline
 & $q_{\lambda}^{II}$ & $-132517.78 (40.23)$ & $50000$ & $0.20$ & $10000$\tabularnewline
 & $q_{\lambda}^{III}$ & $-132116.92 (32.02)$ & $50000$ & $0.17$ & $8500$\tabularnewline
 & $q_{\lambda}^{MF}$ & $-138454.19 (51.74)$ & $50000$ & $0.04$ & $2000$\tabularnewline
\hline
\end{tabular}
\end{table}

Table \ref{tab:Simulated-data.-CPU times} shows the average and standard deviation of the lower bound values for the variational methods for the FSV models with one and four factors. The estimated lower bound values can be used to select the best variational approximations. The $q_{\lambda}^{III}$ method has the highest estimated lower bound value, followed by $q_{\lambda}^{I}$, $q_{\lambda}^{II}$, and $q_{\lambda}^{MF}$. As expected, the $q_{\lambda}^{MF}$ has the lowest lower bound values. Surprisingly, the variational approximation $q_{\lambda}^{I}$ is better than $q_{\lambda}^{II}$ with larger number of variational parameters. The variational approximation $q_{\lambda}^{III}$ also has the smallest standard deviation of the lower bound. The variational methods are potentially much faster than reported  as Figure \ref{fig:Plot-of-Lower bound simulated dataset}
shows that all of the variational approximations already converge after
$20000$ iterations.

\begin{figure}[H]
\caption{Simulated dataset. Scatter plots of the posterior means of the factors log-volatilities
$h_{f,k,t}$, for $k=1,...,4$ and $t=10,20,...,1000$ estimated
using particle MCMC on the x-axis and the variational approximation $q_{\lambda}^{I}$, $q_{\lambda}^{II}$,
$q_{\lambda}^{III}$, and $q_{\lambda}^{MF}$, on the y-axis.
\label{fig:Scatter-plot-of posterior standard deviation of factorlogvolatility}}

\centering{}\includegraphics[width=15cm,height=8cm]{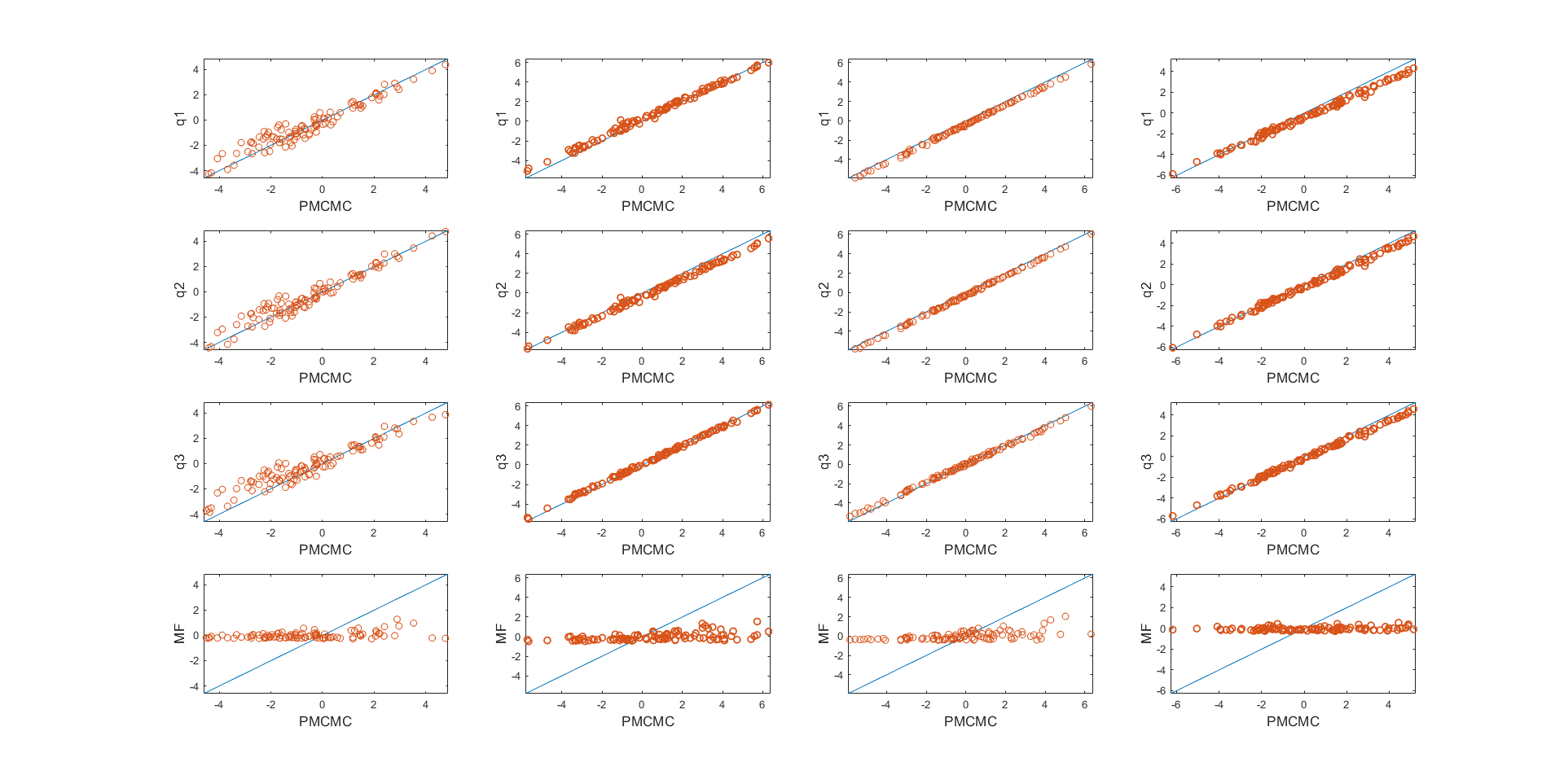}
\end{figure}

\begin{figure}[H]
\caption{Simulated dataset. Scatter plots of the posterior means of the idiosyncratic log-volatility
$h_{\epsilon,s,t}$, for $s=1,...,100$ and $t=10,20,...,1000$ estimated
using particle MCMC on the x-axis and the variational approximation $q_{\lambda}^{III}$, on the y-axis.
\label{fig:Scatter-plot-of posterior means of idiosyncraticlogvolatility q3}}

\centering{}\includegraphics[width=15cm,height=8cm]{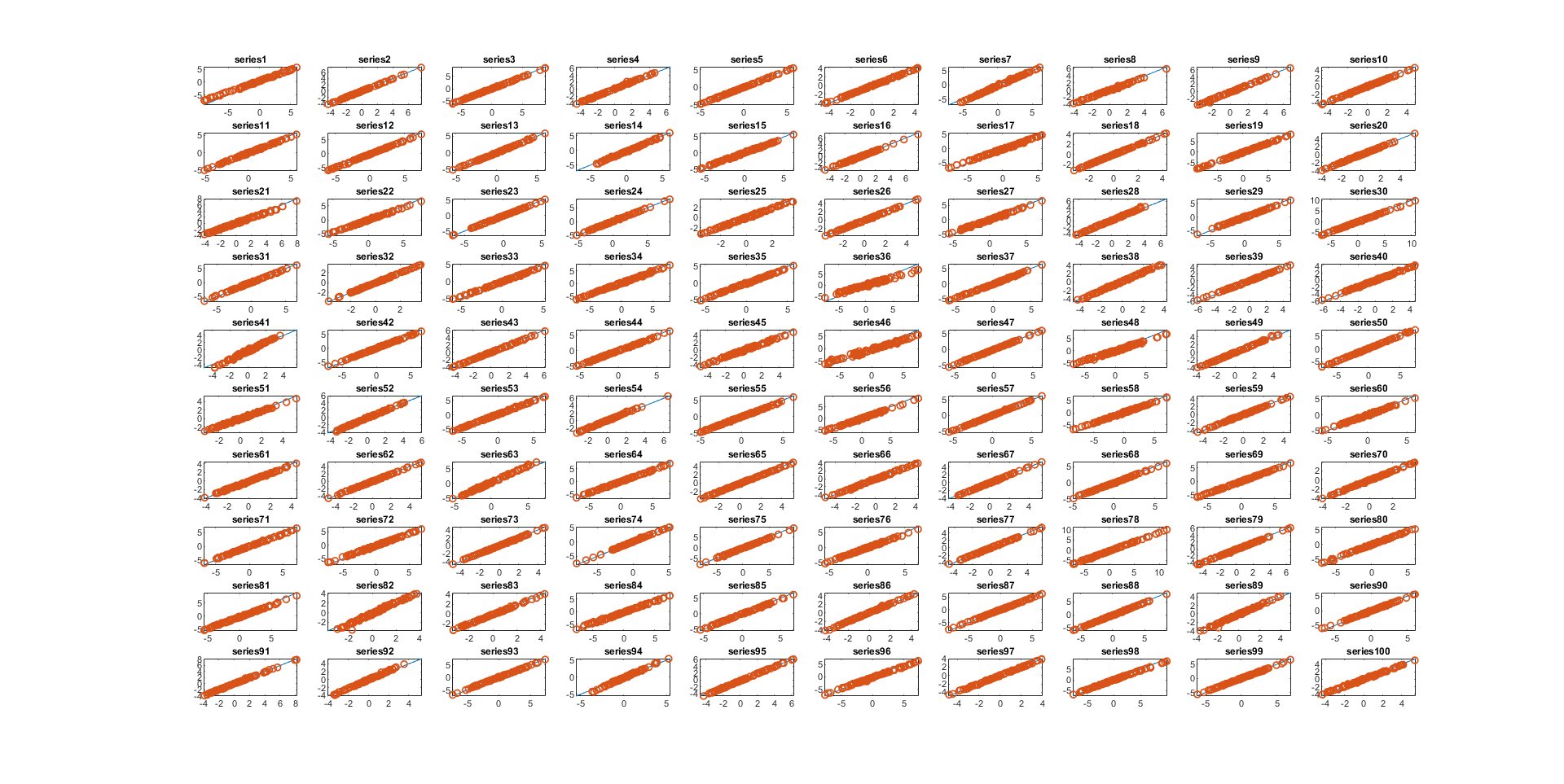}
\end{figure}

Figure \ref{fig:Scatter-plot-of posterior standard deviation of factorlogvolatility} compares the posterior mean estimates of the factor log-volatilities $h_{f,k,t}$, for $k=1,...,4$,
estimated using the different variational approximation methods and the particle MCMC
method. The figure shows that all variational approximations give
posterior mean estimates that are very close to each other and close
to the particle MCMC estimates, except the mean-field variational approximation. Figure~\ref{fig:Scatter-plot-of posterior means of idiosyncraticlogvolatility q3}
and
 Figures \ref{fig:Scatter-plot-of posterior means of idiosyncraticlogvolatility q1}, \ref{fig:Scatter-plot-of posterior means of idiosyncraticlogvolatility q2}, and \ref{fig:Scatter-plot-of posterior means of idiosyncraticlogvolatility MF}
(in Section~\ref{sec:Additional-Figures-for simulation} of the supplement) compare
the variational posterior means of $q_{\lambda}^{III}$, $q_{\lambda}^{I}$, $q_{\lambda}^{II}$, and $q_{\lambda}^{MF}$
with the posterior means computed using the
particle MCMC for all $S=100$ series. The figures show that the variational
approximation $q_{\lambda}^{I}$, $q_{\lambda}^{II}$, and $q_{\lambda}^{III}$ estimate
the posterior means accurately, but the mean-field variational approximation $q_{\lambda}^{MF}$ does not.

\begin{figure}[H]
\caption{Simulated dataset. The marginal posterior density plots of the parameters $\left\{ \psi_{\epsilon,4},\psi_{\epsilon,8},\psi_{\epsilon,20},\psi_{\epsilon,26}\right\} $
estimated using particle MCMC and VB with the variational approximations $q_{\lambda}^{I}$, and
$q_{\lambda}^{II}$, $q_{\lambda}^{III}$, and $q_{\lambda}^{MF}$.
 \label{fig:The-plot-of psisim_4factor}}
\centering{}\includegraphics[width=15cm,height=8cm]{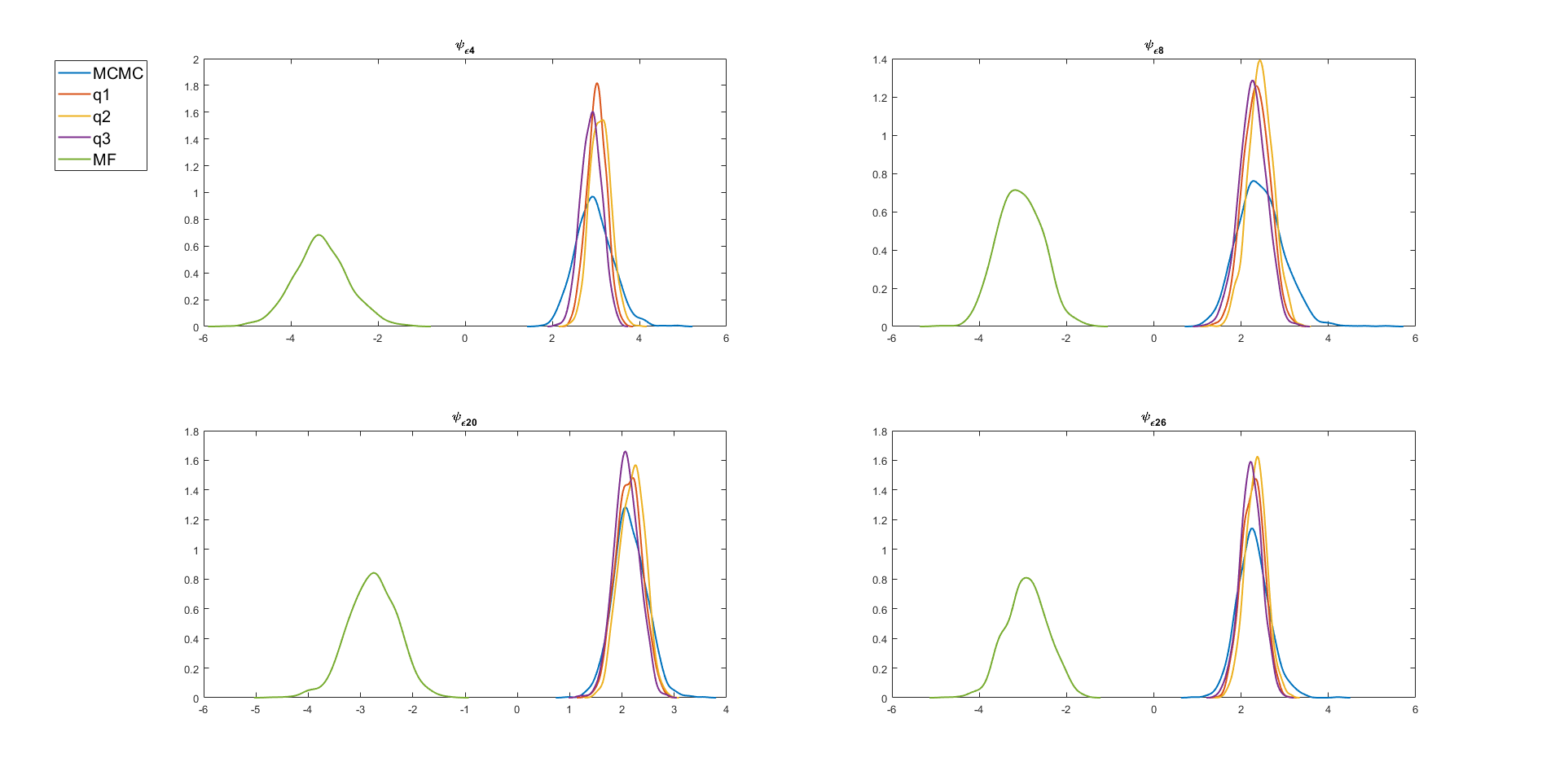}
\end{figure}

\begin{figure}[H]
\caption{Simulated dataset. The marginal posterior density plots of some elements of the $\beta$
estimated using particle MCMC and VB with the variational approximations $q_{\lambda}^{I}$, and
$q_{\lambda}^{II}$, $q_{\lambda}^{III}$, and $q_{\lambda}^{MF}$.
 \label{fig:The-plot-of betaloadingsim_4factor}}

\centering{}\includegraphics[width=15cm,height=8cm]{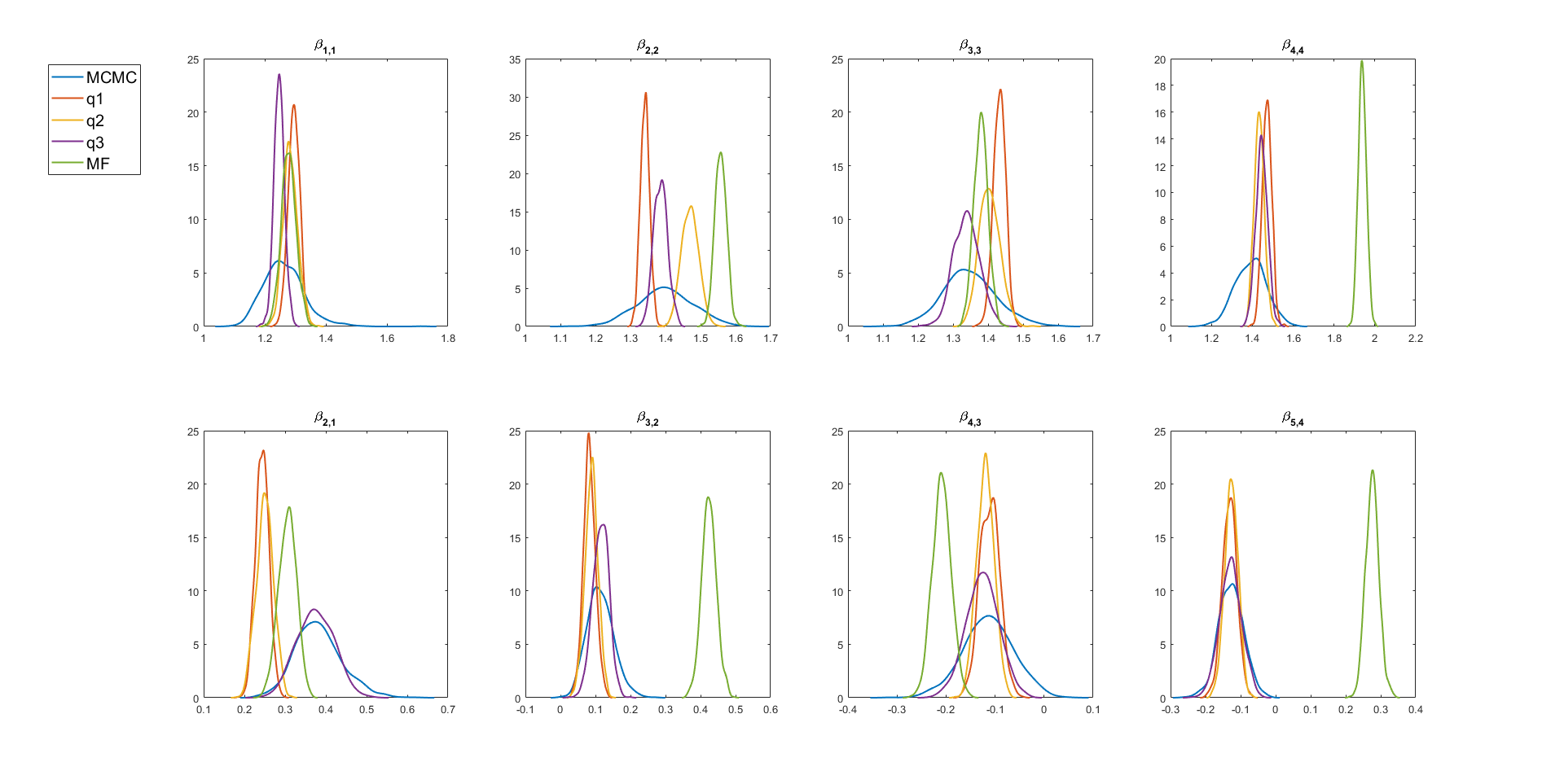}
\end{figure}

Figure \ref{fig:The-plot-of psisim_4factor} shows the
marginal posterior densities of the parameters $\left\{ \psi_{\epsilon,s}\right\} $
for $s=4,8,20,26$. All the variational approximations capture
the posterior means quite well, except for the mean field variational approximation. There is a slight underestimation of
the posterior variances of the $\psi$'s. Figure \ref{fig:The-plot-of betaloadingsim_4factor} shows the marginal posterior densities of some of the parameters in the factor loading matrix $\beta$. The variational approximation $q_{\lambda}^{III}$ captures the posterior means better than $q_{\lambda}^{I}$ and $q_{\lambda}^{II}$, but there is underestimation of the posterior variances of the $\beta$.

We now compare the predictive performance of the variational approximations
with the (exact) particle MCMC method. The minimum variance portfolio implied by the $h$ step-ahead time-varying covariance
matrix $\Sigma_{T+h}$, is considered which can
be used to uniquely defines the optimal portfolio weights
\[
w_{T+h}=\frac{\Sigma_{T+h}^{-1}\mathbb{1}}{\mathbb{1}^{\top}\Sigma_{T+h}^{-1}\mathbb{1}},
\]
where $\mathbb{1}$ denotes an S-variate vector of ones \citep{Bodnar2017}. The optimal portfolio weights guarantee the lowest risk for a given expected portfolio return.

\begin{figure}[H]
\caption{Simulated dataset. Plot of the predictive densities $\widehat{p}\left(y_{T+h}|y_{1:T}\right)$
of an optimally weighted series, for $h=1,...,10$, estimated
using particle MCMC and the variational
approximations $q_{\lambda}^{I}$, $q_{\lambda}^{II}$, $q_{\lambda}^{III}$, and
$q_{\lambda}^{MF}$. \label{fig:Plot-of- predictive densities}}

\begin{centering}
\includegraphics[width=15cm,height=8cm]{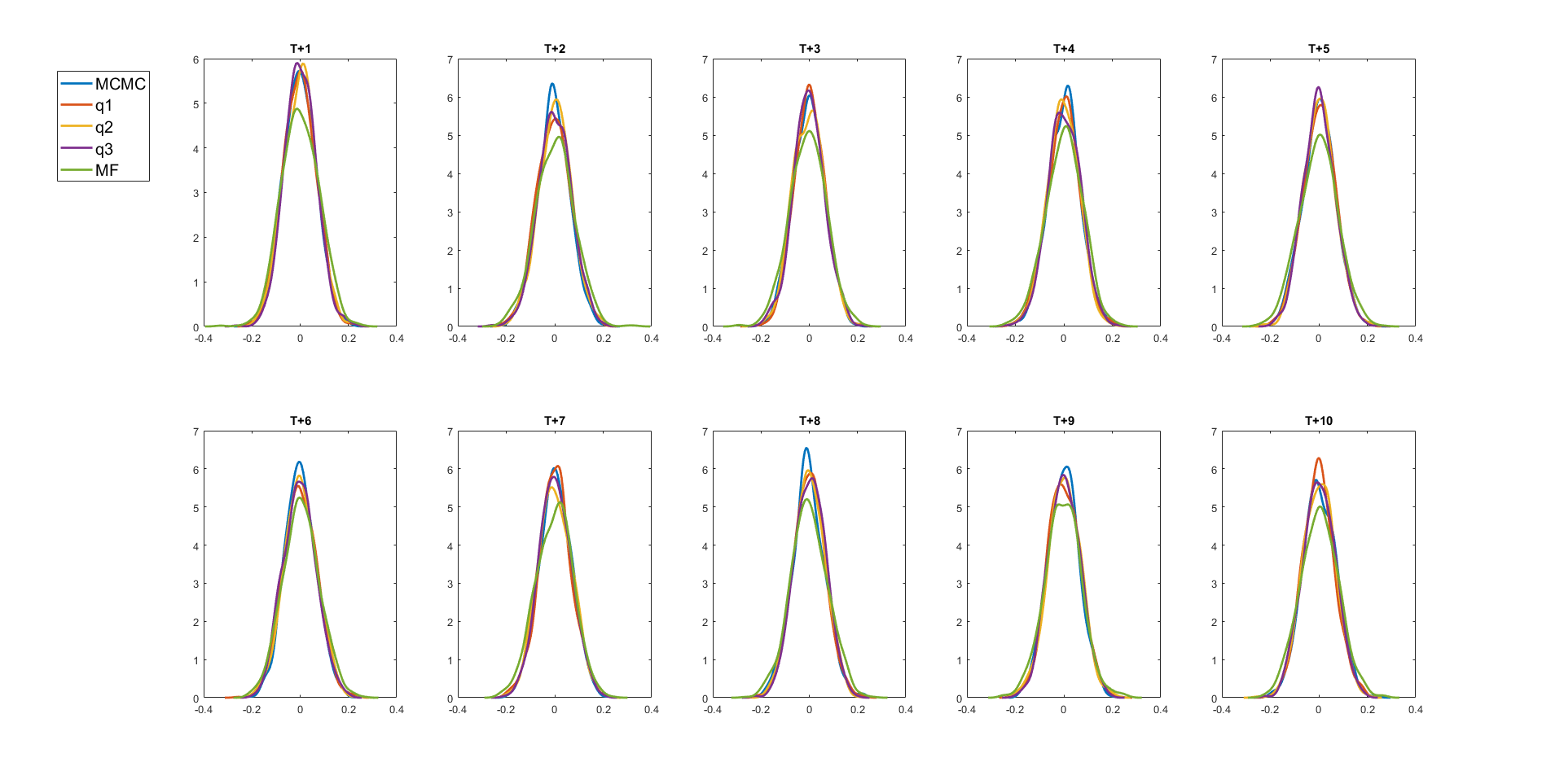}
\par\end{centering}
\end{figure}

\begin{figure}[H]
\caption{Simulated dataset. Plots of the predictive densities of the time-varying correlations
$\widehat{p}\left(\Gamma\left(y_{2,T+h},y_{3,T+h}\right)|y_{1:T}\right), h=1,...,10$,
between series
2 and 3 estimated using particle MCMC and the variational
approximations $q_{\lambda}^{I}$, $q_{\lambda}^{II}$, $q_{\lambda}^{III}$, and
$q_{\lambda}^{MF}$. \label{fig:Plot-of-one step ahead predictive density rho23 }}

\centering{}\includegraphics[width=15cm,height=8cm]{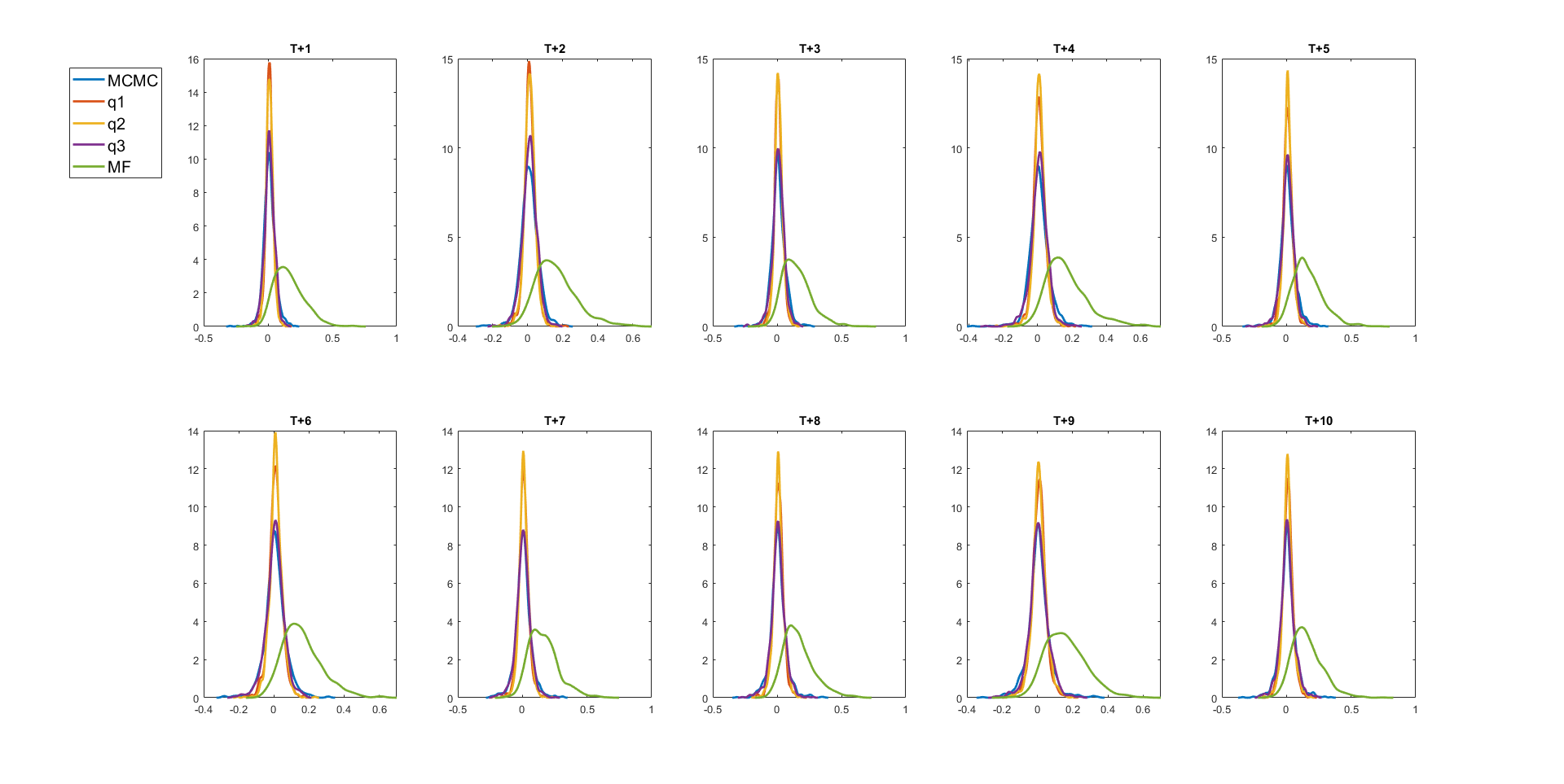}
\end{figure}

Figure~\ref{fig:Plot-of- predictive densities}
shows the multiple-step ahead predictive densities $\widehat{p}\left(y_{T+h}|y_{1:T}\right)$
of an optimally weighted combination of all series, for $h=1,...,10$ obtained
using variational and PMCMC methods. The figure shows that all the variational
predictive densities, except the mean-field variational approximation, are very close to the exact predictive densities
obtained from the particle MCMC method with the $q_{\lambda}^{III}$ the closest to MCMC. Using the multivariate factor SV
model, we can also obtain the predictive densities of the time-varying
correlation  between any two series. Given the time-varying covariance
matrix at time $t$, the correlation matrix at time step $t$ is
\[
\Gamma_{t}=\textrm{diag}\left(\Sigma_{t}\right)^{-\frac{1}{2}}\Sigma_{t}\textrm{diag}\left(\Sigma_{t}\right)^{-\frac{1}{2}}.
\]
Figure~\ref{fig:Plot-of-one step ahead predictive density rho23 } shows
the multiple-step ahead predictive densities of the time-varying correlation
between series 2 and 3 $\widehat{p}\left(\Gamma\left(y_{2,T+h},y_{3,T+h}\right)|y_{1:T}\right)$
for $h=1,...,10$. The figure shows that all the variational predictive, except the mean-field variational approximation,
densities are very close to the exact predictive densities obtained
from the PMCMC method. The estimates from the variational approximation $q_{\lambda}^{III}$ is the closest to MCMC.  Both Figures \ref{fig:Plot-of- predictive densities}
and \ref{fig:Plot-of-one step ahead predictive density rho23 } show
that the variational predictive densities of our variational methods do not underestimate the predictive
variances.

We now investigate the accuracy of the proposed sequential variational method described in Section 3.5 and  consider the sequential versions of $q_{\lambda}^{I}$, $q_{\lambda}^{II}$, and $q_{\lambda}^{III}$. We do not consider the sequential version of the
mean-field variational approximation
because the batch version is
 less accurate than the other variational approximations.
For each of the sequential VB methods,
we consider three different starting points: $t=700$, $800$, and $900$. For example,  the $t=700$ starting point means that  the posterior of the parameters and latent variables is estimated as a batch for the first $t=700$ time points and then  the posteriors are updated as new observations arrive up to $T=1000$;  this method is denoted by $seq-q_{\lambda}-t700$.
The updates for all sequential approaches are done after observing an additional $5$ time points for each series. Figure 8 shows the sequential lower bound for the variational approximation $seq-q_{\lambda}^{III}$ for the last 20 sequential updates with the starting point $t=900$. The figure
 shows that the sequential variational algorithm converges within $2000$ iterations or less for all sequential updates. Figure~\ref{fig:The-plot-of betaloadingsim_4factor_seqVA} and Figure~\ref{fig:The-plot-of psisim_4factor_seqVA} in Section~\ref{sec:Additional-Figures-for simulation} of the supplement show that the estimates from the
 sequential approaches $seq-q_{\lambda}^{I}$ and $seq-q_{\lambda}^{II}$ get worse with earlier starting points $t=700$ and $t=800$, but the estimates from  $seq-q_{\lambda}^{III}$ are still close to $q_{\lambda}^{III}$ and MCMC for all cases.
  Similar observations can be made for the one-step-ahead prediction of the minimum variance portfolio in Figure \ref{fig:Plot-of- predictive densities-seqVA} and one-step-ahead time-varying correlation between series 2 and 3 in Figure \ref{fig:Plot-of-one step ahead predictive density rho23_seqVA} in Section~\ref{sec:Additional-Figures-for simulation} of the supplement.

\begin{figure}[H]
\caption{Simulated dataset. Plot of the sequential lower bound for the variational approximation $seq-q_{\lambda}^{III}$ for the last 20 sequential updates with starting point $t=900$. \label{fig:Plot-of- sequentiallowerboundsimseqq3}}

\begin{centering}
\includegraphics[width=15cm,height=8cm]{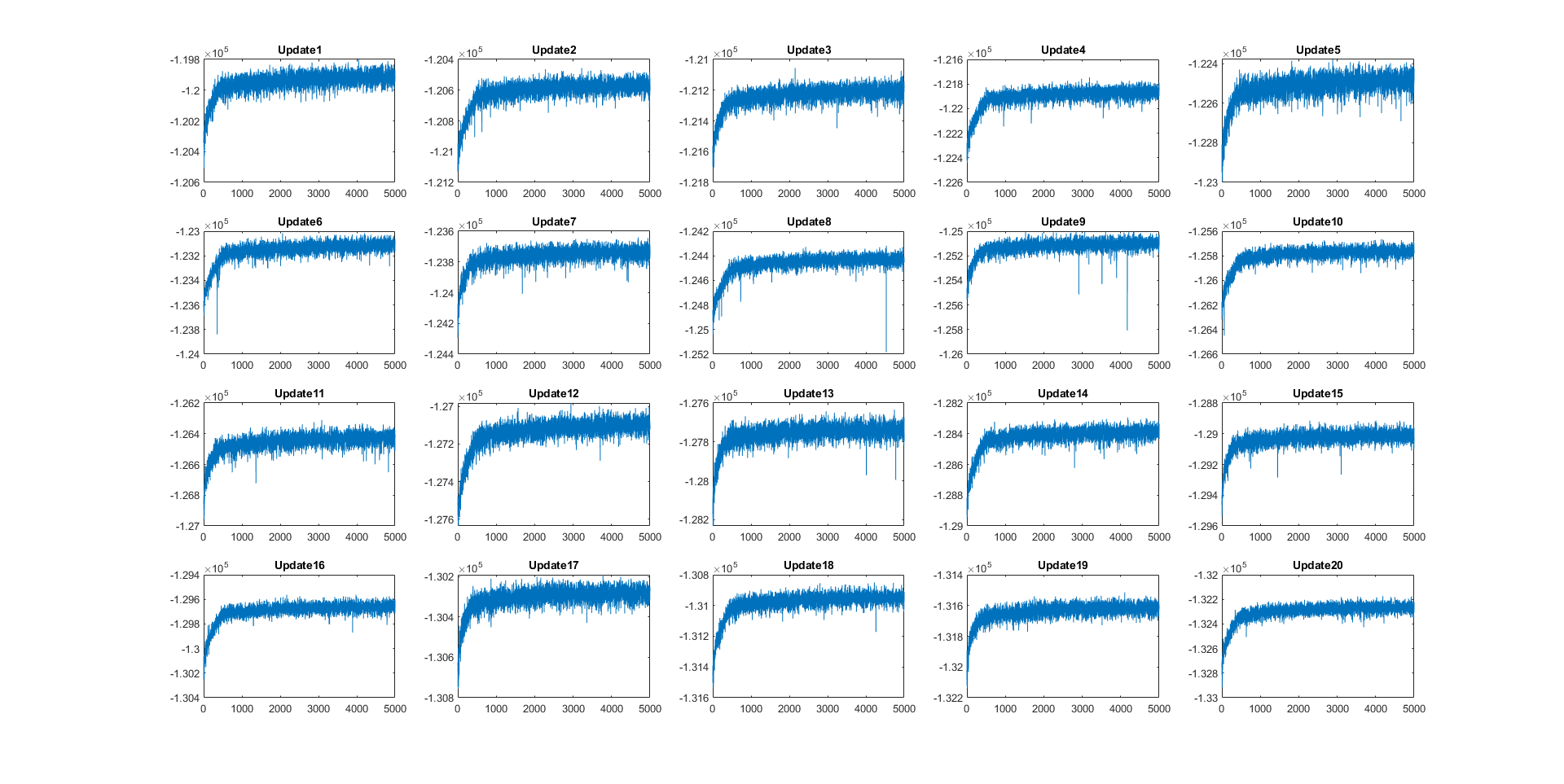}
\par\end{centering}
\end{figure}

\begin{figure}[H]
\caption{Simulated dataset. The marginal posterior density plots of some of the $\beta$
estimated using particle MCMC, VA, and sequential VA.
 \label{fig:The-plot-of betaloadingsim_4factor_seqVA}}

\centering{}\includegraphics[width=15cm,height=8cm]{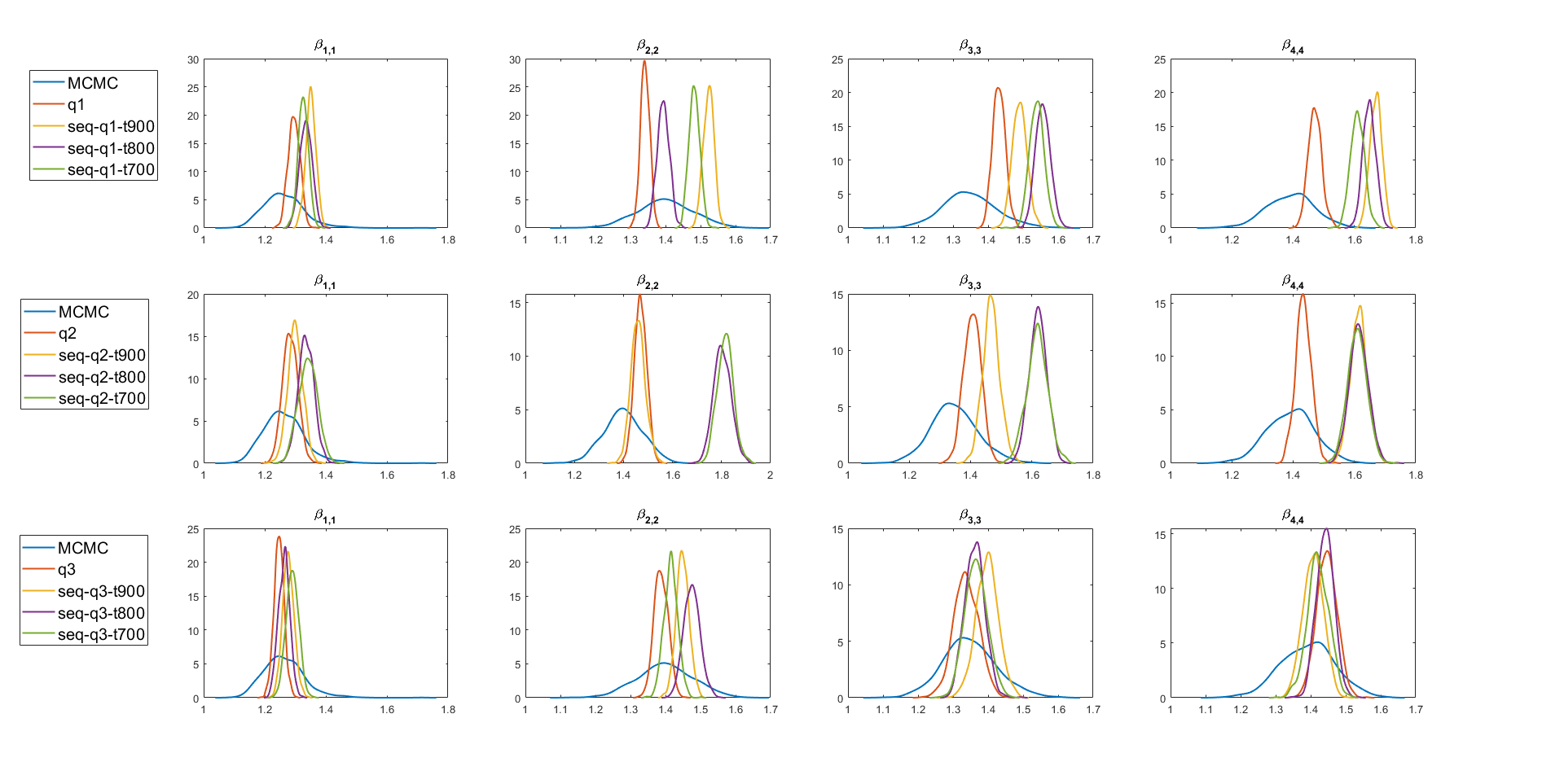}
\end{figure}

We conclude from the simulation study that the variational approximations
are much faster than the exact MCMC approach.  The variational approximations
capture the posterior means of the parameters of the FSV model quite accurately, except for
the mean field variational approximation, but there is some slight underestimation
of the posterior variance for some of the parameters;
The variational approximation $q_{\lambda}^{III}$ is the best.
All the proposed variational
approximations produce predictive densities for the returns
and time-varying correlations between time series that are similar to the exact particle MCMC
method with the $q_{\lambda}^{III}$ being the best; the variational approximation $q_{\lambda}^{I}$
is only slightly faster than $q_{\lambda}^{III}$ and much faster than $q_{\lambda}^{II}$, especially for long time series. The sequential variational approximation $seq-q_{\lambda}^{III}$ is also the most accurate of the sequential approximations. It is therefore important to take into account the posterior dependence between the latent factors and other latent states and parameters in the FSV model. It is important to take into account the dependence between parameters $\theta_{G,\epsilon,s}$ and the idiosyncratic log-volatilities $h_{\epsilon,s,1:T}$ for $s=1,...,S$ and the dependence between parameters $\theta_{G,f,k}$ and the factor log-volatilities $h_{f,k,1:T}$ for $k=1,...,K$. The dependence between $(\theta_{G,\epsilon,s},h_{\epsilon,s,1:T})$ and $(\theta_{G,\epsilon,j},h_{\epsilon,j,1:T})$ for $s\neq j$ and between $(\theta_{G,f,k},h_{f,k,1:T})$ and $(\theta_{G,f,j},h_{f,j,1:T})$ for $k \neq j$ can be possibly ignored.
In terms of accuracy and CPU time, the variational approximation $q_{\lambda}^{III}$ and the sequential approximation $seq-q_{\lambda}^{III}$ are the best.

\subsection{Application to US Stock Returns Data \label{subsec:Application-to-US StockReturnsData}}

\begin{figure}[H]
\caption{SP100 dataset. Plot of the average of the lower bound values for the last 5000 iterations (left) and the CLAPL values (right) for $k=1,...,7$ factors. \label{fig:Plot-of- lower bound CLAPL}}

\begin{centering}
\includegraphics[width=15cm,height=8cm]{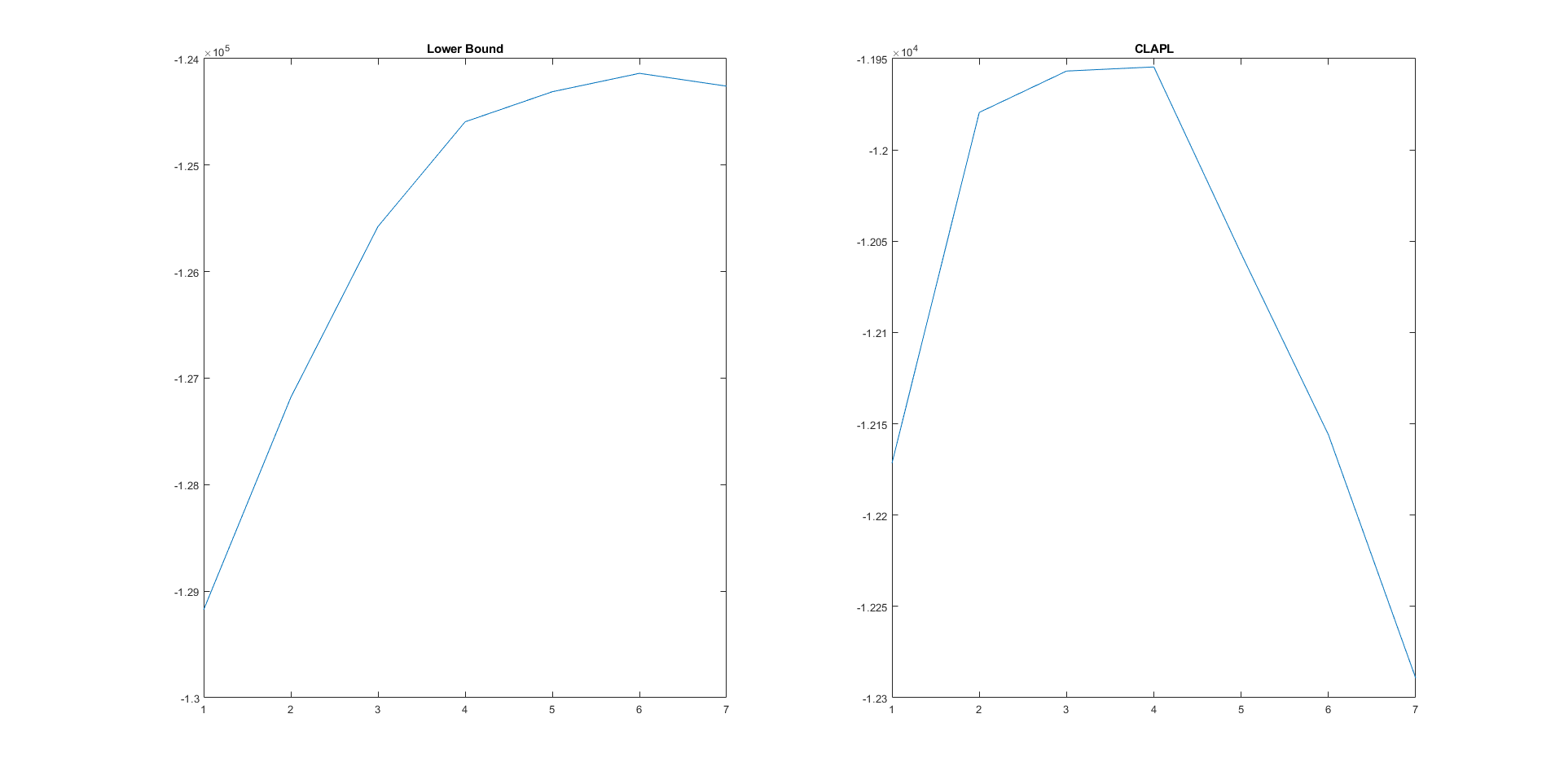}
\par\end{centering}
\end{figure}

Our methods are now applied to a panel of $S=90$ daily return of the Standard and Poor's 100 stocks, from 9th of October 2009 to 30th September 2013 ($T=1000$ demeaned log returns in total). The data cover the following sectors: Consumer Discretionary, Consumer Staples, Energy, Financials, Health Care, Industrials, Information Technology, Materials, Telecommunication Services, and Utilities (see Appendix \ref{tab:TableSP100} for the individual stocks). The number of factors in the FSV model is selected using the lower bound and the cumulative log approximate predictive likelihood (CLAPL) values discussed in Section \ref{subsec:VariationalForecasting} using the variational approximation $q_{\lambda}^{III}$ as  Section \ref{subsec:Simulation-Study} shows that the $q_{\lambda}^{III}$ is the most accurate. The variational approximation is applied to the first $900$ time points and the CLAPL is computed for $t=901,...,1000$ time points. The number of Monte Carlo samples used to estimate the CLAPL at each time point is $M=10000$.
Table \ref{fig:Plot-of- lower bound CLAPL} shows that the CLAPL values increase and then decrease. Similarly, the lower bound values increase significantly from $k=1$ factor to $k=4$ factors, and then flatten out. Therefore, further analysis in this section is based on four-factor FSV model.

We study the performance of $q_{\lambda}^{I}$, $q_{\lambda}^{II}$, $q_{\lambda}^{III}$ and  $q_{\lambda}^{MF}$ on the real data set. The accuracy of the variational approximations for the posterior and predictive densities are compared to the exact particle MCMC method with $N=100$ particles as in the simulation study in Section \ref{subsec:Simulation-Study}. The lower bound values (with standard deviations in brackets) are $-125307.64(47.15)$, $-125303.74(47.07)$, and $-124593.25(32.16)$ for the $q_{\lambda}^{I}$, $q_{\lambda}^{II}$, and $q_{\lambda}^{III}$, respectively. Clearly,  $q_{\lambda}^{III}$ performs the best.

\begin{figure}[H]
\caption{SP100 dataset. The marginal posterior density plots of the parameters $\left\{ \psi_{\epsilon,4},\psi_{\epsilon,8},\psi_{\epsilon,20},\psi_{\epsilon,26}\right\} $
estimated using particle MCMC and VB with the variational approximations $q_{\lambda}^{I}$,
$q_{\lambda}^{II}$, $q_{\lambda}^{III}$, and $q_{\lambda}^{MF}$.
 \label{fig:The-plot-of psiSP100_4factor}}
\centering{}\includegraphics[width=15cm,height=8cm]{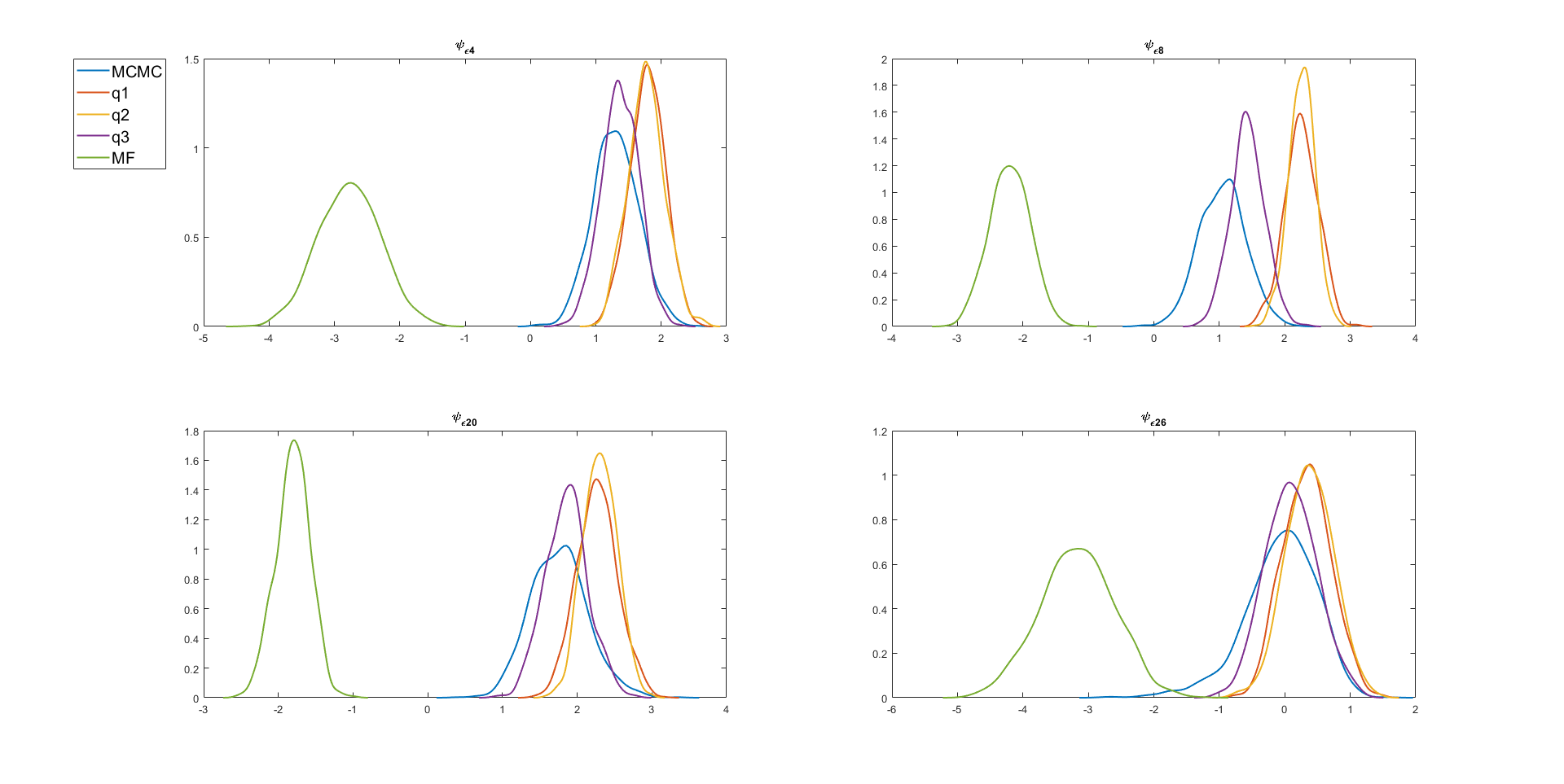}
\end{figure}

\begin{figure}[H]
\caption{SP100 dataset. Scatter plots of the posterior means of the latent factors
$f_{k,t}$, for $k=1,...,4$ and $t=10,20,...,1000$ estimated
using particle MCMC on the x-axis and the variational approximation $q_{\lambda}^{I}$, $q_{\lambda}^{II}$,
$q_{\lambda}^{III}$, and $q_{\lambda}^{MF}$, on the y-axis.
\label{fig:Scatter-plot-of posterior means of latent factors SP100}}

\centering{}\includegraphics[width=15cm,height=8cm]{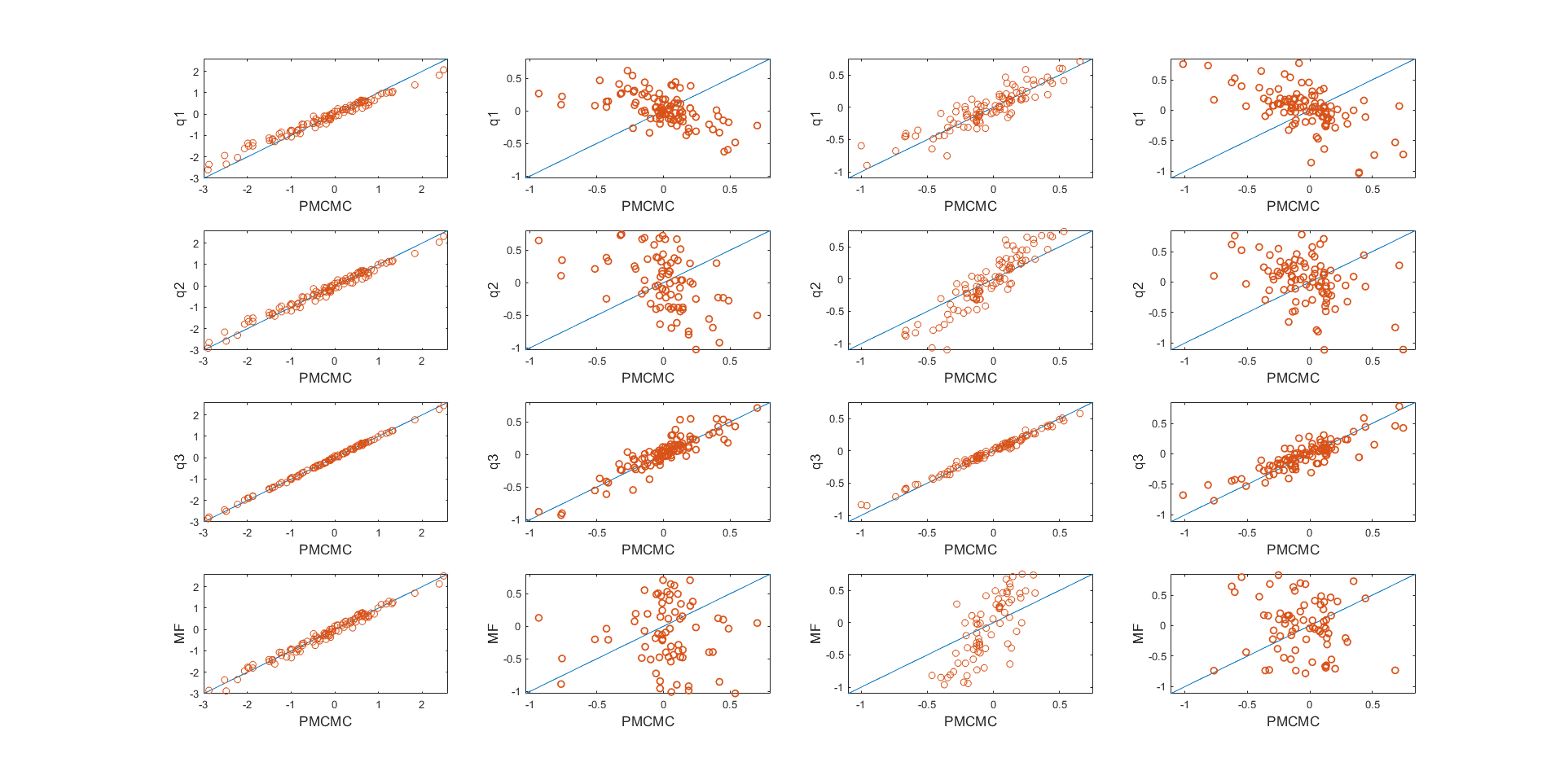}
\end{figure}

Figure \ref{fig:The-plot-of psiSP100_4factor} shows the estimated marginal posterior densities of the parameters $\left\{ \psi_{\epsilon,s}\right\} $ for $s=4,8,20,26$ using the different variational approximations and the exact MCMC method. The $q_{\lambda}^{III}$ approximation is able to capture the posterior means quite well compared to  $q_{\lambda}^{I}$ and  $q_{\lambda}^{II}$. The mean field variational approximation produces the wrong estimates. As in the simulation study, there is some underestimation of the posterior variance of the parameters in all the variational approximations. Similar conclusions can be made from Figures \ref{fig:The-plot-of betaloadingSP100_4factor} and \ref{fig:The-plot-of latentfactorsSP100_4factor} in Appendix \ref{sec:Additional-Figures-for SP100}. Figure \ref{fig:Scatter-plot-of posterior means of latent factors SP100} shows the posterior mean estimates of the latent factors $f_{k,t}$, for $k=1,...,4$. All the variational approximations estimate posterior means of the first latent factor $f_{1,t}$ well, but the variational approximations $q_{\lambda}^{I}$,  $q_{\lambda}^{II}$, and $q_{\lambda}^{MF}$ estimate the second to the fourth latent factors poorly. As expected, the variational approximation $q_{\lambda}^{III}$ estimates all the latent factors well. Similar conclusions can be made from Figures \ref{fig:Scatter-plot-of posterior means of idiosyncraticlogvolatility SP100 q1}, \ref{fig:Scatter-plot-of posterior means of idiosyncraticlogvolatility SP100 q2}, \ref{fig:Scatter-plot-of posterior means of idiosyncraticlogvolatility SP100 q3}, and \ref{fig:Scatter-plot-of posterior means of idiosyncraticlogvolatility SP100 MF} in Appendix \ref{sec:Additional-Figures-for SP100}.  This shows that the performances of the variational
approximation, $q_{\lambda}^{III}$, still performs well when applied to real data.

%

\begin{figure}[H]
\caption{SP100 dataset. Plot of the predictive densities $\widehat{p}\left(y_{T+1}|y_{1:T}\right)$
of the minimum variance portfolio, estimated
using particle MCMC and the variational
approximations $q_{\lambda}^{I}$, $q_{\lambda}^{II}$, $q_{\lambda}^{III}$, and
$q_{\lambda}^{MF}$. \label{fig:Plot-of- predictive densities onestepaheadSP100}}

\begin{centering}
\includegraphics[width=15cm,height=8cm]{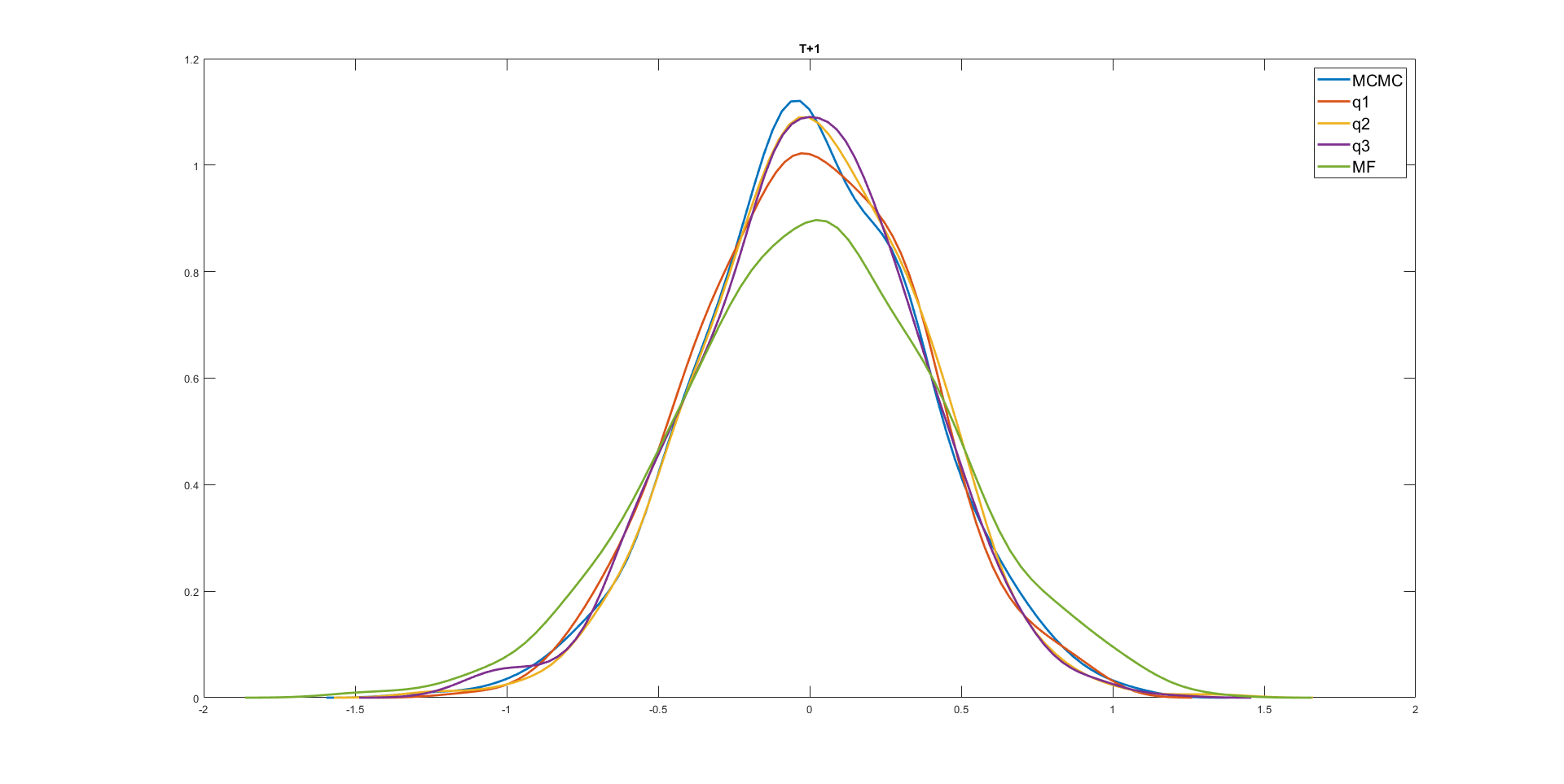}
\par\end{centering}
\end{figure}

\begin{figure}[H]
\caption{SP100 dataset. Plots of the predictive densities of the time-varying correlations
$\widehat{p}\left(\Gamma\left(y_{MSFT,T+h},y_{AIG,T+h}\right)|y_{1:T}\right), h=1,2,3$,
between Microsoft and AIG
estimated using particle MCMC and the variational
approximations $q_{\lambda}^{I}$, $q_{\lambda}^{II}$, $q_{\lambda}^{III}$, and
$q_{\lambda}^{MF}$. \label{fig:Plot-of-one step ahead predictive density rho23 SP100}}

\centering{}\includegraphics[width=15cm,height=8cm]{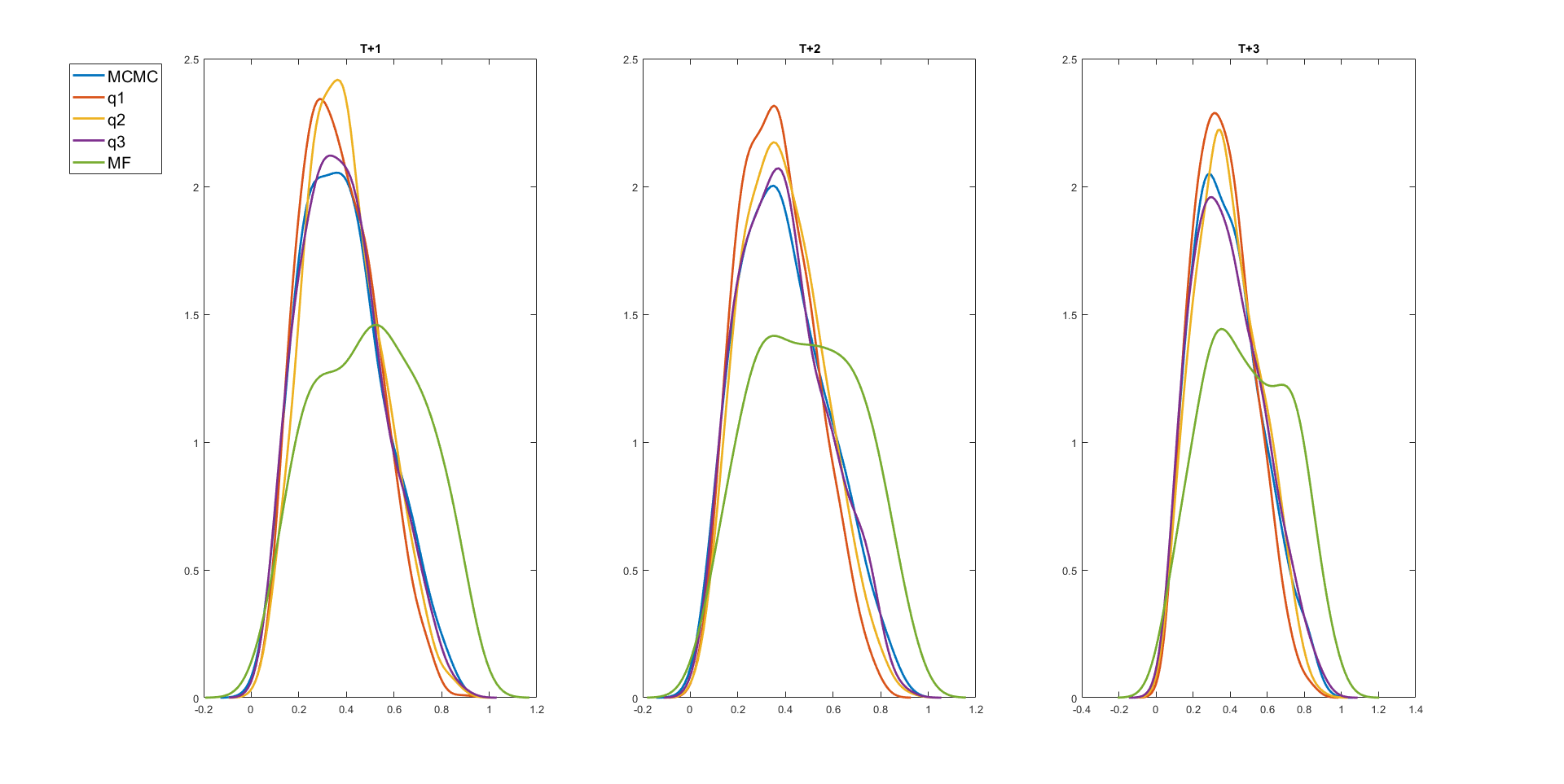}
\end{figure}

The predictive performance of the variational approximations is now compared to that of
the  particle MCMC method using the minimum variance portfolio. Figure \ref{fig:Plot-of- predictive densities onestepaheadSP100} shows the one-step ahead predictive density of the minimum variance portfolio estimated using the MCMC and the variational methods. The figure confirms that the mean-field variational approximation performs poorly. The estimates from $q_{\lambda}^{III}$ is the closest to the MCMC. Figure~\ref{fig:Plot-of-one step ahead predictive density rho23 SP100} plots
the multiple-step ahead predictive densities of the time-varying correlation
between Microsoft and AIG stock returns $\widehat{p}\left(\Gamma\left(y_{MSFT,T+h},y_{AIG,T+h}\right)|y_{1:T}\right)$
for $h=1,...,3$. The figure shows that all the variational predictive approximation densities, except the mean-field  approximation,
 are very close to the exact predictive densities obtained
from PMCMC. Again, the estimates from $q_{\lambda}^{III}$ is closest to MCMC.  Both Figures \ref{fig:Plot-of- predictive densities onestepaheadSP100}
and \ref{fig:Plot-of-one step ahead predictive density rho23 SP100} confirm
that the variational predictive densities of our variational methods do not underestimate the predictive
variances.

\begin{figure}[H]
\caption{SP100 dataset. Plots of the mean estimates of one hundred step ahead  predictive densities of the time-varying correlations
between Microsoft and AIG (left), Amazon and Boeing (Middle), and JP Morgan and Morgan Stanley (Right)
estimated using particle MCMC and the variational
approximations $q_{\lambda}^{I}$, $q_{\lambda}^{II}$, $q_{\lambda}^{III}$, and
$q_{\lambda}^{MF}$. \label{fig:Plot-of-one step ahead predictive density rho23 SP100meanestimates}}

\centering{}\includegraphics[width=15cm,height=8cm]{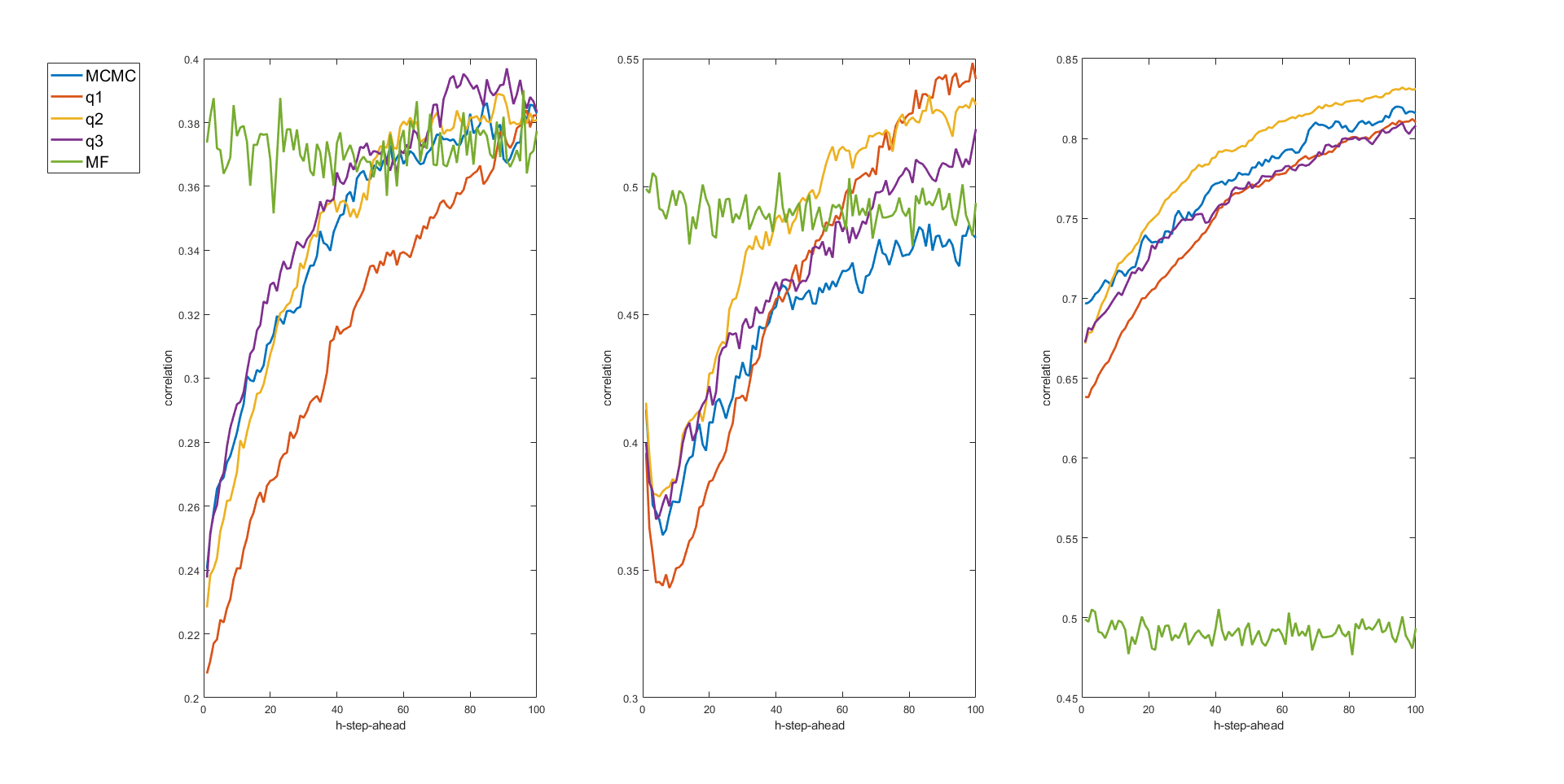}
\end{figure}

Figure \ref{fig:Plot-of-one step ahead predictive density rho23 SP100meanestimates} shows the mean estimates of one hundred step ahead  predictive densities of the time-varying correlation between Microsoft and AIG, Amazon and Boeing, and JP Morgan and Morgan Stanley.  The correlations between Microsoft and AIG, Amazon and Boeing, and  JP Morgan and Morgan Stanley industries tend to increase over time.
The mean-field variational approximation clearly does not  capture this time-varying correlation obtained from MCMC and the other variational approximations.

\begin{figure}[H]
\caption{SP100 dataset. The marginal posterior density plots of the parameters $\left\{ \psi_{\epsilon,4},\psi_{\epsilon,8},\psi_{\epsilon,20},\psi_{\epsilon,26}\right\} $
estimated using particle MCMC, $q_{\lambda}^{III}$, and $seq-q_{\lambda}^{III}$.
 \label{fig:The-plot-of psiSP100_4factor_seqVA}}
\centering{}\includegraphics[width=15cm,height=8cm]{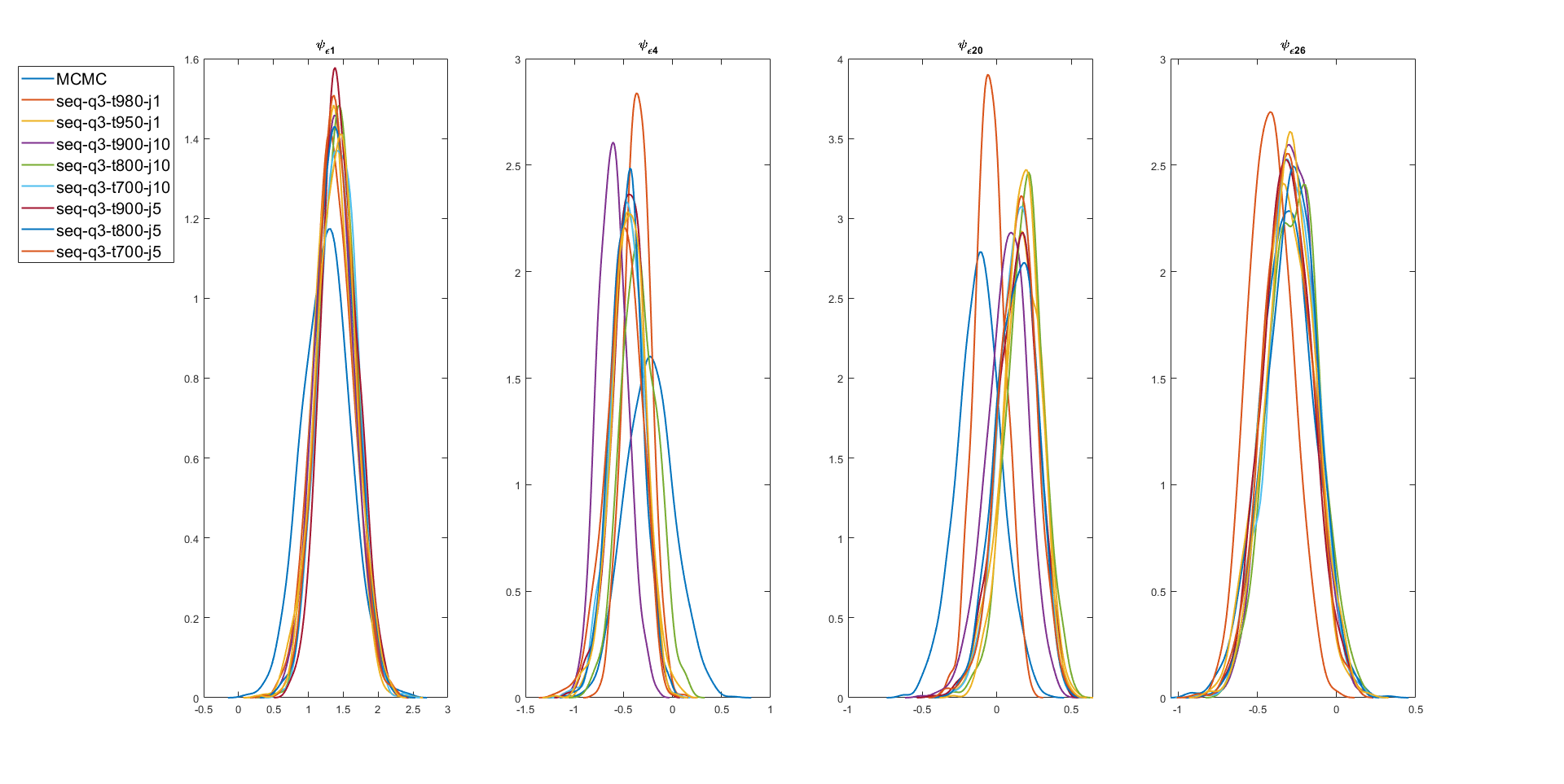}
\end{figure}

\begin{figure}[H]
\caption{SP100 dataset. Plots of the sequential one-step-ahead predictive densities of the minimum variance portfolio estimated using the variational
approximations $seq-q_{\lambda}^{III}$ from 19th July 2013 to 30th September 2013. \label{fig:Plot-of-one step ahead predictive densityseqVA}}

\centering{}\includegraphics[width=15cm,height=8cm]{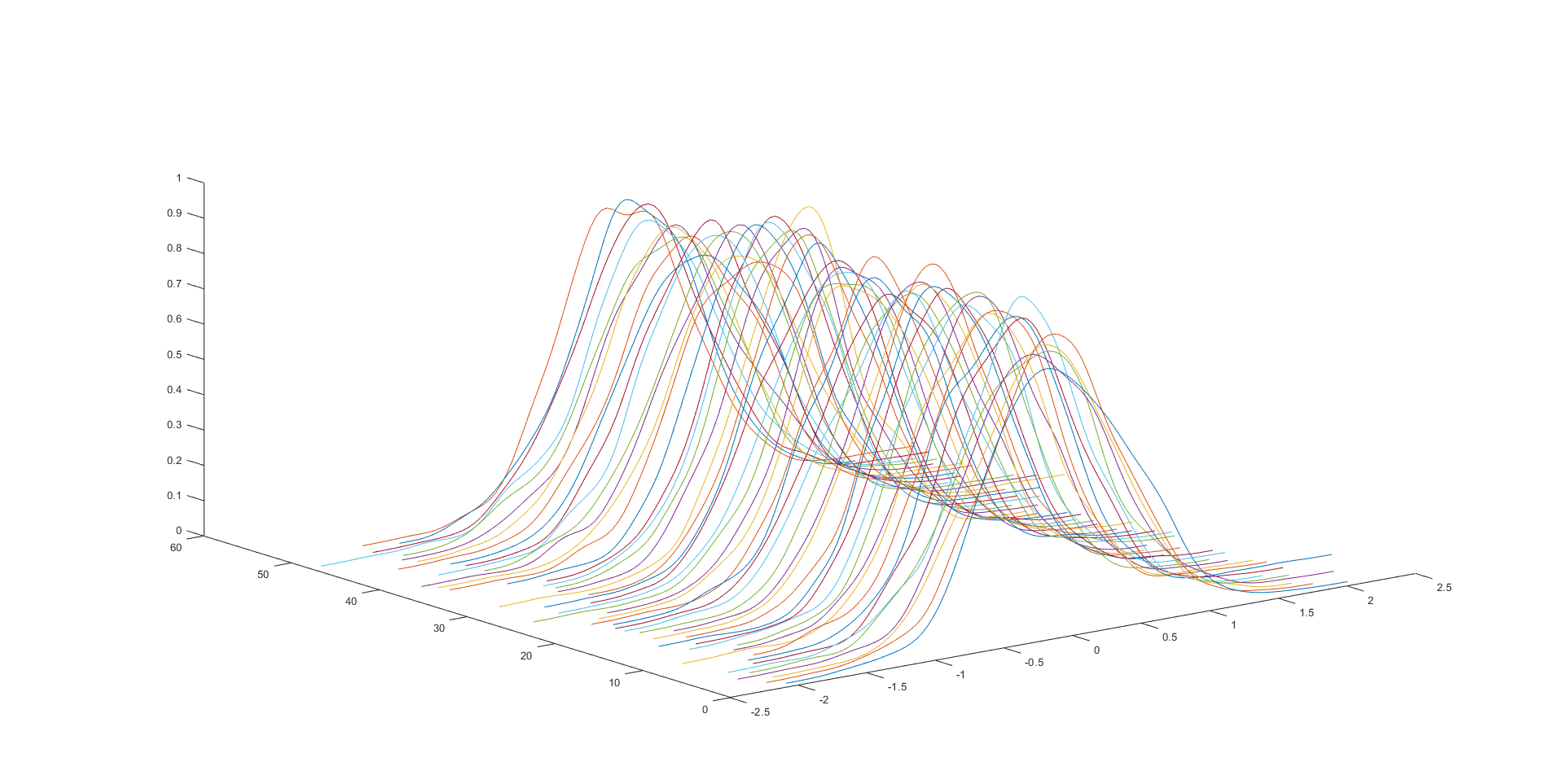}
\end{figure}

We now investigate the accuracy of the sequential variational method, in particular $seq-q_{\lambda}^{III}$ to the real data and compare it to standard $q_{\lambda}^{III}$ and particle MCMC. Different starting points and update frequencies are considered to evaluate the robustness of the sequential approach. The scenarios considered are: (1) starting points $t=950,980$ and updates after observing $j=1$ new observation until $T=1000$, (2) $t=700,800,900$ and $j=5$, and (3) $t=700,800,900$ and $j=10$.  Figure \ref{fig:The-plot-of psiSP100_4factor_seqVA} shows the estimated marginal posterior densities of the parameters $\left\{ \psi_{\epsilon,s}\right\} $ for $s=1,4,20,26$ using the different batch variational approximations, sequential variational approximations, and particle MCMC. All the estimates from the different versions of the sequential approximations are very close to the standard variational approximation and the exact MCMC method. Similar conclusions can be made from Figure \ref{fig:The-plot-of betaloadingSP100_4factor_seqVA} in Appendix \ref{sec:Additional-Figures-for SP100}.
This shows that the performance of  $seq-q_{\lambda}^{III}$ does not deteriorate
when applied to the real data and is robust to
 the starting points and update frequencies. Another important advantage of the sequential variational approximation
$seq-q_{\lambda}^{III}$ is that it can provide
sequential one-step ahead predictive densities of each stock return
and any weighted portfolio as well as the time-varying correlation between
industry stock returns; these are difficult to obtain using MCMC or
particle MCMC approaches. For example, Figure \ref{fig:Plot-of-one step ahead predictive densityseqVA}
plots the sequential one-step ahead predictive densities of the minimum variance portfolio
of the SP100 stock returns from 19th July 2013 to 30th September 2013.

\begin{figure}[H]
\caption{SP100 dataset. Plots of the one-step-ahead predictive densities of Apple, JP Morgan, Morgan Stanley, and the minimum variance portfolio estimated using variational
approximation $q_{\lambda}^{III}$ for FSV models with normal and t distributions for the idiosyncratic errors. \label{fig:Plot-of-one step ahead predictive densityNormalt}}

\centering{}\includegraphics[width=15cm,height=8cm]{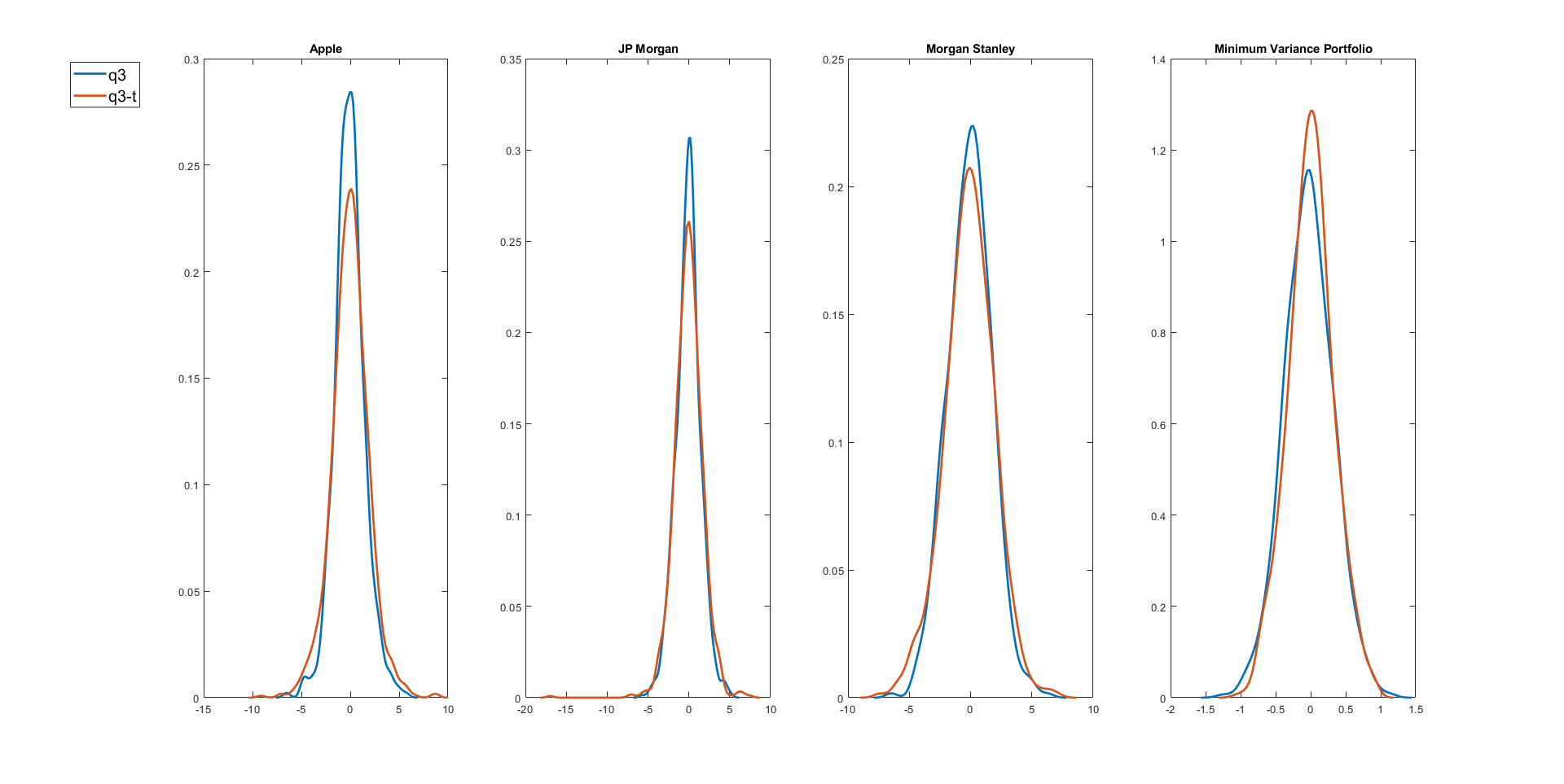}
\end{figure}

Finally, we compare the one-step-ahead predictive densities of some of the individual stock returns and the minimum variance portfolio from the FSV model with normal and t-distributions for the idiosyncratic and latent factor errors estimated using the variational approximations. Figure \ref{fig:Plot-of-one step ahead predictive densityNormalt} shows the one-step-ahead predictive densities of Apple, JP Morgan, and Morgan Stanley industries as well as the minimum variance portfolio. It is clear that the predictive densities from the FSV model with t-distribution has some characteristics of financial stock returns, such as heavy tails, particularly in the left tails, and also sharper peaks in the middle of the distributions. Figure \ref{fig:dof} in Appendix \ref{sec:Additional-Figures-for SP100} plots the degree of freedom parameters $v_{f,k}$ for $k=1,...,K$ and $v_{\epsilon,s}$ for $s=1,...,S$.

\section{Conclusions\label{sec:Conclusions}}

This paper proposes fast and efficient variational methods for estimating
posterior densities for  the high dimensional FSV model in 
Section~\ref{sec:Description-of-Factor SV model} with normal and t distributions for the idiosyncratic and latent factor errors. It also extends these batch methods to sequential variational
updating that allows for one-step and multi-step ahead variational forecast distributions as new data arrives.

We show that the methods work
well for both simulated and real data and are much faster than exact MCMC approaches. The data analyses  suggest that a) Variational approaches
are much faster than MCMC and particle MCMC approaches,
especially for high dimensional time series. b)  The
variational approximations capture the posterior means of the parameters
of the multivariate factor SV model quite accurately, except for the mean-field
variational approximation; however, there is some slight underestimation
in the posterior variances for some of the parameters. The variational approximation $q_{\lambda}^{III}$ is the most accurate for both the simulated and real datasets.
It is therefore important to take into account the posterior dependence between the latent factors and other latent states and the parameters in the FSV model.
The dependence between different idiosyncratic and factor log-volatilities can be ignored.
c) The variational approximations $q_{\lambda}^{III}$ also
produces the best one-step and multiple-step ahead predictive densities
for the individual stock returns, the minimum variance portfolio, and
the time-varying correlation between any pair of stock returns when compared to the exact
particle MCMC method, without the problem of underestimating the prediction variance.
d) The mean-field variational approximation appears to perform poorly in both batch and sequential modes. (e) The sequential approach $seq-q_{\lambda}^{III}$ is robust against the starting points and update frequencies and produces state and parameter estimates that are close to batch $q_{\lambda}^{III}$ and exact particle MCMC. It also accurately estimates the predictive densities of the individual stock returns, the minimum variance portfolio, and
the time-varying correlation between any pair of stock returns.

The FSV model in the paper assumes that the factor loadings are static with uncorrelated latent factors and idiosyncratic errors. Therefore, it may require more factors to properly explain the comovement of multivariate time series than in more complex factor models which allow for time-varying factor loadings \citep{Lopes2007,Barigozzi2019}, and correlated latent factors \citep{Zhou2014} and idiosyncratic errors \citep{Bai2002}. The FSV model in our article is still widely used in practice, for example \cite{Kastner2019}, \cite{Kastner:2017}, and \cite{Li:scharth}.
It is also straightforward to incorporate other distributions, such as t, skew normal, or skew-t distributions for the idiosyncratic errors and latent factors.
Section~\ref{sec:Multivariate-FSV-Model t-distribution} of the supplement discusses the FSV model with t-distribution for the idiosyncratic errors and latent factors.
An immediate extension of our work would be to such more complex factor models.

Future work will consider developing variational approaches for other complex and high-dimensional time series models, such as the vector autoregressive model with stochastic volatility (VAR) of \citet{Chan2020} and multivariate financial time series model with recurrent neural network type architectures, e.g. the Long-Short term Memory model of \citet{Hochreiter1997}.


%



\bibliographystyle{apalike}
\bibliography{references_v1}

\pagebreak
\renewcommand{\thesection}{S\arabic{section}}
\renewcommand{\thetable}{S\arabic{table}}
\renewcommand{\thefigure}{S\arabic{figure}}
\setcounter{section}{0}
\setcounter{table}{0}
\setcounter{figure}{0}

\section*{Online Supplement for \textquotedblleft Fast Variational Approximation for Multivariate Factor Stochastic Volatility Model\textquotedblright}

\section{Additional Tables and Figures for the simulated dataset \label{sec:Additional-Figures-for simulation}}

\begin{figure}[H]
\caption{Simulated dataset. Scatter plots of the posterior means of the latent factors
$f_{k,t}$, for $k=1,...,4$ and $t=10,20,...,1000$ estimated
using particle MCMC on the x-axis and the variational approximation $q_{\lambda}^{I}$, $q_{\lambda}^{II}$,
$q_{\lambda}^{III}$, and $q_{\lambda}^{MF}$, on the y-axis.
\label{fig:Scatter-plot-of posterior means of latent factors}}

\centering{}\includegraphics[width=15cm,height=8cm]{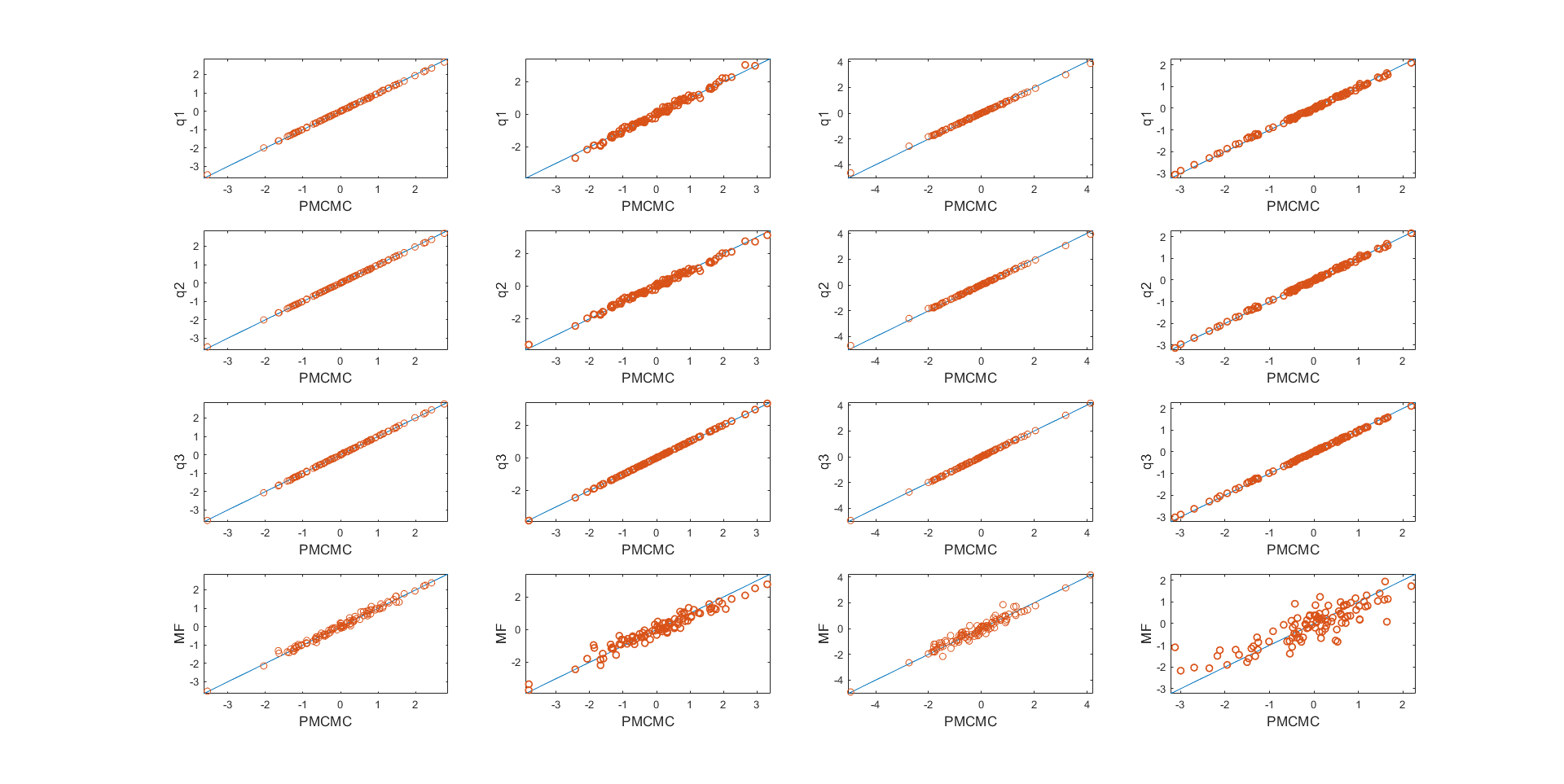}
\end{figure}

\begin{figure}[H]
\caption{Simulated dataset. Scatter plots of the posterior standard deviation of the latent factors
$f_{k,t}$, for $k=1,...,4$ and $t=10,20,...,1000$ estimated
using particle MCMC on the x-axis and the variational approximation $q_{\lambda}^{I}$, $q_{\lambda}^{II}$,
$q_{\lambda}^{III}$, and $q_{\lambda}^{MF}$, on the y-axis.
\label{fig:Scatter-plot-of posterior standard deviation of latent factors}}

\centering{}\includegraphics[width=15cm,height=8cm]{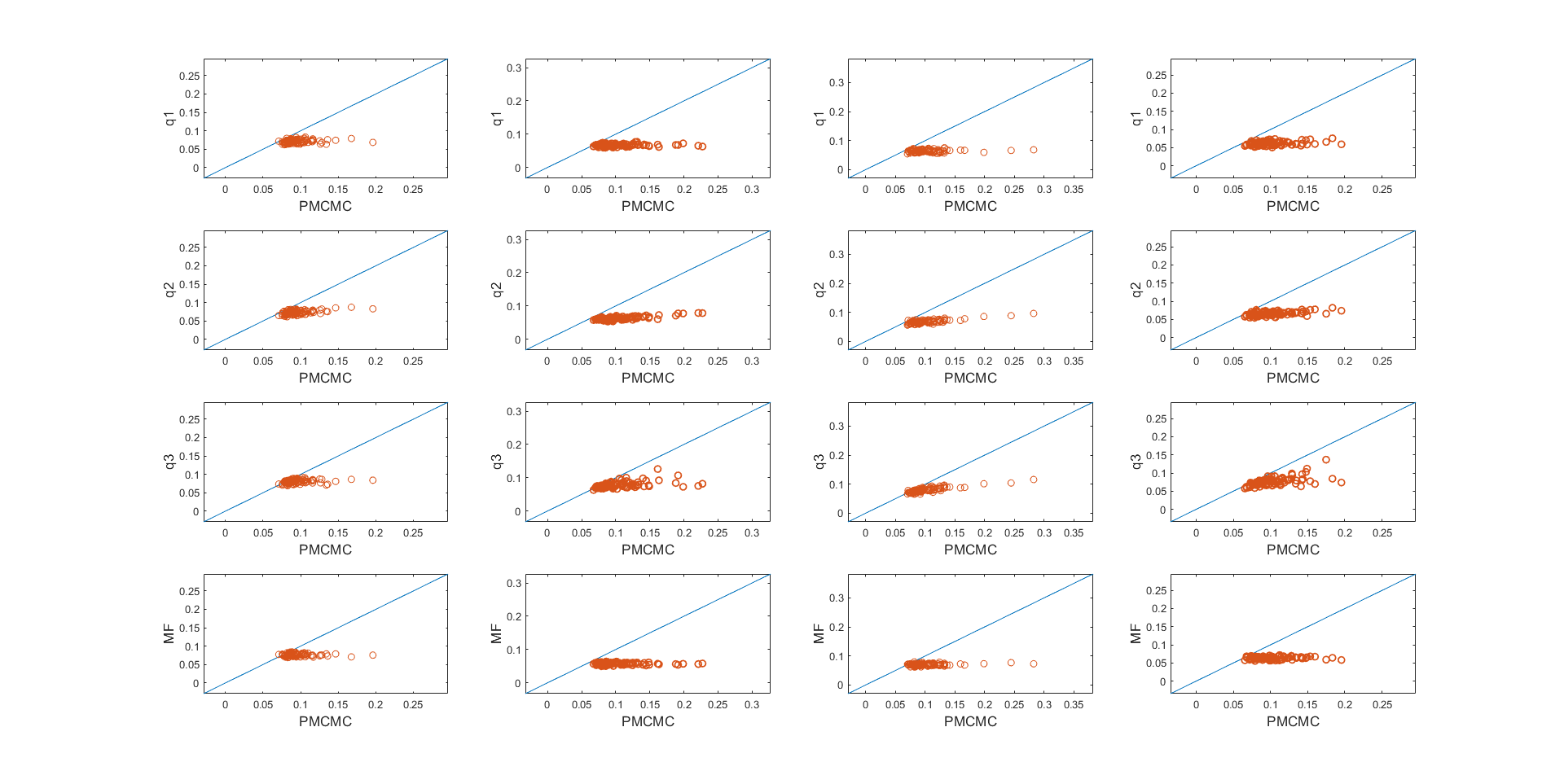}
\end{figure}

\begin{figure}[H]
\caption{Simulated dataset. Scatter plots of the posterior means of the factors log-volatilities
$h_{f,k,t}$, for $k=1,...,4$ and $t=10,20,...,1000$ estimated
using particle MCMC on the x-axis and the variational approximation $q_{\lambda}^{I}$, $q_{\lambda}^{II}$,
$q_{\lambda}^{III}$, and $q_{\lambda}^{MF}$, on the y-axis.
\label{fig:Scatter-plot-of posterior standard deviation of factorlogvolatility}}

\centering{}\includegraphics[width=15cm,height=8cm]{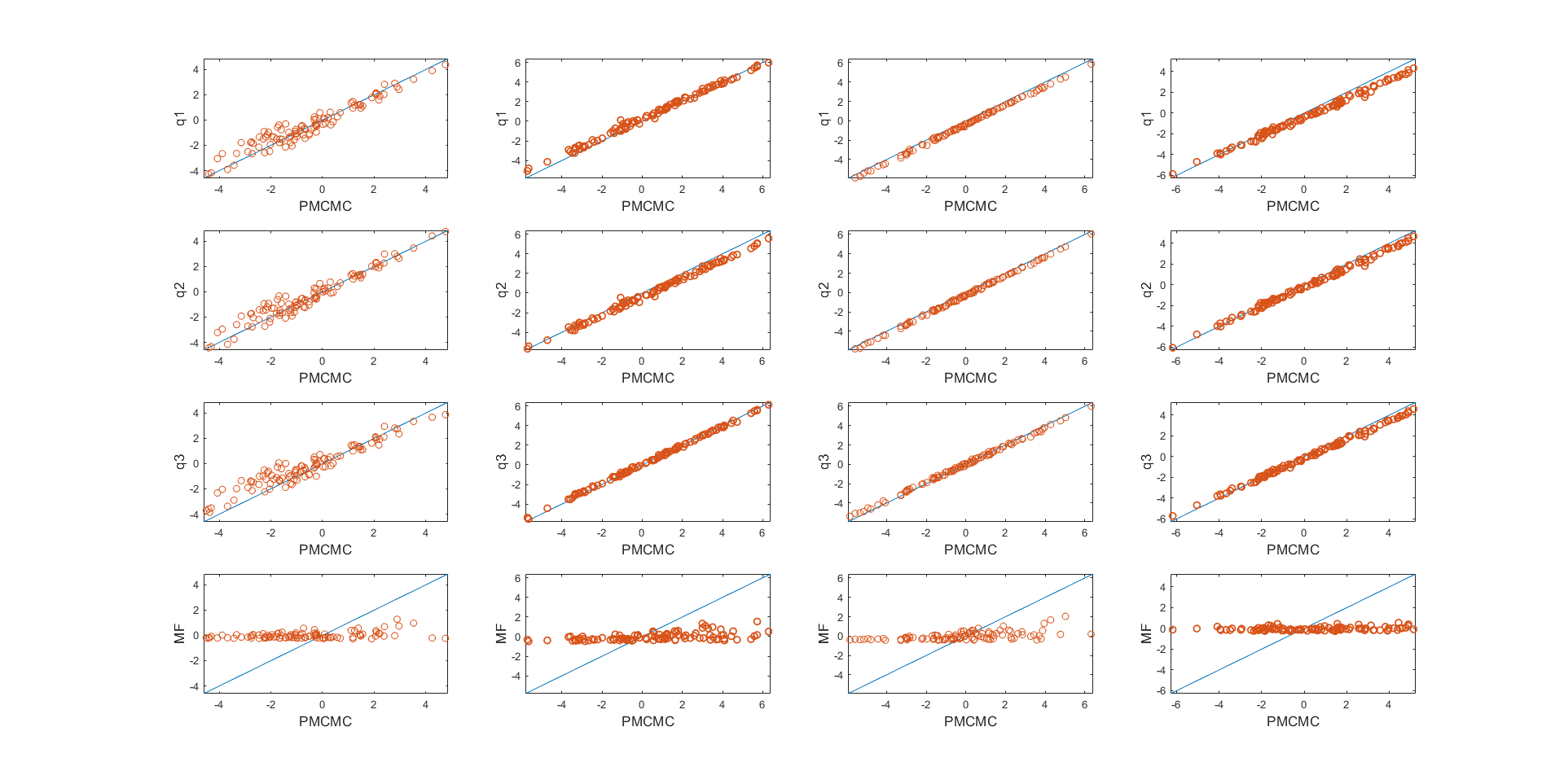}
\end{figure}

\begin{figure}[H]
\caption{Simulated dataset. Scatter plots of the posterior means of the idiosyncratic log-volatilities
$h_{\epsilon,s,t}$, for $s=1,...,100$ and $t=10,20,...,1000$ estimated
using particle MCMC on the x-axis and the variational approximation $q_{\lambda}^{I}$ on the y-axis.
\label{fig:Scatter-plot-of posterior means of idiosyncraticlogvolatility q1}}

\centering{}\includegraphics[width=15cm,height=8cm]{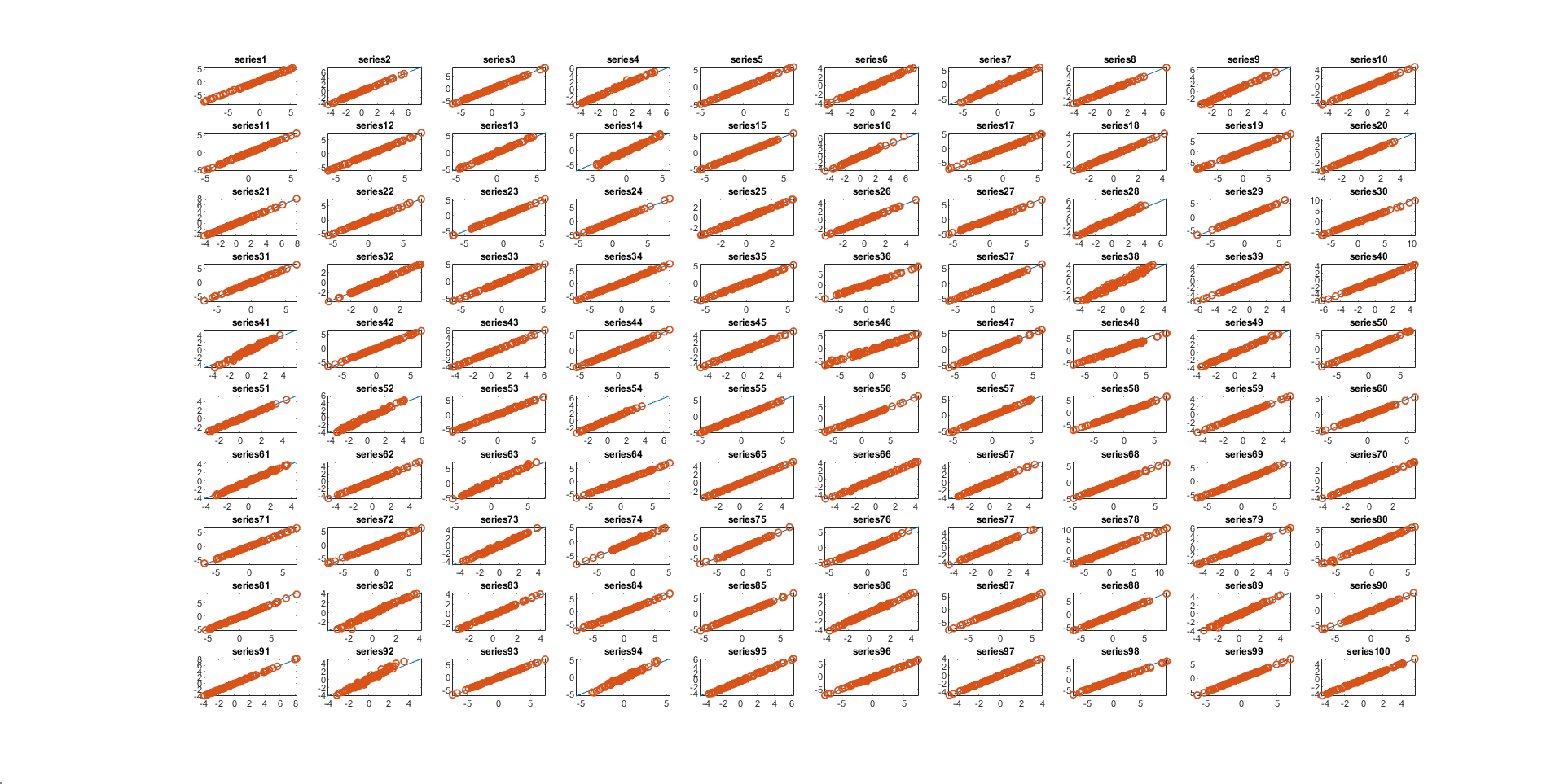}
\end{figure}

\begin{figure}[H]
\caption{Simulated dataset. Scatter plots of the posterior means of the idiosyncratic log-volatilities
$h_{\epsilon,s,t}$, for $s=1,...,100$ and $t=10,20,...,1000$ estimated
using particle MCMC on the x-axis and the variational approximation $q_{\lambda}^{II}$ on the y-axis.
\label{fig:Scatter-plot-of posterior means of idiosyncraticlogvolatility q2}}

\centering{}\includegraphics[width=15cm,height=8cm]{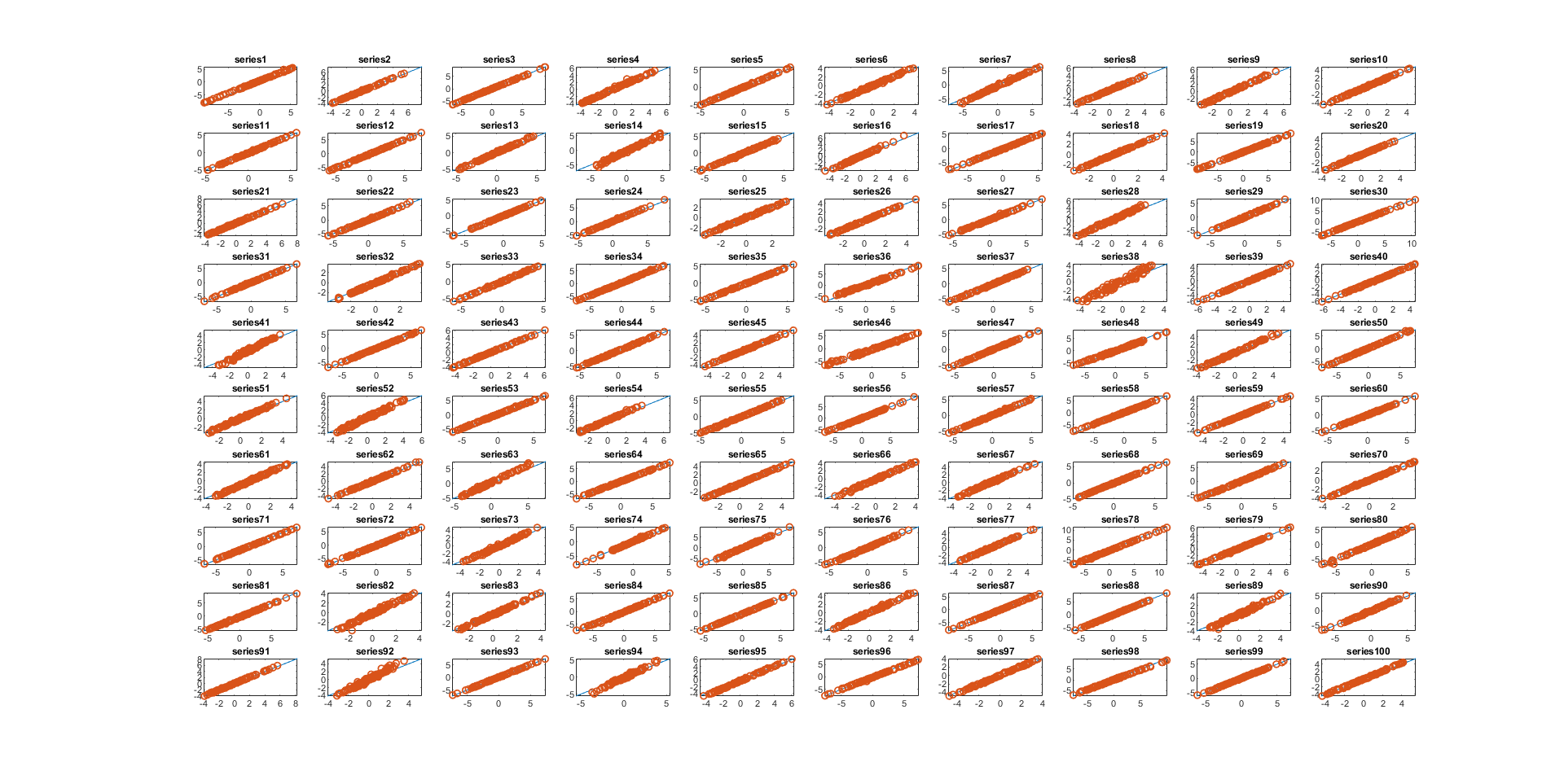}
\end{figure}

\begin{figure}[H]
\caption{Simulated dataset. Scatter plots of the posterior means of the idiosyncratic log-volatilities
$h_{\epsilon,s,t}$, for $s=1,...,100$ and $t=10,20,...,1000$ estimated
using particle MCMC on the x-axis and the variational approximation  $q_{\lambda}^{MF}$, on the y-axis.
\label{fig:Scatter-plot-of posterior means of idiosyncraticlogvolatility MF}}

\centering{}\includegraphics[width=15cm,height=8cm]{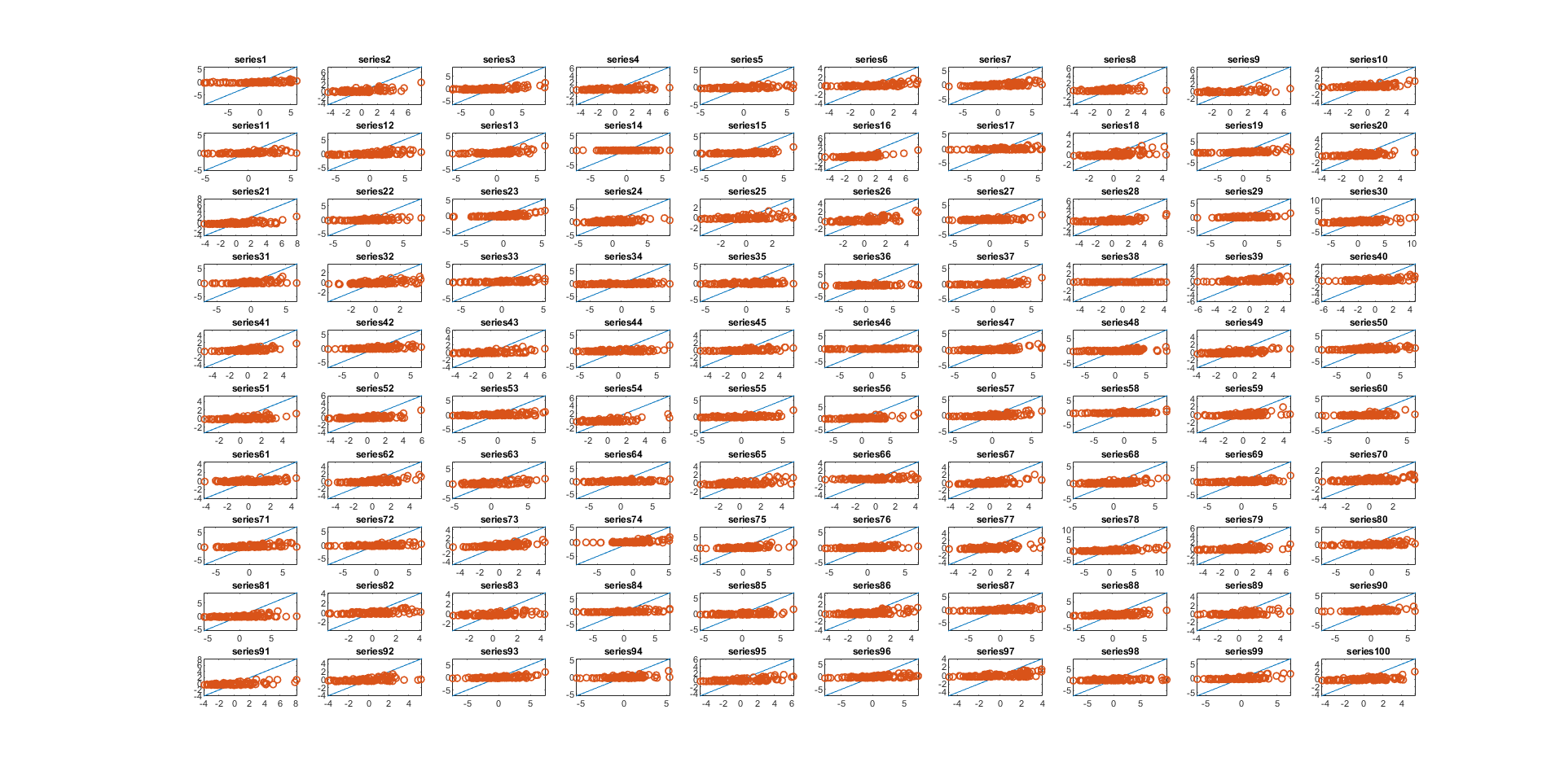}
\end{figure}

\begin{figure}[H]
\caption{Simulated dataset. Plots of the predictive densities of the time-varying correlations
$\widehat{p}\left(\Gamma\left(y_{2,T+h},y_{4,T+h}\right)|y_{1:T}\right), h=1,...,10$,
between series
2 and 4 estimated using particle MCMC and the variational
approximations $q_{\lambda}^{I}$, $q_{\lambda}^{II}$, $q_{\lambda}^{III}$, and
$q_{\lambda}^{MF}$. \label{fig:Plot-of-one step ahead predictive density rho }}

\centering{}\includegraphics[width=15cm,height=8cm]{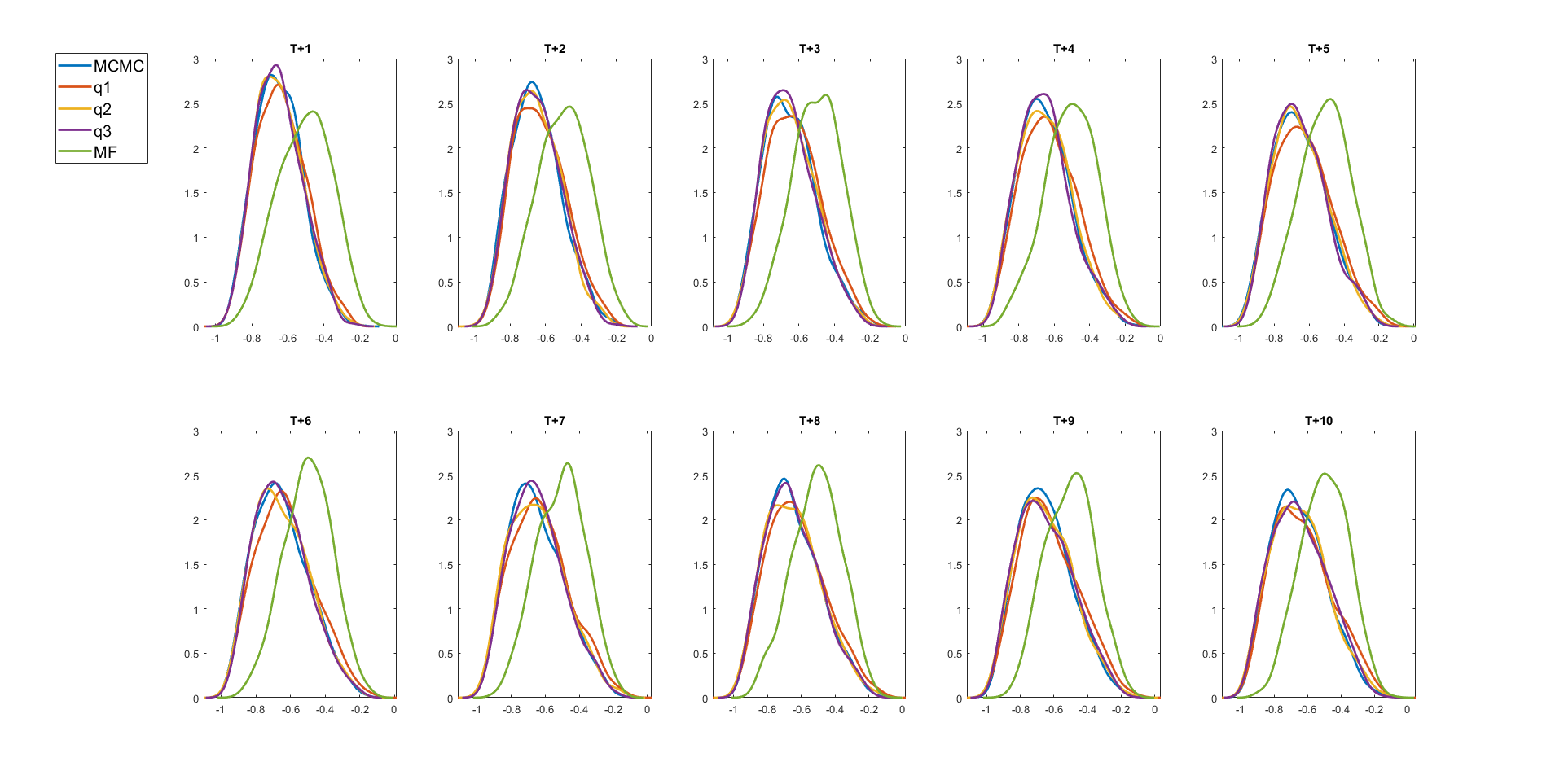}
\end{figure}

\begin{figure}[H]
\caption{Simulated dataset. The marginal posterior density plots of the parameters $\left\{ \psi_{\epsilon,4},\psi_{\epsilon,8},\psi_{\epsilon,20},\psi_{\epsilon,26}\right\}$
estimated using particle MCMC, batch VA, and sequential VA.
 \label{fig:The-plot-of psisim_4factor_seqVA}}
\centering{}\includegraphics[width=15cm,height=8cm]{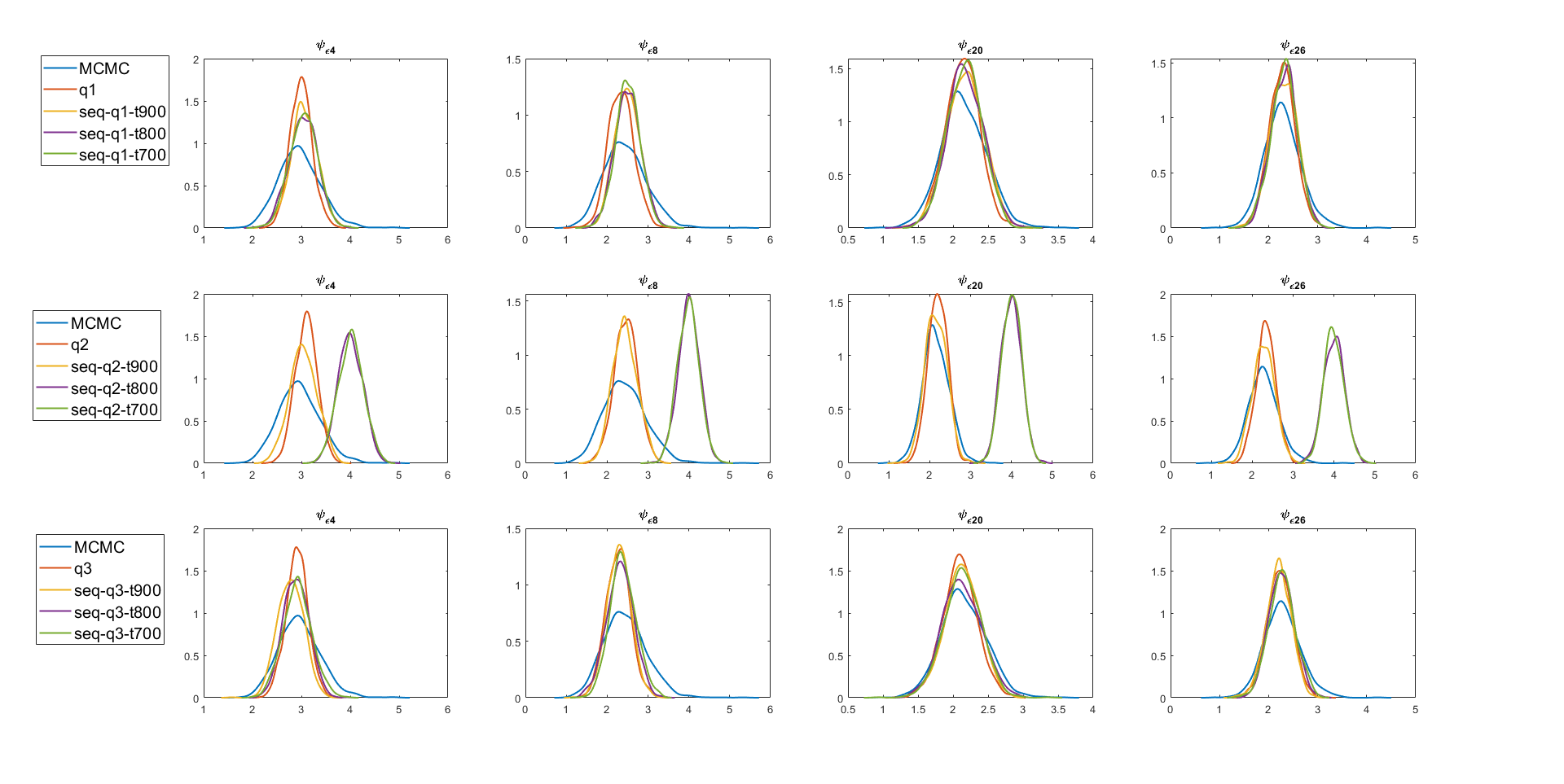}
\end{figure}

\begin{figure}[H]
\caption{Simulated dataset. Plot of the predictive densities $\widehat{p}\left(y_{T+1}|y_{1:T}\right)$
of an optimally weighted series, estimated
using particle MCMC, VA, and sequential VA. \label{fig:Plot-of- predictive densities-seqVA}}

\begin{centering}
\includegraphics[width=15cm,height=8cm]{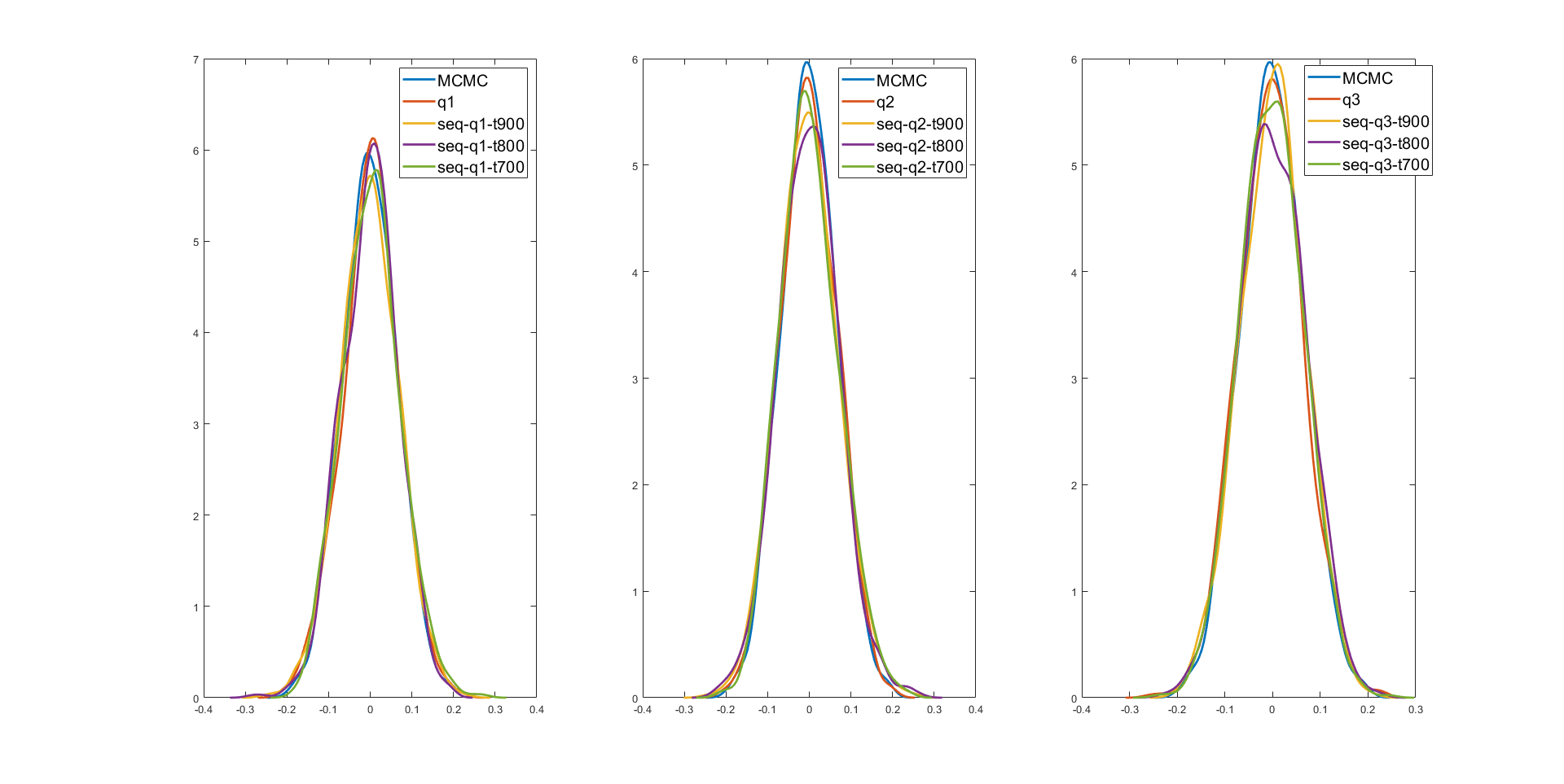}
\par\end{centering}
\end{figure}

\begin{figure}[H]
\caption{Simulated dataset. Plots of the predictive densities of the time-varying correlations
$\widehat{p}\left(\Gamma\left(y_{2,T+1},y_{3,T+1}\right)|y_{1:T}\right)$,
between series
2 and 3 estimated using particle MCMC, VA, and the sequential VA. \label{fig:Plot-of-one step ahead predictive density rho23_seqVA}}

\centering{}\includegraphics[width=15cm,height=8cm]{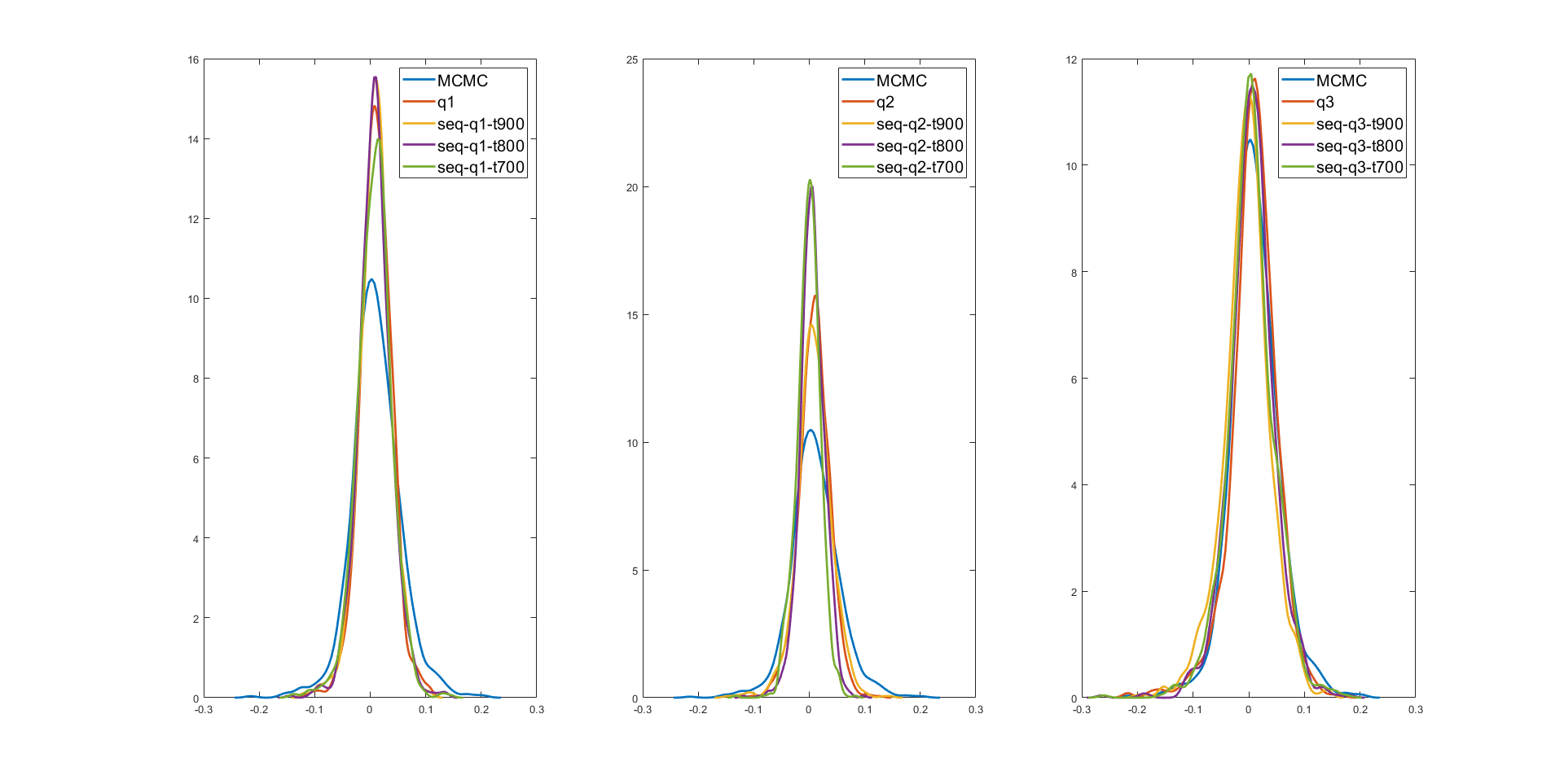}
\end{figure}

\section{Additional Tables and Figures for the SP100 dataset \label{sec:Additional-Figures-for SP100}}

\begin{figure}[H]
\caption{SP100 dataset. The marginal posterior density plots of some of the elements of $\beta$
estimated using particle MCMC,   $q_{\lambda}^{I}$,
$q_{\lambda}^{II}$, $q_{\lambda}^{III}$, and $q_{\lambda}^{MF}$.
 \label{fig:The-plot-of betaloadingSP100_4factor}}

\centering{}\includegraphics[width=15cm,height=8cm]{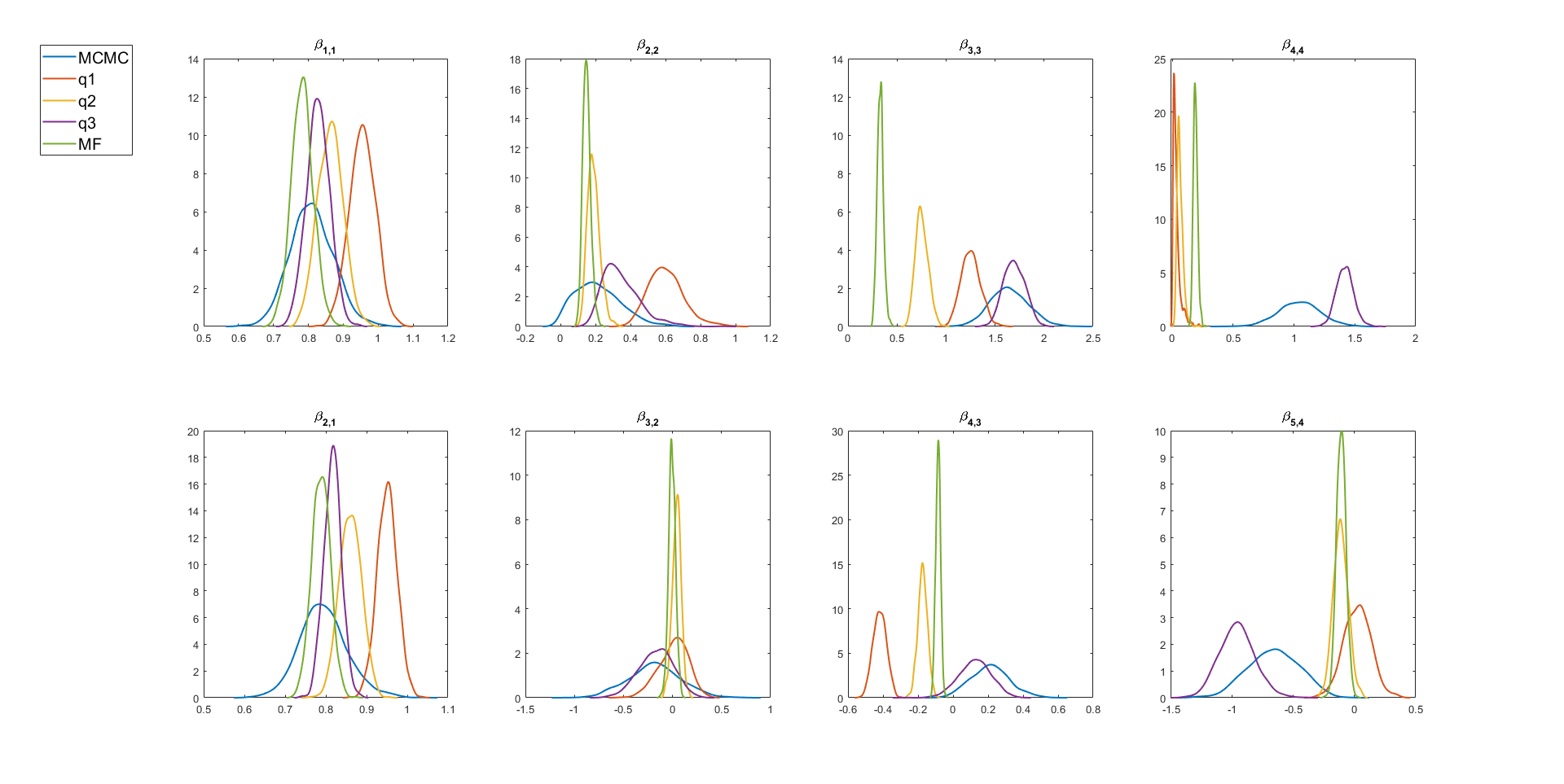}
\end{figure}

\begin{figure}[H]
\caption{SP100 dataset. The marginal posterior density plots of some of the 1st latent factors
estimated using particle MCMC,  $q_{\lambda}^{I}$,
$q_{\lambda}^{II}$, $q_{\lambda}^{III}$ and $q_{\lambda}^{MF}$.
 \label{fig:The-plot-of latentfactorsSP100_4factor}}

\centering{}\includegraphics[width=15cm,height=8cm]{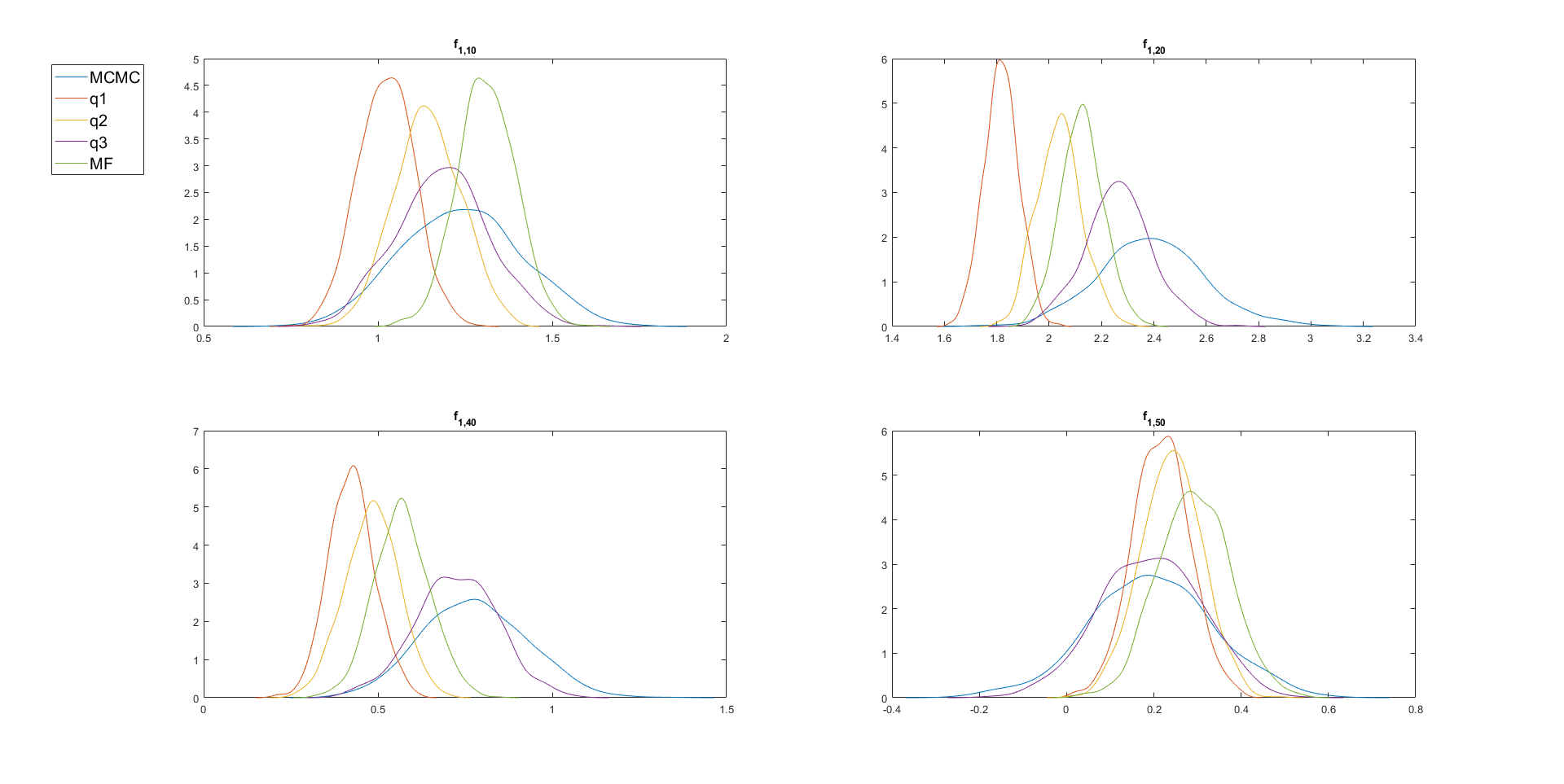}
\end{figure}

\begin{figure}[H]
\caption{SP100 dataset. Scatter plots of the posterior means of the idiosyncratic log-volatilities
$h_{\epsilon,s,t}$, for $s=1,...,90$ and $t=10,20,...,1000$ estimated
using particle MCMC on the x-axis and the variational approximation $q_{\lambda}^{I}$ on the y-axis.
\label{fig:Scatter-plot-of posterior means of idiosyncraticlogvolatility SP100 q1}}

\centering{}\includegraphics[width=15cm,height=8cm]{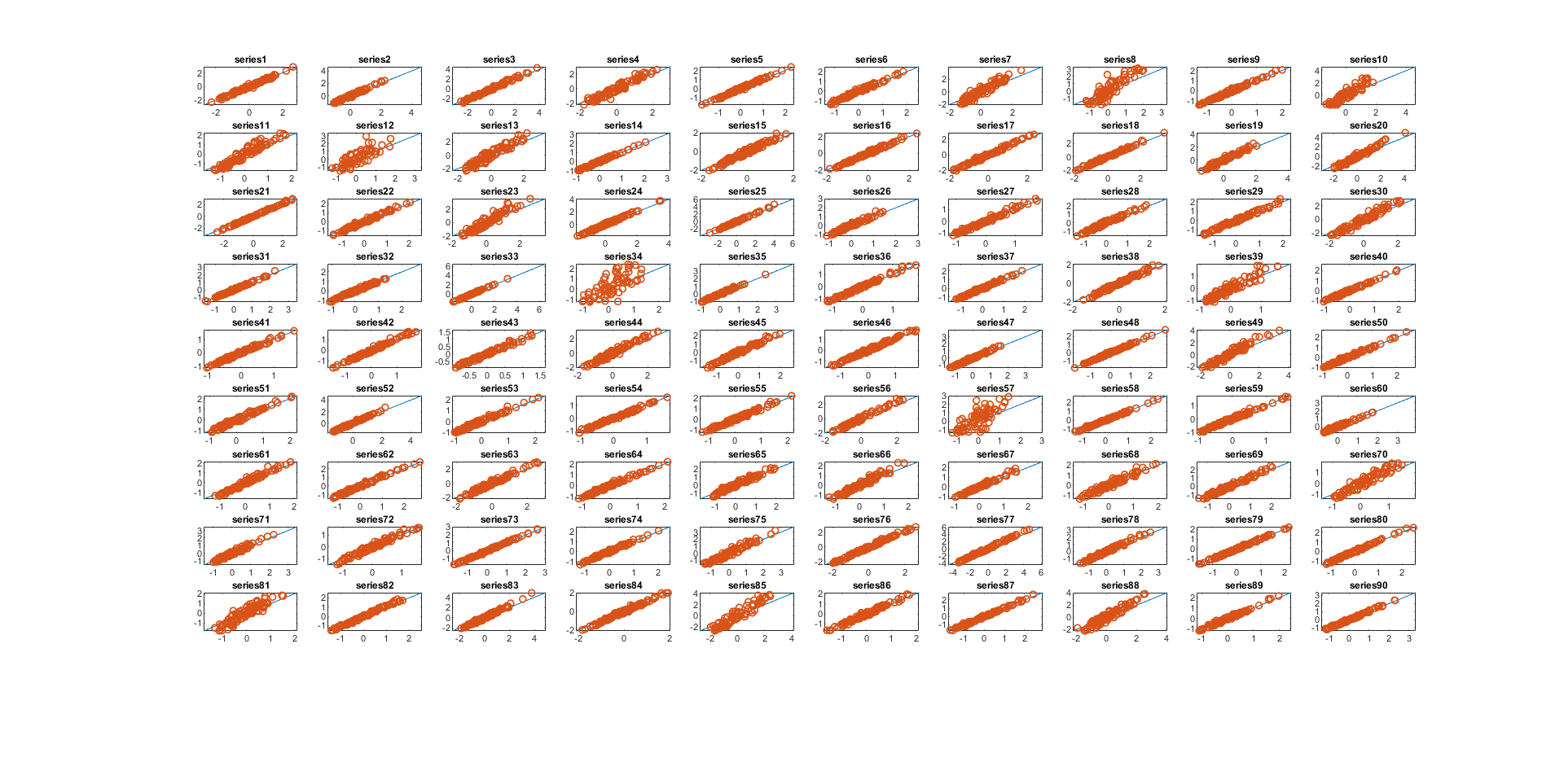}
\end{figure}

\begin{figure}[H]
\caption{SP100 dataset. Scatter plots of the posterior means of the idiosyncratic log-volatilities
$h_{\epsilon,s,t}$, for $s=1,...,90$ and $t=10,20,...,1000$ estimated
using particle MCMC on the x-axis and the variational approximation $q_{\lambda}^{II}$ on the y-axis.
\label{fig:Scatter-plot-of posterior means of idiosyncraticlogvolatility SP100 q2}}

\centering{}\includegraphics[width=15cm,height=8cm]{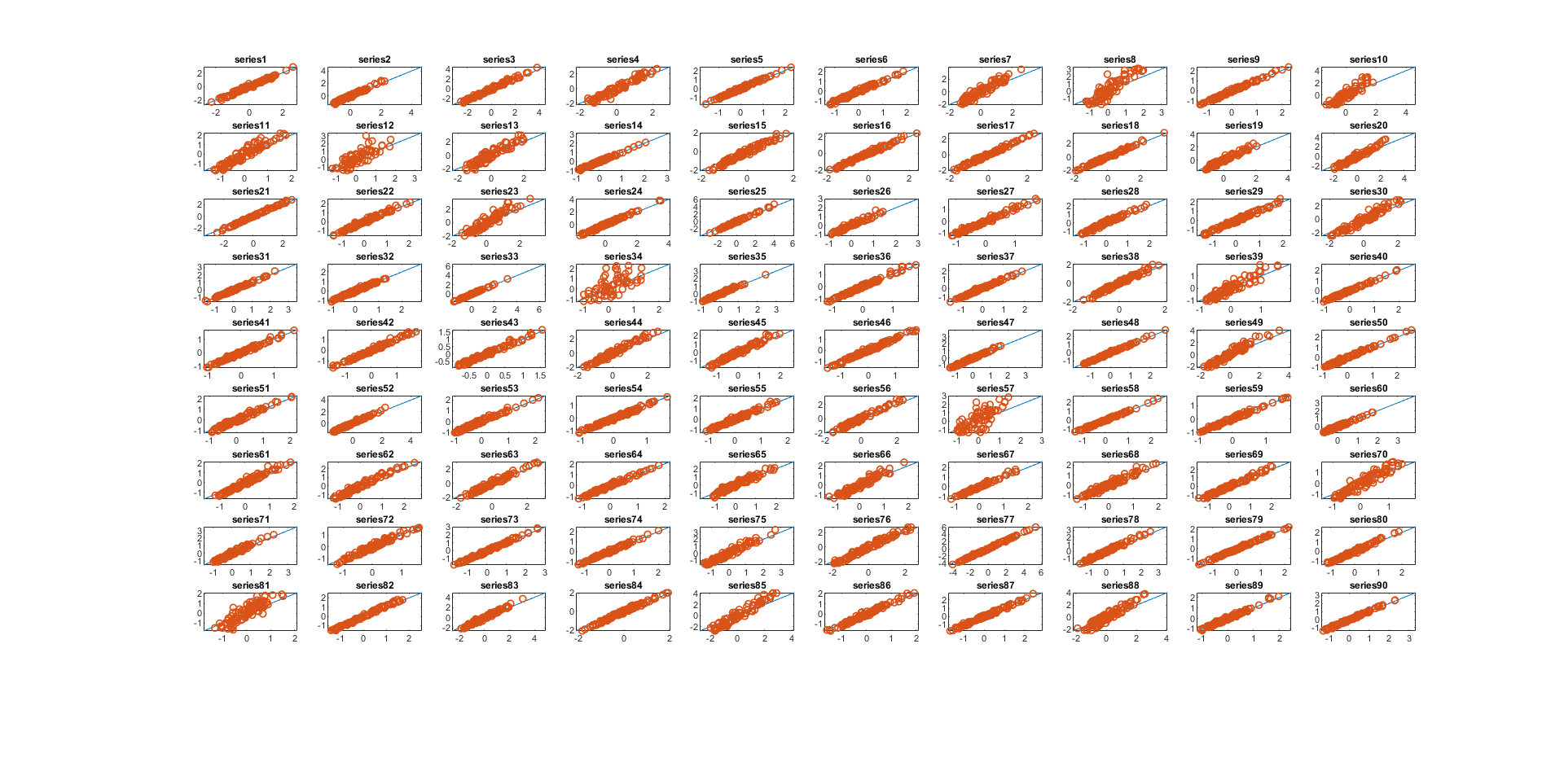}
\end{figure}

\begin{figure}[H]
\caption{SP100 dataset. Scatter plots of the posterior means of the idiosyncratic log-volatilities
$h_{\epsilon,s,t}$, for $s=1,...,90$ and $t=10,20,...,1000$ estimated
using particle MCMC on the x-axis and the variational approximation $q_{\lambda}^{III}$, on the y-axis.
\label{fig:Scatter-plot-of posterior means of idiosyncraticlogvolatility SP100 q3}}

\centering{}\includegraphics[width=15cm,height=8cm]{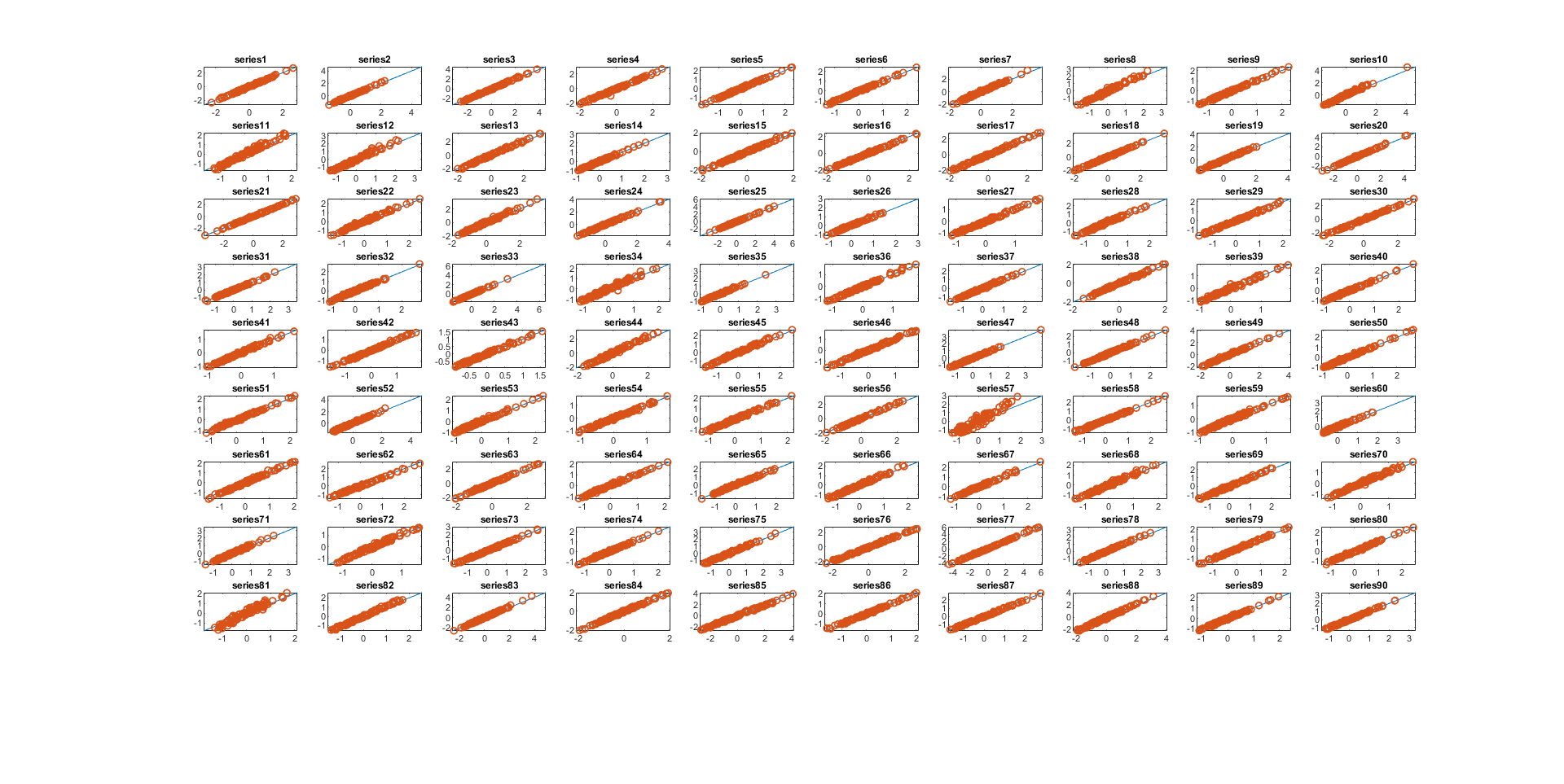}
\end{figure}

\begin{figure}[H]
\caption{SP100 dataset. Scatter plots of the posterior means of the idiosyncratic log-volatilities
$h_{\epsilon,k,t}$, for $s=1,...,90$ and $t=10,20,...,1000$ estimated
using particle MCMC on the x-axis and the variational approximation  $q_{\lambda}^{MF}$, on the y-axis.
\label{fig:Scatter-plot-of posterior means of idiosyncraticlogvolatility SP100 MF}}

\centering{}\includegraphics[width=15cm,height=8cm]{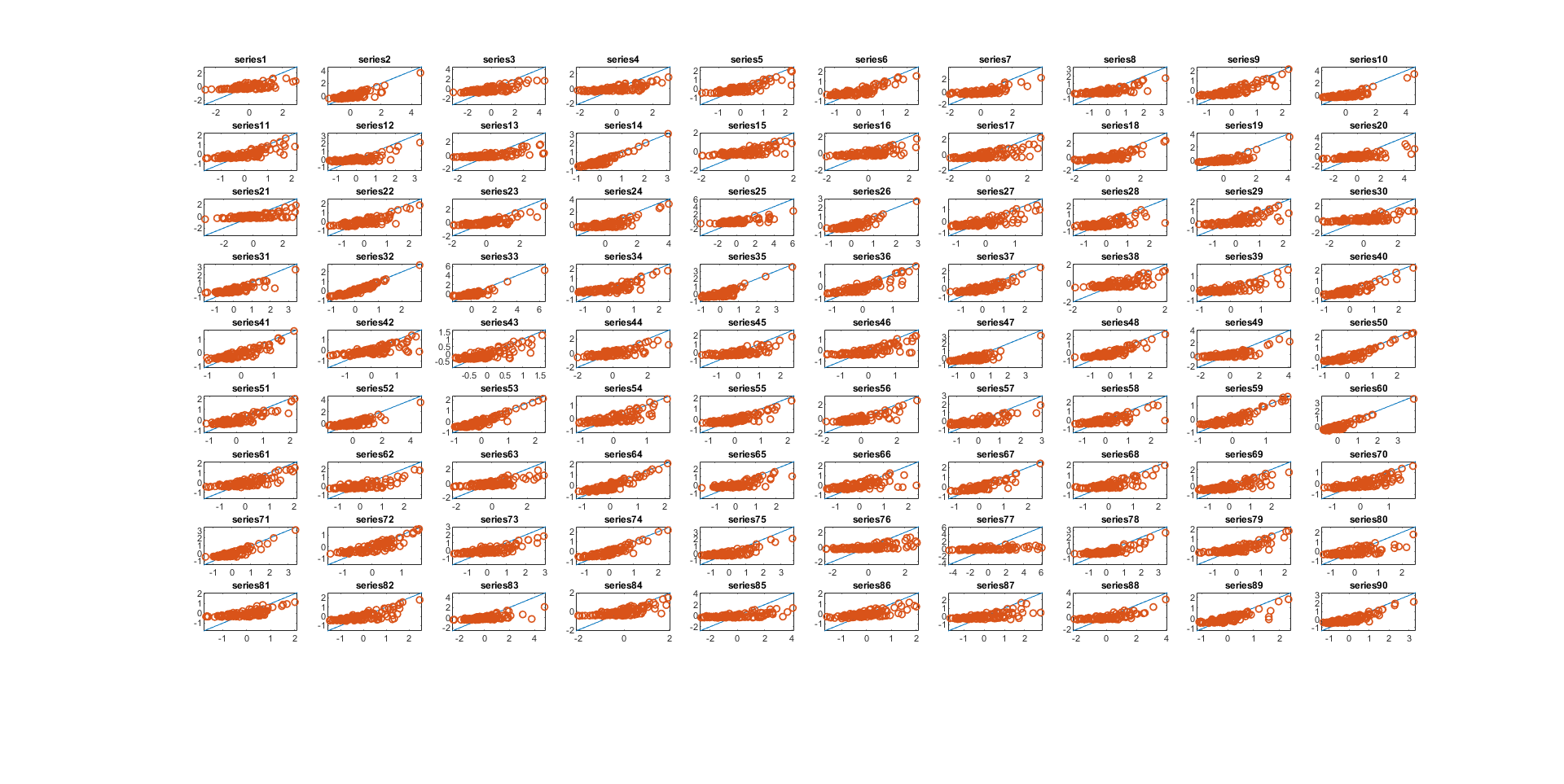}
\end{figure}

\begin{figure}[H]
\caption{SP100 dataset. The marginal posterior density plots of some of the $\beta$
estimated using particle MCMC, $q_{\lambda}^{III}$ and $seq-q_{\lambda}^{III}$.
 \label{fig:The-plot-of betaloadingSP100_4factor_seqVA}}

\centering{}\includegraphics[width=15cm,height=8cm]{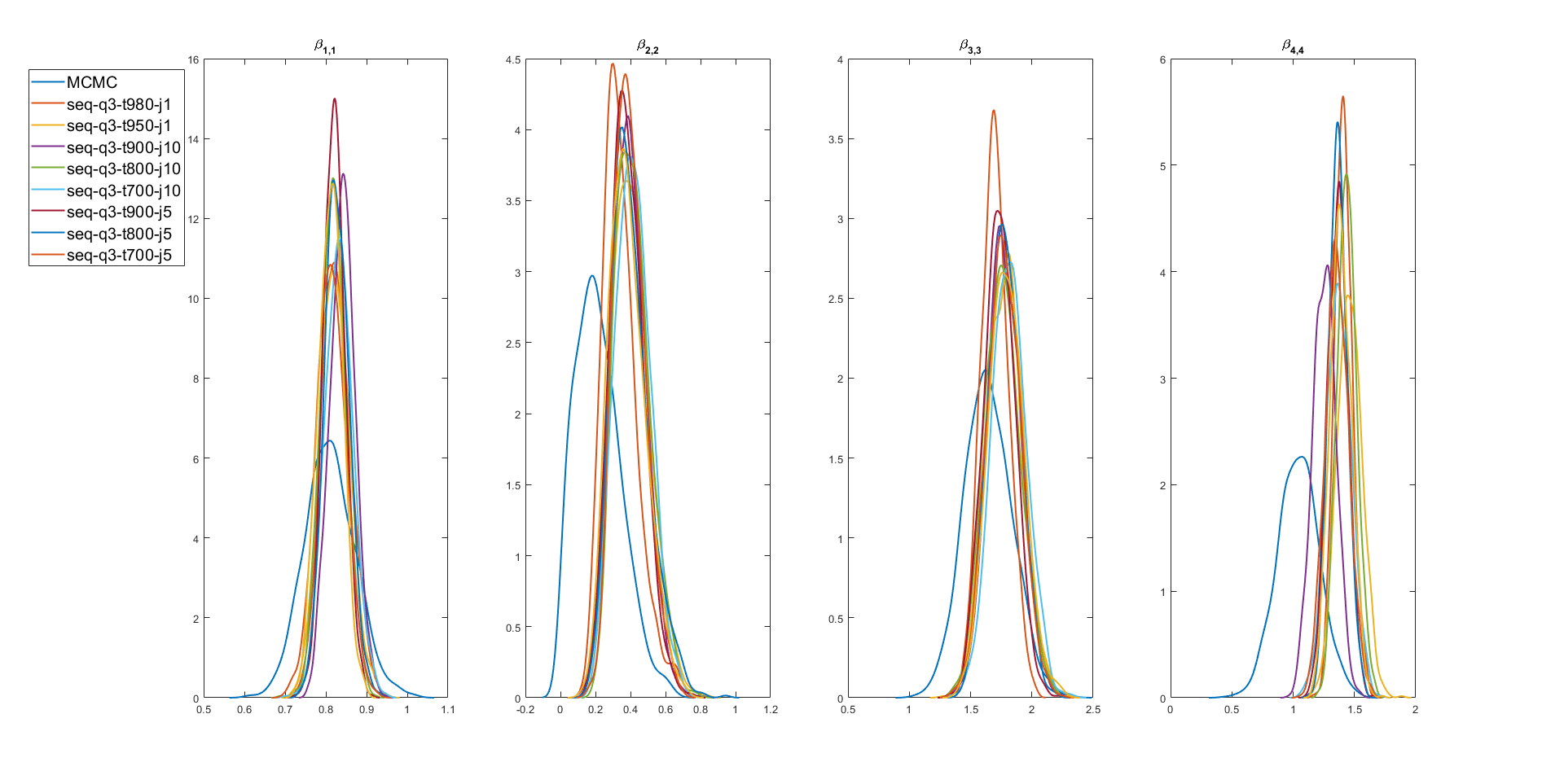}
\end{figure}

\begin{figure}[H]
\caption{SP100 dataset. The posterior mean estimates of the degree of freedom parameters, $v_{f,k}$ for $k=1,...,4$ and $v_{\epsilon,s}$ for $s=1,...,90$
estimated using $q_{\lambda}^{III}$ (sorted in ascending order).
 \label{fig:dof}}

\centering{}\includegraphics[width=15cm,height=8cm]{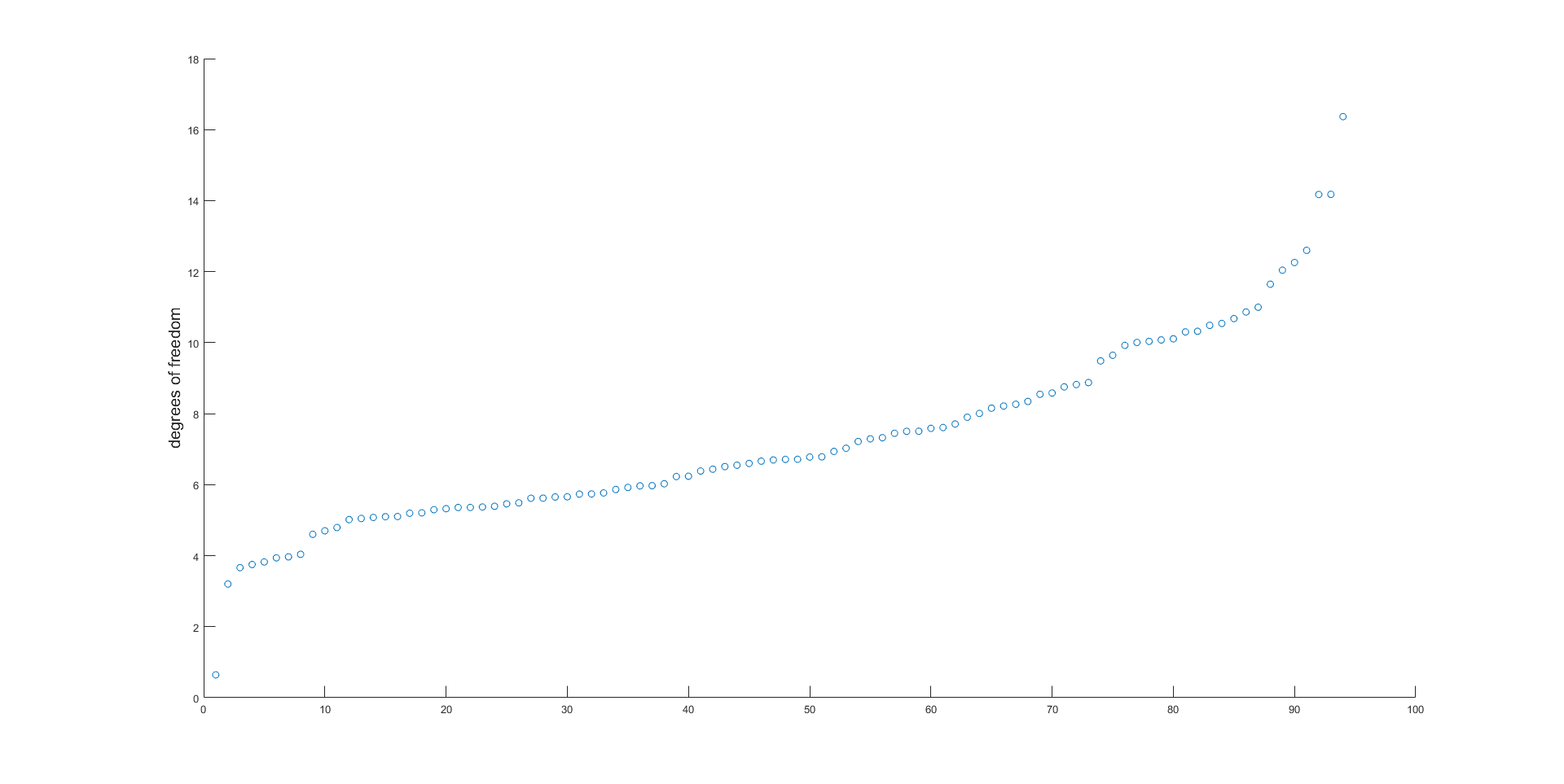}
\end{figure}

\section{List of SP100 Stock Returns\label{tab:TableSP100}}
\begin{table}[H]

\caption{S\&P100 constituents}

\begin{centering}
\begin{tabular}{cccc}
\hline
{\small{}Ticker} & {\small{}Name} &  & \tabularnewline
\hline
{\small{}AAPL} & {\small{}Apple Inc.} & {\small{}HPQ} & {\small{}Hewlett Packard Co.}\tabularnewline
{\small{}ABT} & {\small{}Abbott Lab.} & {\small{}IBM} & {\small{}International Business Machines}\tabularnewline
{\small{}AEP} & {\small{}American Electric Power Co.} & {\small{}INTC} & {\small{}Intel Corporation}\tabularnewline
{\small{}AIG} & {\small{}American International Group Inc.} & {\small{}JNJ} & {\small{}Johnson and Johnson Inc.}\tabularnewline
{\small{}ALL} & {\small{}Allstate Corp.} & {\small{}JPM} & {\small{}JP Morgan Chase \& Co.}\tabularnewline
{\small{}AMGN} & {\small{}Amgen Inc.} & {\small{}KO} & {\small{}The Coca-Cola Company}\tabularnewline
{\small{}AMZN} & {\small{}Amazon.com} & {\small{}LLY} & {\small{}Eli Lilly and Company}\tabularnewline
{\small{}APA} & {\small{}Apache Corp.} & {\small{}LMT} & {\small{}Lockheed-Martin}\tabularnewline
{\small{}APC} & {\small{}Anadarko Petroleum Corp.} & {\small{}LOW} & {\small{}Lowe's}\tabularnewline
{\small{}AXP} & {\small{}American Express Inc.} & {\small{}MCD} & {\small{}McDonald's Corp.}\tabularnewline
{\small{}BA} & {\small{}Boeing Co.} & {\small{}MDT} & {\small{}Medtronic Inc.}\tabularnewline
{\small{}BAC} & {\small{}Bank of America Corp.} & {\small{}MMM} & {\small{}3M Company}\tabularnewline
{\small{}BAX} & {\small{}Baxter International Inc. } & {\small{}MO} & {\small{}Altria Group}\tabularnewline
{\small{}BK} & {\small{}Bank of New York} & {\small{}MRK} & {\small{}Merck \& Co.}\tabularnewline
{\small{}BMY} & {\small{}Bristol-Myers Squibb} & {\small{}MS} & {\small{}Morgan Stanley}\tabularnewline
{\small{}BRK.B} & {\small{}Berkshire Hathaway} & {\small{}MSFT} & {\small{}Microsoft}\tabularnewline
{\small{}C} & {\small{}Citigroup Inc.} & {\small{}NKE} & {\small{}Nike}\tabularnewline
{\small{}CAT} & {\small{}Caterpillar Inc.} & {\small{}NOV} & {\small{}National Oilwell Varco}\tabularnewline
{\small{}CL} & {\small{}Colgate-Palmolive Co.} & {\small{}NSC} & {\small{}Norfolk Southern Corp.}\tabularnewline
{\small{}CMCSA} & {\small{}Comcast Corp.} & {\small{}ORCL} & {\small{}Oracle Corporation}\tabularnewline
{\small{}COF} & {\small{}Capital One Financial Corp.} & {\small{}OXY} & {\small{}Occidental Petroleum Corp.}\tabularnewline
{\small{}COP} & {\small{}ConocoPhillips} & {\small{}PEP} & {\small{}Pepsico Inc.}\tabularnewline
{\small{}COST} & {\small{}Costco} & {\small{}PFE} & {\small{}Pfizer Inc.}\tabularnewline
{\small{}CSCO} & {\small{}Cisco Systems} & {\small{}PG} & {\small{}Procter and Gambel Co.}\tabularnewline
{\small{}CVS} & {\small{}CVS Caremark} & {\small{}QCOM} & {\small{}Qualcomm Inc.}\tabularnewline
{\small{}CVX} & {\small{}Chevron} & {\small{}RTN} & {\small{}Raytheon Co.}\tabularnewline
{\small{}DD} & {\small{}DuPont} & {\small{}SBUX} & {\small{}Starbucks Corporation}\tabularnewline
{\small{}DELL} & {\small{}Dell} & {\small{}SLB} & {\small{}Schlumberger}\tabularnewline
{\small{}DIS} & {\small{}The Walt Disney Company} & {\small{}SO} & {\small{}Southern Company}\tabularnewline
{\small{}DOW} & {\small{}Dow Chemical} & {\small{}SPG} & {\small{}Simon Property Group, Inc.}\tabularnewline
{\small{}DVN} & {\small{}Devon Energy} & {\small{}T} & {\small{}AT\&T Inc.}\tabularnewline
{\small{}EBAY} & {\small{}eBay Inc.} & {\small{}TGT} & {\small{}Target Corp.}\tabularnewline
{\small{}EMC} & {\small{}EMC Corporation} & {\small{}TWX} & {\small{}Time Warner Inc.}\tabularnewline
{\small{}EMR} & {\small{}Emerson Electric Co.} & {\small{}TXN} & {\small{}Texas Instruments}\tabularnewline
{\small{}EXC} & {\small{}Exelon} & {\small{}UNH} & {\small{}United Health Group Inc.}\tabularnewline
{\small{}F} & {\small{}Ford Motor} & {\small{}UNP} & {\small{}Union Pacific Corp.}\tabularnewline
{\small{}FCX} & {\small{}Freeport-McMoran} & {\small{}UPS} & {\small{}Uniter Parcel Service Inc.}\tabularnewline
{\small{}FDX} & {\small{}FedEx} & {\small{}USB} & {\small{}US Bancorp}\tabularnewline
{\small{}GD} & {\small{}General Dynamics} & {\small{}UTX} & {\small{}United Technologies Corp.}\tabularnewline
{\small{}GE} & {\small{}General Electric Co.} & {\small{}VZ} & {\small{}Verizon Communication Inc.}\tabularnewline
{\small{}GILD} & {\small{}Gilead Sciences} & {\small{}WAG} & {\small{}Walgreens}\tabularnewline
{\small{}GS} & {\small{}Goldman Sachs} & {\small{}WFC} & {\small{}Wells Fargo}\tabularnewline
{\small{}HAL} & {\small{}Haliburton} & {\small{}WMB} & {\small{}Williams Companies}\tabularnewline
{\small{}HS} & {\small{}Home Depot} & {\small{}WMT} & {\small{}Wal-Mart}\tabularnewline
{\small{}HON} & {\small{}HoneyWell} & {\small{}XOM} & {\small{}Exxon Mobil Corp.}\tabularnewline
\hline
\end{tabular}
\par\end{centering}
\end{table}

%
%

\section{The factor loading matrix and the latent factors \label{sec:factorloadingmat}}
This section discusses the parameterisation of the factor loading matrix $\beta$ and the latent factors $f_{t}$.
The factor loading matrix  $\beta$ is unidentified without further constraints \citep{Geweke1996}; we follow \citeauthor{Geweke1996} and assume that
$\beta$
is  lower triangular, i. e., $\beta_{s,k}=0$ for $k>s$. The restriction on the leading
diagonal elements $\beta_{s,s}>0$ is also imposed. The main drawback of the lower
triangular assumption on the factor loading matrix $\beta$ is that the
resulting inference can depend on the order in which the components
of $y_{t}$ are chosen \citep{Chan:2017}. We follow \citet{Conti2014} and \citet{Kastner:2017} to obtain an appropriate ordering of the returns for the real data.
We run and post-process the factor loading estimates from the unrestricted sampler by choosing from column $1$, the stock $s$ with the largest value of $|\beta_{s,1}|$. We repeat this for columns $2$ to $7$. By an unrestricted sampler we mean that we do not
restrict $\beta$ to be lower triangular.

\section{Particle MCMC for FSV Models \label{sec:Particle-MCMC-for FSV Models}}

The key to making the estimation of the FSV model feasible is that
given the latent factors $f_{1:T}$ and factor loading matrix $\beta$,
the FSV model in Section \ref{sec:Description-of-Factor SV model}
becomes $S$ independent univariate SV models for the idiosyncratic
errors and $K$ independent univariate SV models for the latent factors.
Conditional on the latent factors $f_{1:T}$ and factor loading matrix
$\beta$, the $S$ univariate SV models with $\epsilon_{st}$ the
$t$th observation on the $s$th SV model and the $K$ univariate
SV models with $f_{kt}$ the $t$th observation on the $k$th univariate
SV model can be estimated independently and in parallel. \citet{Gunawan2020} proposed a particle hybrid sampler for general state space models and they applied their method to estimate FSV model. They showed that the particle hybrid sampler is much more efficient than other PMCMC samplers available in the literature.

\section{Lower Bound and Reparameterisation Gradients for the Variational
Approximation $q_{\lambda}^{III}\left(\theta,x_{1:T}\right)$ \label{sec:Reparameterisation-Gradients-for q3}}

This section discusses the lower bound and reparameterisation gradients
for the variational approximation $q_{\lambda}^{III}\left(\theta,x_{1:T}\right)$
for the FSV model. The set of parameters
and latent variables in the FSV model is  $\theta=\left(\left\{ \kappa_{\epsilon,s},\alpha_{\epsilon,s},\psi_{\epsilon,s}\right\} _{s=1}^{S},\left\{ \alpha_{f,k},\psi_{f,k}\right\} _{k=1}^{K},\beta\right)$
and $x_{1:T}=\left(\left\{ h_{\epsilon,s,1:T}\right\} _{s=1}^{S},\left\{ h_{f,k,1:T}\right\} _{k=1}^{K},\left\{ f_{k,1:T}\right\} _{k=1}^{K}\right)$. We approximate the augmented posterior density
$p\left(\theta,x_{1:T}|y\right)$ by
\begin{equation}
q_{\lambda}^{III}\left(\theta,x_{1:T}\right)=p\left(f_{1:T}|\theta,x_{1:T,-f_{1:T}},y\right)q_{\lambda}^{*}\left(\theta,x_{1:T,-f_{1:T}}|y\right),
\end{equation}
where
\begin{equation}
q_{\lambda}^{*}\left(\theta,x_{1:T,-f_{1:T}}\right)=\prod_{s=1}^{S}q_{\lambda_{\epsilon,s}}^{TBN}\left(\theta_{\epsilon,s},h_{\epsilon,s,1:T}\right)\prod_{k=1}^{K}q_{\lambda_{f,k}}^{TBN}\left(\theta_{G,f,k},h_{f,k,1:T}\right)q_{\lambda_{\beta}}^{SRN}\left(\beta\right).
\end{equation}
It is easy to generate the latent factors directly from their full conditional distribution $p\left(f_{1:T}|\theta,x_{1:T,-f_{1:T}},y\right)$ (see Section \ref{subsec:VB-approach-for factor SV model}).
After some simple algebra, the lower bound is
\begin{multline}
\mathcal{L}\left(\lambda\right)=E_{q}\left(\log p\left(y|f,h_{\epsilon},h_{f},\theta\right)+\log p\left(f|h_{\epsilon},h_{f},\theta\right)+\log p\left(h_{\epsilon},h_{f}|\theta\right)+\log p\left(\theta\right)-\right.\\
\left.\log p\left(f|\theta,h_{\epsilon},h_{f},y\right)-\log q_{\lambda}^{*}\left(\theta,h_{\epsilon},h_{f}|y\right)\right);\label{eq:LB_q3}
\end{multline}
 $h_{\epsilon}=\left\{ h_{\epsilon,s,1:T}\right\} _{s=1}^{S}$,
$h_{f}=\left\{ h_{f,k,1:T}\right\} _{k=1}^{K}$, and $f=\left\{ f_{k,1:T}\right\} _{k=1}^{K}$.
From Bayes theorem,
\[
p\left(f_{1:T}|\theta,x_{1:T,-f_{1:T}},y\right)
=\frac{p\left(y|f,h_{\epsilon},h_{f},\theta\right)
p\left(f|h_{\epsilon},h_{f},\theta\right)}{p\left(y|h_{\epsilon},h_{f},\theta\right)};
\]
substituting this into Equation \eqref{eq:LB_q3} gives
\begin{equation}
\mathcal{L}\left(\lambda\right)=E_{q}\left(\log p\left(y|h_{\epsilon},h_{f},\theta\right)+\log p\left(h_{\epsilon},h_{f}|\theta\right)+\log p\left(\theta\right)-\log q_{\lambda}^{*}\left(\theta,h_{\epsilon},h_{f}|y\right)\right).\label{eq:LB_q3-1}
\end{equation}

A similar derivation to \citet{Maya2020} can be used to obtain the
reparameterisation gradient. Let $\widetilde{\eta}=\left(\eta,f^{\top}\right)^{\top}$
with $\eta=\left(\left(\eta_{\theta_{G,\epsilon,s}}^{\top},
\eta_{h_{1:T,\epsilon,s}}^{\top}\right)^{\top},\left(\eta_{\theta_{G,f,k}}^{\top},
\eta_{h_{1:T,f,k}}^{\top}\right)^{\top},\eta_{\beta}^{\top}\right)$,
have the product density $p_{\widetilde{\eta}}\left(\widetilde{\eta}\right)
=p_{\eta}\left(\eta\right)p\left(f|u\left(\eta,\lambda\right),y\right)$,
where $p_{\eta}\left(\eta\right)$ does not depend on the variational
parameter $\lambda$, and\\


\[
u\left(\eta,\lambda\right)=\left(u\left(\eta_{\beta},\lambda_{\beta}\right),\left\{ u\left(\eta_{\theta_{G,\epsilon,s}},\eta_{h_{1:T,\epsilon,s}},\lambda_{\epsilon,s}\right)\right\} _{s=1}^{S},\left\{ u\left(\eta_{\theta_{G,f,k}},\eta_{h_{1:T,f,k}},\lambda_{f,k}\right)\right\} _{k=1}^{K}\right),
\]
so that $\left(\theta,x_{1:T}\right)=\left(u\left(\eta,\lambda\right),f\right)$.
The reparameterisation gradient used to implement SGA is
\begin{multline*}
\nabla_{\lambda}\mathcal{L}\left(\lambda\right)=E_{p_{\widetilde{\eta}}}\left(\nabla_{\lambda}u\left(\eta,\lambda\right)\left(\nabla_{\theta,x_{1:T,-f_{1:T}}}\log h\left(\theta,x_{1:T}\right)-\nabla_{\theta,x_{1:T,-f_{1:T}}}\log q_{\lambda}^{*}\left(\theta,h_{\epsilon},h_{f}|y\right)\right)\right),
\end{multline*}
where $h\left(\theta,x_{1:T}\right)=p\left(y|f,
h_{\epsilon},h_{f},\theta\right)p\left(f|h_{\epsilon},h_{f},\theta\right)
p\left(h_{\epsilon},h_{f}|\theta\right)p\left(\theta\right)$.
The term $\nabla_{\theta,x_{1:T,-f_{1:T}}}\log p\left(f|\theta,h_{\epsilon},h_{f},y\right)$
is not needed, nor the derivatives with respect to the latent factors
$f_{1:T}$, which simplifies the computation required for the reparameterisation
gradient. We only need a single draw from the posterior distribution
of the latent factors.


We now present the algorithm for the variational approximation $q_{\lambda}^{III}\left(\theta,x_{1:T}\right)$

\subsubsection*{Algorithm 2}

Initialise all the variational parameters $\lambda$. At each iteration
$j$, (1) Generate Monte Carlo samples for all the parameters and
latent states from their variational distributions, (2) Compute the
unbiased estimates of gradient of the lower bound with respect to
each of the variational parameter and update the variational parameters
using Stochastic Gradient methods.

%

\subsection*{STEP 1 }
\begin{itemize}
\item \textbf{For} $s=1:S$,
\begin{itemize}
\item Generate Monte Carlo samples for $\theta_{G,\epsilon,s}$ and $h_{\epsilon,s,1:T}$
from the variational distribution $q_{\lambda_{\epsilon,s}}^{TBN}\left(\theta_{G,\epsilon,s},h_{\epsilon,s,1:T}\right)$
\begin{enumerate}
\item Generate $\eta_{\theta_{G,\epsilon,s}}\sim N\left(0,I_{G_{\epsilon}}\right)$
and $\eta_{h_{1:T,\epsilon,s}}\sim N\left(0,I_{T}\right)$, where
$G_{\epsilon}$ is the number of parameters of idiosyncratic log-volatilities
\item Generate $\theta_{G,\epsilon,s}=\mu_{G,\epsilon,s}^{\left(j\right)}+\left(C_{G,\epsilon,s}^{\left(j\right)}\right)^{-\top}\eta_{\theta_{G,\epsilon,s}}$
and $h_{1:T,\epsilon,s}=\mu_{L,\epsilon,s}^{\left(j\right)}+\left(C_{L,\epsilon,s}^{\left(j\right)}\right)^{-\top}\eta_{x_{1:T,\epsilon,s}}$.
\end{enumerate}
\end{itemize}
\item \textbf{For} $k=1:K$,
\begin{itemize}
\item Generate Monte Carlo samples for $\theta_{G,f,k}$ and $h_{f,k,1:T}$
from the variational distribution $q_{\lambda_{f,k}}^{TBN}\left(\theta_{G,f,k},h_{f,k,1:T}\right)$
\begin{enumerate}
\item Generate $\eta_{\theta_{G,f,k}}\sim N\left(0,I_{G_{f}}\right)$ and
$\eta_{h_{1:T,f,k}}\sim N\left(0,I_{T}\right)$, where $G_{f}$ is
the number of parameters of factor log-volatilities
\item Generate $\theta_{G,f,k}=\mu_{G,f,k}^{\left(j\right)}+\left(C_{G,f,k}^{\left(j\right)}\right)^{-\top}\eta_{\theta_{G,f,k}}$
and $h_{1:T,f,k}=\mu_{L,f,k}^{\left(j\right)}+\left(C_{L,f,k}^{\left(j\right)}\right)^{-\top}\eta_{x_{1:T,f,k}}$.
\end{enumerate}
\end{itemize}
\item Generate Monte Carlo samples for the factor loading $\beta$ from
the variational distribution
\begin{enumerate}
\item Generate $z_{\beta}\sim N\left(0,I_{p}\right)$ and $\eta_{\beta}\sim N\left(0,I_{R_{\beta}}\right)$
calculating $\xi_{\beta}=\mu_{\xi_{\beta}}^{\left(j\right)}+B^{\left(j\right)}_{\xi_{\beta}}z_{\beta}+d^{\left(j\right)}_{\xi_{\beta}}\circ\eta_{\beta}$,
and $R_{\beta}$ is the total number of parameters in the factor loading
matrix $\beta$.
\item Generate $\beta_{i}=t_{\gamma_{\beta_{i}}}^{-1(j)}\left(\xi_{\beta_{i}}\right)$,
for $i=1,...,R_{\beta}$.
\end{enumerate}
\item Generate Monte Carlo samples for the latent factors $f_{1:T}$ by
sampling from its conditional distribution $p\left(f_{1:T}|\theta,x_{1:T,-f_{1:T}},y\right)$.
\end{itemize}

\subsection*{STEP 2}
\begin{itemize}
\item \textbf{For} $s=1:S$, Update the variational parameters $\lambda_{\epsilon,s}$
of the variational distributions $q_{\lambda_{\epsilon,s}}^{TBN}\left(\theta_{\epsilon,s},h_{\epsilon,s,1:T}\right)$
\begin{enumerate}
\item Construct unbiased estimates of gradients $\widehat{\nabla_{\lambda_{\epsilon,s}}\mathcal{L}\left(\lambda_{\epsilon,s}\right)}$.
\item Set adaptive learning rates $a_{t,\epsilon,s}$ using ADAM method.
\item Set $\lambda_{\epsilon,s}^{\left(j+1\right)}=
    \lambda_{\epsilon,s}^{\left(j\right)}+a_{j,\epsilon,s}
    \widehat{\nabla_{\lambda_{\epsilon,s}}
    \mathcal{L}\left(\lambda_{\epsilon,s}\right)}^{\left(j\right)}$,

where $\lambda_{\epsilon,s}=\left\{ \mu_{G,\epsilon,s}^{\top},v\left(C_{G,\epsilon,s}^{*}\right)^{\top},d_{\epsilon,s}^{\top},\textrm{vec}\left(D_{\epsilon,s}\right)^{\top},f_{\epsilon,s}^{*\top},\textrm{vec}\left(F_{\epsilon,s}\right)^{\top}\right\} ^{\top}$
as defined in Section \ref{subsec:Sparse-Cholesky-factorTan2019}.
\end{enumerate}
\item For $k=1:K$,
\begin{itemize}
\item Update the variational parameters $\lambda_{f,k}$ of the variational
distributions $q_{\lambda_{f,k}}^{TBN}\left(\theta_{G,f,k},x_{f,k,1:T}\right)$
\begin{enumerate}
\item Construct unbiased estimates of gradients $\widehat{\nabla_{\lambda_{f,k}}\mathcal{L}\left(\lambda_{f,k}\right)}$.
\item Set adaptive learning rates $a_{j,f,k}$ using ADAM method.
\item Set $\lambda_{f,k}^{\left(j+1\right)}=\lambda_{f,k}^{\left(j\right)}+a_{j,f,k}\widehat{\nabla_{\lambda_{f,k}}\mathcal{L}\left(\lambda_{f,k}\right)}^{\left(j\right)}$,

where $\lambda_{f,k}=\left\{ \mu_{G,f,k}^{\top},v\left(C_{G,f,k}^{*}\right)^{\top},d_{f,k}^{\top},\textrm{vec}\left(D_{f,k}\right)^{\top},f_{f,k}^{*\top},\textrm{vec}\left(F_{f,k}\right)^{\top}\right\} ^{\top}$.
\end{enumerate}
\end{itemize}
\item Update the variational parameters $\lambda_{\beta}$ from the variational
distribution $q_{\lambda_{\beta}}^{SRN}\left(\beta\right)$.
\begin{enumerate}
\item Construct unbiased estimates of gradients $\widehat{\nabla_{\lambda_{\beta}}\mathcal{L}\left(\lambda_{\beta}\right)}$.
\item Set adaptive learning rates $a_{t,\beta}$ using the ADAM method.
\item Set $\lambda_{\beta}^{\left(j+1\right)}=\lambda_{\beta}^{\left(j\right)}+a_{j,\beta}\widehat{\nabla_{\lambda_{\beta}}\mathcal{L}\left(\lambda_{\beta}\right)}^{\left(j\right)}$,
where $\lambda_{\beta}=\left(\gamma_{\beta_{1}}^{\top},...,\gamma_{\beta_{R_{\beta}}}^{\top},\mu_{\xi_{\beta}}^{\top},vech(B_{\xi_{\beta}})^{\top},d_{\xi_{\beta}}^{\top}\right)^{\top}$.
\end{enumerate}
\end{itemize}

\section{Deriving the Gradients for the FSV Models \label{sec:Deriving-the-Gradients for FSV Models}}

This section derives the gradients for the FSV models discussed in
Section \ref{sec:Description-of-Factor SV model}. The set of parameters
and latent variables in the FSV model is given by $\theta=\left(\left\{ \kappa_{\epsilon,s},\alpha_{\epsilon,s},\psi_{\epsilon,s}\right\} _{s=1}^{S},\left\{ \alpha_{f,k},\psi_{f,k}\right\} _{k=1}^{K},\beta\right)$
and $x_{1:T}=\left(\left\{ h_{\epsilon,s,1:T}\right\} _{s=1}^{S},\left\{ h_{f,k,1:T}\right\} _{k=1}^{K},\left\{ f_{k,1:T}\right\} _{k=1}^{K}\right)$.
The required gradients are:
\begin{itemize}
\item For $s=1,...,S$,
\begin{multline*}
\nabla_{\alpha_{\epsilon,s}}\log p\left(y,\theta,x_{1:T}\right)=\frac{1}{2}\sum_{t=1}^{T}\left(h_{\epsilon,s,t}\left(y_{t}-\beta f_{t}\right)^{2}\exp\left(-\tau_{\epsilon,s}h_{\epsilon,s,t}-\kappa_{\epsilon,s}\right)-h_{\epsilon,s,t}\right)\left(1-\exp\left(-\tau_{\epsilon,s}\right)\right)\\
-\frac{2\tau_{\epsilon,s}}{\left(1+\tau_{\epsilon,s}^{2}\right)}\frac{\exp\left(\alpha_{\epsilon,s}\right)}{1+\exp\left(\alpha_{\epsilon,s}\right)}+1-\frac{\exp\left(\alpha_{\epsilon,s}\right)}{1+\exp\left(\alpha_{\epsilon,s}\right)};
\end{multline*}
\end{itemize}
\begin{multline*}
\nabla_{\kappa_{\epsilon,s}}\log p\left(y,\theta,x_{1:T}\right)=\frac{1}{2}\left(\sum_{t=1}^{T}\left(y_{t}-\beta f_{t}\right)^{2}\exp\left(-\tau_{\epsilon,s}h_{\epsilon,s,t}-\kappa_{\epsilon,s}\right)-T\right)-\kappa_{\epsilon,s}/\sigma_{\kappa}^{2};
\end{multline*}

\begin{multline*}
\nabla_{\psi_{\epsilon,s}}\log p\left(y,\theta,x_{1:T}\right)=\left(\sum_{t=2}^{T}\left(h_{\epsilon,s,t}-\phi_{\epsilon,s}h_{\epsilon,s,t-1}\right)h_{\epsilon,s,t-1}+h_{\epsilon,s,1}^{2}\phi_{\epsilon,s}-\frac{\phi_{\epsilon,s}}{1-\phi_{\epsilon,s}^{2}}\right)\left(\phi_{\epsilon,s}\left(1-\phi_{\epsilon,s}\right)\right)+\\
\left(\frac{\left(a_{0}-1\right)}{\left(1+\phi_{\epsilon,s}\right)}-\frac{\left(b_{0}-1\right)}{\left(1-\phi_{\epsilon,s}\right)}\right)\frac{\exp\left(\psi_{\epsilon,s}\right)}{\left(1+\exp\left(\psi_{\epsilon,s}\right)\right)^{2}}+\frac{\left(1-\exp\left(\psi_{\epsilon,s}\right)\right)}{\left(1+\exp\left(\psi_{\epsilon,s}\right)\right)};
\end{multline*}

\begin{multline*}
\nabla_{h_{\epsilon,s,1}}\log p\left(y,\theta,x_{1:T}\right)=\frac{\tau_{\epsilon,s}}{2}\left(\left(y_{1}-\beta f_{1}\right)^{2}\exp\left(-\tau_{\epsilon,s}h_{\epsilon,s,1}-\kappa_{\epsilon,s}\right)-1\right)+\\
\phi_{\epsilon,s}\left(h_{\epsilon,s,2}-\phi_{\epsilon,s}h_{\epsilon,s,1}\right)-h_{\epsilon,s,1}\left(1-\phi_{\epsilon,s}^{2}\right);
\end{multline*}

\begin{multline*}
\nabla_{h_{\epsilon,s,T}}\log p\left(y,\theta,x_{1:T}\right)=\frac{\tau_{\epsilon,s}}{2}\left(\left(y_{T}-\beta f_{T}\right)^{2}\exp\left(-\tau_{\epsilon,s}h_{\epsilon,s,T}-\kappa_{\epsilon,s}\right)-1\right)-\\
\left(h_{\epsilon,s,T}-\phi_{\epsilon,s}h_{\epsilon,s,T-1}\right);
\end{multline*}

\begin{multline*}
\nabla_{h_{\epsilon,s,t}}\log p\left(y,\theta,x_{1:T}\right)=\frac{\tau_{\epsilon,s}}{2}\left(\left(y_{t}-\beta f_{t}\right)^{2}\exp\left(-\tau_{\epsilon,s}h_{\epsilon,s,t}-\kappa_{\epsilon,s}\right)-1\right)+\\
\phi_{\epsilon,s}\left(h_{\epsilon,s,t+1}-\phi_{\epsilon,s}h_{\epsilon,s,t}\right)-\left(h_{\epsilon,s,t}-\phi_{\epsilon,s}h_{\epsilon,s,t-1}\right)\;\textrm{for}\;t=2,...,T-1;
\end{multline*}

\begin{itemize}
\item For $k=1,...,K,$
\begin{multline*}
\nabla_{\alpha_{f,k}}\log p\left(y,\theta,x_{1:T}\right)=\frac{1}{2}\sum_{t=1}^{T}\left(h_{f,k,t}\left(f_{k,t}\right)^{2}\exp\left(-\tau_{f,k}h_{f,k,t}\right)-h_{f,k,t}\right)\left(1-\exp\left(-\tau_{f,k}\right)\right)\\
-\frac{2\tau_{f,k}}{\left(1+\tau_{f,k}^{2}\right)}\frac{\exp\left(\alpha_{f,k}\right)}{1+\exp\left(\alpha_{f,k}\right)}+1-\frac{\exp\left(\alpha_{f,k}\right)}{1+\exp\left(\alpha_{f,k}\right)};
\end{multline*}
\end{itemize}

\begin{multline*}
\nabla_{\psi_{f,k}}\log p\left(y,\theta,x_{1:T}\right)=\left(\sum_{t=2}^{T}\left(h_{f,k,t}-\phi_{f,k}h_{f,k,t-1}\right)h_{f,k,t-1}+h_{f,k,1}^{2}\phi_{f,k}-\frac{\phi_{f,k}}{1-\phi_{f,k}^{2}}\right)\left(\phi_{f,k}\left(1-\phi_{f,k}\right)\right)+\\
\left(\frac{\left(a_{0}-1\right)}{\left(1+\phi_{f,k}\right)}-\frac{\left(b_{0}-1\right)}{\left(1-\phi_{f,k}\right)}\right)\frac{\exp\left(\psi_{f,k}\right)}{\left(1+\exp\left(\psi_{f,k}\right)\right)^{2}}+\frac{\left(1-\exp\left(\psi_{f,k}\right)\right)}{\left(1+\exp\left(\psi_{f,k}\right)\right)};
\end{multline*}

\begin{multline*}
\nabla_{h_{f,k,1}}\log p\left(y,\theta,x_{1:T}\right)=\frac{\tau_{f,k}}{2}\left(\left(f_{k,1}\right)^{2}\exp\left(-\tau_{f,k}h_{f,k,1}\right)-1\right)+\\
\phi_{f,k}\left(h_{f,k,2}-\phi_{f,k}h_{f,k,1}\right)-h_{f,k,1}\left(1-\phi_{f,k}^{2}\right);
\end{multline*}

\begin{multline*}
\nabla_{h_{f,k,T}}\log p\left(y,\theta,x_{1:T}\right)=\frac{\tau_{f,k}}{2}\left(\left(f_{k,T}\right)^{2}\exp\left(-\tau_{f,k}h_{f,k,T}\right)-1\right)-\\
\left(h_{f,k,T}-\phi_{f,k}h_{f,k,T-1}\right);
\end{multline*}

\begin{multline*}
\nabla_{h_{f,k,t}}\log p\left(y,\theta,x_{1:T}\right)=\frac{\tau_{f,k}}{2}\left(\left(f_{k,t}\right)^{2}\exp\left(-\tau_{f,k}h_{f,k,t}\right)-1\right)+\\
\phi_{f,k}\left(h_{f,k,t+1}-\phi_{f,k}h_{f,k,t}\right)-\left(h_{f,k,t}-\phi_{f,k}h_{f,k,t-1}\right)\;\textrm{for}\;t=2,...,T-1.
\end{multline*}

\begin{itemize}
\item The gradients with respect to the latent factors $f_{t}$, for $t=1,...,T$,

\[
\nabla_{f_{t}}\log p\left(y,\theta,x_{1:T}\right)=\left(y_{t}-\beta f_{t}\right)^{'}V_{t}^{-1}\beta-f_{t}^{'}D_{t}^{-1};
\]
where
\[
V_{t}=\textrm{diag}\left(\exp\left(\tau_{\epsilon,1}h_{\epsilon,1,t}+\kappa_{\epsilon,1}\right),...,\exp\left(\tau_{\epsilon,S}h_{\epsilon,S,t}+\kappa_{\epsilon,S}\right)\right);
\]
\[
D_{t}=\textrm{diag}\left(\exp\left(\tau_{f,1}h_{f,1,t}\right),...,\exp\left(\tau_{f,K}h_{f,K,t}+\kappa_{f,K}\right)\right).
\]

\item The gradients with respect to the factor loading $\beta_{s,.}=\left(\beta_{s,1},...,\beta_{s,k_{s}}\right)^{'}$
for $s=1,...,S$:
\[
\nabla_{\beta_{s,.}}\log p\left(y,\theta,x_{1:T}\right)=\left(y_{s,1:T}-F_{s}\beta_{s,.}\right)^{'}\widetilde{V}_{s}^{-1}F_{s}-\nabla_{\beta_{s,.}}\log p\left(\beta_{s,.}\right),
\]
where $y_{s,1:T}=\left(y_{s,1},...,y_{s,T}\right)^{'},$
\[
F_{s}=\left[\begin{array}{ccc}
f_{11} & \cdots & f_{k_{s}1}\\
\vdots &  & \vdots\\
f_{1T} & \cdots & f_{k_{s}T}
\end{array}\right];
\]
\end{itemize}
\[
\widetilde{V}_{s}=\textrm{diag}\left(\exp\left(\tau_{\epsilon,s}h_{\epsilon,s,1}+\kappa_{\epsilon,s}\right),...,\exp\left(\tau_{\epsilon,s}h_{\epsilon,s,T}+\kappa_{\epsilon,s}\right)\right).
\]

\section{Multivariate t-distributed FSV Model \label{sec:Multivariate-FSV-Model t-distribution}}

Suppose that $P_{t}$ is a $S\times1$ vector of daily stock prices
and define $y_{t}\coloneqq\log P_{t}-\log P_{t-1}$ as the vector
of stock returns of the stocks. The $t$-distributed FSV model is expressed as
\begin{equation}
y_{t}=\beta f_{t}+\sqrt{W_{\epsilon,s,t}}\epsilon_{t},\:\left(t=1,...,T\right);\label{eq:factor model-1}
\end{equation}
 $f_{t}$ is a $K\times1$ vector of latent factors (with $K\lll S$),
$\beta$ is a $S\times K$ factor loading matrix of the unknown parameters.
The latent factors $f_{k,t}$, $k=1,...,K$ are assumed independent
with $f_{t}\sim N\left(0,\sqrt{W_{f,k,t}}D_{t}\right)$. The time-varying
diagonal variance matrix $D_{t}$ with $k$th diagonal
element $\exp\left(\tau_{f,k}h_{f,k,t}\right)$. Each $h_{f,k,t}$
 follows an independent autoregressive process
\begin{equation}
h_{f,k,1}\sim N\left(0,\frac{1}{1-\phi_{f,k}^{2}}\right),\;h_{f,k,t}=\phi_{f,k}h_{f,k,t-1}+\eta_{f,k,t}\;\eta_{f,k,t}\sim N\left(0,1\right),k=1,...,K.\label{eq:statetransitionfactor-1}
\end{equation}
We model the idiosyncratic error as $\epsilon_{t}\sim N\left(0,V_{t}\right)$;
$V_{t}$ is a diagonal time-varying variance matrix with  $s$th
diagonal element $\exp\left(\tau_{\epsilon,s}h_{\epsilon,s,t}+\kappa_{\epsilon,s}\right)$.
Each $h_{\epsilon,s,t}$  follows an independent autoregressive
process
\begin{equation}
h_{\epsilon,s,1}\sim N\left(0,\frac{1}{1-\phi_{\epsilon s}^{2}}\right),\;h_{\epsilon st}=\phi_{\epsilon,s}h_{\epsilon,s,t-1}+\eta_{\epsilon,s,t}\;\eta_{\epsilon,s,t}\sim N\left(0,1\right),s=1,...,S.\label{eq:statetransitionidiosyncratic-1-1}
\end{equation}
Each of $W_{\epsilon,s,t}$ is Inverse Gamma $IG\left(\frac{v_{\epsilon,s}}{2},\frac{v_{\epsilon,s}}{2}\right)$ distributed,
with $v_{\epsilon,s}$ the degree of freedom, for $s=1,...,S$. Similarly,
each of $W_{f,k,t}$ is Inverse Gamma $IG\left(\frac{v_{f,k}}{2},\frac{v_{f,k}}{2}\right)$
distributed,
with $v_{f,k}$ the degree of freedom, for $k=1,...,K$. The log
transformation is used to map the constrained degrees of freedom parameters
$\nu_{\epsilon,s}^{*}=\log\left(v_{\epsilon,s}\right)$ and $\nu_{f,k}^{*}=\log\left(v_{f,k}\right)$
to the real line $\R$. The prior for each of $v_{\epsilon,s}$ and $v_{f,k}$
is  Gamma $G\left(a_{\nu},b_{v}\right)$ distributed with $a_{v}=20$
and $b_{v}=1.25$. The same transformations and priors are used for the other
parameters as in the FSV model discussed in Section \ref{sec:Description-of-Factor SV model}.


The set of variables in the t-distributed  factor SV model  is
$$\theta\coloneqq\left(\left\{ \kappa_{\epsilon,s},\alpha_{\epsilon,s},\psi_{\epsilon,s},v_{\epsilon,s}\right\} _{s=1}^{S},\left\{ \alpha_{f,k},\psi_{f,k},v_{f,k}\right\} _{k=1}^{K},\beta\right)$$
and
$$x_{1:T}\coloneqq\left(\left\{ h_{\epsilon,s,1:T},W_{\epsilon,s,1:T}\right\} _{s=1}^{S},\left\{ h_{f,k,1:T},W_{f,k,1:T}\right\} _{k=1}^{K},\left\{ f_{k,1:T}\right\} _{k=1}^{K}\right)$$.
The posterior distribution of the multivariate factor SV model is
given by
\begin{multline}
p\left(\theta,x_{1:T}|y\right)=\left\{ \prod_{s=1}^{S}\prod_{t=1}^{T}p\left(y_{s,t}|\beta_{s},f_{t},h_{\epsilon,s,t},\mu_{\epsilon,s},\alpha_{\epsilon,s},\psi_{\epsilon,s},W_{\epsilon,s,1:T}\right)\right.\\
\left.p\left\{ h_{\epsilon,s,t}|h_{\epsilon,s,t-1},\mu_{\epsilon,s},\alpha_{\epsilon,s},\psi_{\epsilon,s}\right\} p\left(W_{\epsilon,s,1:T}|v_{\epsilon,s}\right)\right\} \\
\left\{ \prod_{k=1}^{K}\prod_{t=1}^{T}p\left(f_{k,t}|h_{f,k,t},\alpha_{f,k},\psi_{f,k},W_{f,k,1:T}\right)p\left(h_{f,k,t}|h_{f,k,t-1},\alpha_{f,k},\psi_{f,k}\right)p\left(W_{f,k,1:T}|v_{f,k}\right)\right\} \\
\left\{ \prod_{s=1}^{S}p\left(\mu_{\epsilon,s}\right)p\left(\phi_{\epsilon,s}\right)\bigg|\frac{\partial\phi_{\epsilon,s}}{\partial\psi_{\epsilon,s}}\bigg|p\left(\tau_{\epsilon,s}\right)\bigg|\frac{\partial\tau_{\epsilon,s}}{\partial\alpha_{\epsilon,s}}\bigg|p\left(v_{\epsilon,s}\right)\bigg|\frac{\partial v_{\epsilon,s}}{\partial v_{\epsilon,s}^{*}}\bigg|\right\} \\
\left\{ \prod_{k=1}^{K}p\left(\phi_{f,k}\right)\bigg|\frac{\partial\phi_{f,k}}{\partial\psi_{f,k}}\bigg|p\left(\tau_{f,k}\right)\bigg|\frac{\partial\tau_{f,k}}{\partial\alpha_{f,k}}\bigg|p\left(v_{f,k}\right)\bigg|\frac{\partial v_{f,k}}{\partial v_{f,k}^{*}}\bigg|\right\} p\left(\beta\right)\prod_{k=1}^{K}\bigg|\frac{\partial\beta_{k,k}}{\partial\delta_{k}}\bigg|.\label{eq:posteriorfactormodel-1}
\end{multline}

\section{Variational Approximations for the t-distributed FSV Models \label{variational approximation and gradient FSV with t distribution}}

This section discusses the variational approximations for the t-distributed FSV
models  discussed in Section \ref{sec:Multivariate-FSV-Model t-distribution}.
The posterior distribution is
\begin{equation}
p\left(\theta,x_{1:T}|y\right)=p\left(f,W_{\epsilon},W_{f}|\theta,h_{\epsilon},h_{f},y\right)p\left(\theta,h_{\epsilon},h_{f}|y\right),\label{eq:posteriortdist}
\end{equation}
The conditional distributions $p\left(W_{f}|\theta,h_{\epsilon},h_{f},f,W_{\epsilon},y\right)$,
$p\left(W_{\epsilon}|\theta,h_{\epsilon},h_{f},f,W_{f},y\right)$,
and $p\left(f|\theta,h_{\epsilon},h_{f},W_{\epsilon},W_{f},y\right)$
are available in closed form and are given by
\[
p\left(W_{\epsilon,s,t}|\theta,h_{\epsilon},h_{f},f,W_{f},y\right)=IG\left(\frac{v_{\epsilon,s}+1}{2},\frac{v_{\epsilon,s}}{2}+\frac{1}{2}\left(y_{s,t}-\beta_{s,.}f_{t}\right)^{2}\exp\left(-\tau_{\epsilon,s}h_{\epsilon,s,t}-\kappa_{\epsilon,s}\right)\right)\; \textrm{for }s=1,...,S,
\]
\[
p\left(W_{f,k,t}|\theta,h_{\epsilon},h_{f},f,W_{\epsilon},y\right)=IG\left(\frac{v_{f,k}+1}{2},\frac{v_{f,k}}{2}+\frac{1}{2}\left(f_{k,t}\right)^{2}\exp\left(-\tau_{f,ks}h_{f,k,t}\right)\right)\;\textrm{for }k=1,...,K,
\]
and
\[
p\left(f_{k,t}|\theta,h_{\epsilon},h_{f},W_{\epsilon},W_{f}\right)=N\left(\mu_{f_{k,t}},\Sigma_{f_{k,t}}\right)\;\textrm{for }k=1,...,K\;\textrm{and}\;\textrm{for }t=1,...,T,
\]
where $\mu_{f_{k,t}}=\Sigma_{f_{k,t}}\beta^{'}\left(\widetilde{V}_{t}^{-1}y_{t}\right)$,
$\Sigma_{f_{k,t}}=\left(\beta\widetilde{V}_{t}^{-1}\beta^{'}+\widetilde{D}_{t}^{-1}\right)^{-1}$,
and

\[
\widetilde{V}_{t}=\textrm{diag}\left(W_{\epsilon,1,t}\exp\left(\tau_{\epsilon,1}h_{\epsilon,1,t}+\kappa_{\epsilon,1}\right),...,W_{\epsilon,S,t}\exp\left(\tau_{\epsilon,S}h_{\epsilon,S,t}+\kappa_{\epsilon,S}\right)\right);
\]
\[
\widetilde{D}_{t}=\textrm{diag}\left(W_{f,1,t}\exp\left(\tau_{f,1}h_{f,1,t}\right),...,W_{f,K,t}\exp\left(\tau_{f,K}h_{f,K,t}\right)\right).
\]
Because it is easy to sample from the full conditional distribution
$p\left(\left\{ f_{k,1:T}\right\} ,\left\{ W_{\epsilon,s}\right\} _{s=1}^{S},\left\{ W_{f,k}\right\} _{k=1,}^{K}|\theta,x_{1:T,-W_{\epsilon},W_{f},f},y\right)$,
it is unnecessary to approximate it. We propose the following variational
approximation for FSV models with t-distribution:
\begin{equation}
q_{\lambda}^{t}\left(\theta,x_{1:T}\right)=p\left(f,W_{\epsilon},W_{f}|\theta,x_{1:T,-W_{\epsilon},W_{f},f},y\right)q_{\lambda}^{*}\left(\theta,x_{1:T,-f,W_{\epsilon},W_{f}}\right),
\end{equation}
where
\begin{multline*}
q_{\lambda}^{*}\left(\theta,x_{1:T,-f,W_{\epsilon},W_{f}}\right)=\prod_{s=1}^{S}q_{\lambda_{\epsilon,s}}^{TBN}\left(\theta_{\epsilon,s},h_{\epsilon,s,1:T}\right)\prod_{k=1}^{K}q_{\lambda_{f,k}}^{TBN}\left(\theta_{G,f,k},h_{f,k,1:T}\right)\\
q_{\lambda_{\beta}}^{SRN}\left(\beta\right)\prod_{s=1}^{S}q_{\lambda_{v_{\epsilon,s}}}^{SRN}\left(v_{\epsilon,s}\right)\prod_{k=1}^{K}q_{\lambda_{v_{f,k}}}^{SRN}\left(v_{f,k}\right).
\end{multline*}
We now derive the gradients for the FSV models discussed in Section
\ref{sec:Multivariate-FSV-Model t-distribution}. The required gradients
are:
\begin{itemize}
\item For $s=1,...,S$,
\begin{multline*}
\nabla_{\alpha_{\epsilon,s}}\log p\left(y,\theta,x_{1:T}\right)=\frac{1}{2}\sum_{t=1}^{T}\left(h_{\epsilon,s,t}\left(y_{t}-\beta f_{t}\right)^{2}W_{\epsilon,s,t}^{-1}\exp\left(-\tau_{\epsilon,s}h_{\epsilon,s,t}-\kappa_{\epsilon,s}\right)-h_{\epsilon,s,t}\right)\left(1-\exp\left(-\tau_{\epsilon,s}\right)\right)\\
-\frac{2\tau_{\epsilon,s}}{\left(1+\tau_{\epsilon,s}^{2}\right)}\frac{\exp\left(\alpha_{\epsilon,s}\right)}{1+\exp\left(\alpha_{\epsilon,s}\right)}+1-\frac{\exp\left(\alpha_{\epsilon,s}\right)}{1+\exp\left(\alpha_{\epsilon,s}\right)};
\end{multline*}
\end{itemize}
\begin{multline*}
\nabla_{\kappa_{\epsilon,s}}\log p\left(y,\theta,x_{1:T}\right)=\frac{1}{2}\left(\sum_{t=1}^{T}\left(y_{t}-\beta f_{t}\right)^{2}W_{\epsilon,s,t}^{-1}\exp\left(-\tau_{\epsilon,s}h_{\epsilon,s,t}-\kappa_{\epsilon,s}\right)-T\right)-\kappa_{\epsilon,s}/\sigma_{\kappa}^{2};
\end{multline*}

\begin{multline*}
\nabla_{\psi_{\epsilon,s}}\log p\left(y,\theta,x_{1:T}\right)=\left(\sum_{t=2}^{T}\left(h_{\epsilon,s,t}-\phi_{\epsilon,s}h_{\epsilon,s,t-1}\right)h_{\epsilon,s,t-1}+h_{\epsilon,s,1}^{2}\phi_{\epsilon,s}-\frac{\phi_{\epsilon,s}}{1-\phi_{\epsilon,s}^{2}}\right)\left(\phi_{\epsilon,s}\left(1-\phi_{\epsilon,s}\right)\right)+\\
\left(\frac{\left(a_{0}-1\right)}{\left(1+\phi_{\epsilon,s}\right)}-\frac{\left(b_{0}-1\right)}{\left(1-\phi_{\epsilon,s}\right)}\right)\frac{\exp\left(\psi_{\epsilon,s}\right)}{\left(1+\exp\left(\psi_{\epsilon,s}\right)\right)^{2}}+\frac{\left(1-\exp\left(\psi_{\epsilon,s}\right)\right)}{\left(1+\exp\left(\psi_{\epsilon,s}\right)\right)};
\end{multline*}

\begin{multline*}
\nabla_{v_{\epsilon,s}}\log p\left(y,\widetilde{\theta}\right)=\left\{ \frac{T}{2}\log\left(\frac{v_{\epsilon,s}}{2}\right)+\frac{T}{2}-\frac{T}{2}\Psi\left(\frac{v_{\epsilon,s}}{2}\right)\right.\\
\left.-\frac{1}{2}\sum_{t=1}^{T}\log\left(W_{\epsilon,s,t}\right)-\frac{1}{2}\sum_{t=1}^{T}\frac{1}{W_{\epsilon,s,t}}+1+v_{\epsilon,s}\left(\left(a_{v}-1\right)\frac{1}{v_{\epsilon,s}}-\frac{1}{b_{v}}\right)\right\} ;
\end{multline*}
where $\Psi\left(\cdot\right)$ is the digamma function.

\begin{multline*}
\nabla_{h_{\epsilon,s,1}}\log p\left(y,\theta,x_{1:T}\right)=\frac{\tau_{\epsilon,s}}{2}\left(\left(y_{1}-\beta f_{1}\right)^{2}W_{\epsilon,s,1}^{-1}\exp\left(-\tau_{\epsilon,s}h_{\epsilon,s,1}-\kappa_{\epsilon,s}\right)-1\right)+\\
\phi_{\epsilon,s}\left(h_{\epsilon,s,2}-\phi_{\epsilon,s}h_{\epsilon,s,1}\right)-h_{\epsilon,s,1}\left(1-\phi_{\epsilon,s}^{2}\right);
\end{multline*}

\begin{multline*}
\nabla_{h_{\epsilon,s,T}}\log p\left(y,\theta,x_{1:T}\right)=\frac{\tau_{\epsilon,s}}{2}\left(\left(y_{T}-\beta f_{T}\right)^{2}W_{\epsilon,s,T}^{-1}\exp\left(-\tau_{\epsilon,s}h_{\epsilon,s,T}-\kappa_{\epsilon,s}\right)-1\right)-\\
\left(h_{\epsilon,s,T}-\phi_{\epsilon,s}h_{\epsilon,s,T-1}\right);
\end{multline*}

\begin{multline*}
\nabla_{h_{\epsilon,s,t}}\log p\left(y,\theta,x_{1:T}\right)=\frac{\tau_{\epsilon,s}}{2}\left(\left(y_{t}-\beta f_{t}\right)^{2}W_{\epsilon,s,t}^{-1}\exp\left(-\tau_{\epsilon,s}h_{\epsilon,s,t}-\kappa_{\epsilon,s}\right)-1\right)+\\
\phi_{\epsilon,s}\left(h_{\epsilon,s,t+1}-\phi_{\epsilon,s}h_{\epsilon,s,t}\right)-\left(h_{\epsilon,s,t}-\phi_{\epsilon,s}h_{\epsilon,s,t-1}\right)\;\textrm{for}\;t=2,...,T-1;
\end{multline*}

\begin{itemize}
\item For $k=1,...,K,$
\begin{multline*}
\nabla_{\alpha_{f,k}}\log p\left(y,\theta,x_{1:T}\right)=\frac{1}{2}\sum_{t=1}^{T}\left(h_{f,k,t}\left(f_{k,t}\right)^{2}W_{f,k,t}^{-1}\exp\left(-\tau_{f,k}h_{f,k,t}\right)-h_{f,k,t}\right)\left(1-\exp\left(-\tau_{f,k}\right)\right)\\
-\frac{2\tau_{f,k}}{\left(1+\tau_{f,k}^{2}\right)}\frac{\exp\left(\alpha_{f,k}\right)}{1+\exp\left(\alpha_{f,k}\right)}+1-\frac{\exp\left(\alpha_{f,k}\right)}{1+\exp\left(\alpha_{f,k}\right)};
\end{multline*}
\end{itemize}
\begin{multline*}
\nabla_{\psi_{f,k}}\log p\left(y,\theta,x_{1:T}\right)=\left(\sum_{t=2}^{T}\left(h_{f,k,t}-\phi_{f,k}h_{f,k,t-1}\right)h_{f,k,t-1}+h_{f,k,1}^{2}\phi_{f,k}-\frac{\phi_{f,k}}{1-\phi_{f,k}^{2}}\right)\left(\phi_{f,k}\left(1-\phi_{f,k}\right)\right)+\\
\left(\frac{\left(a_{0}-1\right)}{\left(1+\phi_{f,k}\right)}-\frac{\left(b_{0}-1\right)}{\left(1-\phi_{f,k}\right)}\right)\frac{\exp\left(\psi_{f,k}\right)}{\left(1+\exp\left(\psi_{f,k}\right)\right)^{2}}+\frac{\left(1-\exp\left(\psi_{f,k}\right)\right)}{\left(1+\exp\left(\psi_{f,k}\right)\right)};
\end{multline*}

\begin{multline*}
\nabla_{v_{f,k}}\log p\left(y,\theta,x_{1:T}\right)=\left\{ \frac{T}{2}\log\left(\frac{v_{f,k}}{2}\right)+\frac{T}{2}-\frac{T}{2}\Psi\left(\frac{v_{f,k}}{2}\right)\right.\\
\left.-\frac{1}{2}\sum_{t=1}^{T}\log\left(W_{f,k,t}\right)-\frac{1}{2}\sum_{t=1}^{T}\frac{1}{W_{f,k,t}}+1+v_{f,k}\left(\left(a_{v}-1\right)\frac{1}{v_{f,k}}-\frac{1}{b_{v}}\right)\right\} ;
\end{multline*}

\begin{multline*}
\nabla_{h_{f,k,1}}\log p\left(y,\theta,x_{1:T}\right)=\frac{\tau_{f,k}}{2}\left(\left(f_{k,1}\right)^{2}W_{f,k,1}^{-1}\exp\left(-\tau_{f,k}h_{f,k,1}\right)-1\right)+\\
\phi_{f,k}\left(h_{f,k,2}-\phi_{f,k}h_{f,k,1}\right)-h_{f,k,1}\left(1-\phi_{f,k}^{2}\right);
\end{multline*}

\begin{multline*}
\nabla_{h_{f,k,T}}\log p\left(y,\theta,x_{1:T}\right)=\frac{\tau_{f,k}}{2}\left(\left(f_{k,T}\right)^{2}W_{f,k,T}^{-1}\exp\left(-\tau_{f,k}h_{f,k,T}\right)-1\right)-\\
\left(h_{f,k,T}-\phi_{f,k}h_{f,k,T-1}\right);
\end{multline*}

\begin{multline*}
\nabla_{h_{f,k,t}}\log p\left(y,\theta,x_{1:T}\right)=\frac{\tau_{f,k}}{2}\left(\left(f_{k,t}\right)^{2}W_{f,k,t}^{-1}\exp\left(-\tau_{f,k}h_{f,k,t}\right)-1\right)+\\
\phi_{f,k}\left(h_{f,k,t+1}-\phi_{f,k}h_{f,k,t}\right)-\left(h_{f,k,t}-\phi_{f,k}h_{f,k,t-1}\right)\;\textrm{for}\;t=2,...,T-1.
\end{multline*}

\begin{itemize}
\item The gradients with respect to the factor loading $\beta_{s,.}=\left(\beta_{s,1},...,\beta_{s,k_{s}}\right)^{'}$
for $s=1,...,S$:
\[
\nabla_{\beta_{s,.}}\log p\left(y,\theta,x_{1:T}\right)=\left(y_{s,1:T}-F_{s}\beta_{s,.}\right)^{'}\widetilde{V}_{s}^{-1}F_{s}-\nabla_{\beta_{s,.}}\log p\left(\beta_{s,.}\right),
\]
where $y_{s,1:T}=\left(y_{s,1},...,y_{s,T}\right)^{'},$
\[
F_{s}=\left[\begin{array}{ccc}
f_{11} & \cdots & f_{k_{s}1}\\
\vdots &  & \vdots\\
f_{1T} & \cdots & f_{k_{s}T}
\end{array}\right];
\]
\end{itemize}
\[
\widetilde{V}_{s}=\textrm{diag}\left(W_{\epsilon,s,1}\exp\left(\tau_{\epsilon,s}h_{\epsilon,s,1}+\kappa_{\epsilon,s}\right),...,W_{\epsilon,s,T}\exp\left(\tau_{\epsilon,s}h_{\epsilon,s,T}+\kappa_{\epsilon,s}\right)\right).
\]

\subsubsection*{Algorithm 3 (Variational Approximation for the t-distributed FSV model )}

Initialise all the variational parameters $\lambda$. At each iteration
$j$, (1) Generate Monte Carlo samples for all the parameters and
latent states from their variational distributions, (2) Compute the
unbiased estimates of gradient of the lower bound with respect to
each of the variational parameter and update the variational parameters
using Stochastic Gradient methods.

\subsection*{STEP 1 }
\begin{itemize}
\item \textbf{For} $s=1:S$,
\begin{itemize}
\item Generate Monte Carlo samples for $\theta_{G,\epsilon,s}$ and $h_{\epsilon,s,1:T}$
from the variational distribution $q_{\lambda_{\epsilon,s}}^{TBN}\left(\theta_{G,\epsilon,s},h_{\epsilon,s,1:T}\right)$
\begin{enumerate}
\item Generate $\eta_{\theta_{G,\epsilon,s}}\sim N\left(0,I_{G_{\epsilon}}\right)$
and $\eta_{h_{1:T,\epsilon,s}}\sim N\left(0,I_{T}\right)$, where
$G_{\epsilon}$ is the number of parameters of idiosyncratic log-volatilities
\item Generate $\theta_{G,\epsilon,s}=\mu_{G,\epsilon,s}^{\left(j\right)}+\left(C_{G,\epsilon,s}^{\left(j\right)}\right)^{-\top}\eta_{\theta_{G,\epsilon,s}}$
and $h_{1:T,\epsilon,s}=\mu_{L,\epsilon,s}^{\left(j\right)}+\left(C_{L,\epsilon,s}^{\left(j\right)}\right)^{-\top}\eta_{h_{1:T,\epsilon,s}}$.
\end{enumerate}
\item Generate Monte Carlo samples for $v_{\epsilon,s}$ from the variational
distribution $q_{\lambda_{v_{\epsilon,s}}}^{SRN}\left(v_{\epsilon,s}\right)$,
\begin{enumerate}
\item Generate $\left(\eta_{v_{\epsilon,s}}\right)\sim N\left(0,I_{T}\right)$
and calculating $\xi_{v_{\epsilon,s}}=\mu_{\xi_{v_{\epsilon,s}}}^{\left(j\right)}+d_{\xi_{v_{\epsilon,s}}}^{\left(j\right)}\circ\eta_{v_{\epsilon,s}}$.
\item Generate $v_{\epsilon,s}=t_{\gamma_{v_{\epsilon,s}}}^{-1(j)}\left(\xi_{v_{\epsilon,s}}\right)$,
for $t=1,...,T$.
\end{enumerate}
\end{itemize}
\item \textbf{For} $k=1:K$,
\begin{itemize}
\item Generate Monte Carlo samples for $\theta_{G,f,k}$ and $h_{f,k,1:T}$
from the variational distribution $q_{\lambda_{f,k}}^{TBN}\left(\theta_{G,f,k},h_{f,k,1:T}\right)$
\begin{enumerate}
\item Generate $\eta_{\theta_{G,f,k}}\sim N\left(0,I_{G_{f}}\right)$ and
$\eta_{h_{1:T,f,k}}\sim N\left(0,I_{T}\right)$, where $G_{f}$ is
the number of parameters of factor log-volatilities
\item Generate $\theta_{G,f,k}=\mu_{G,f,k}^{\left(j\right)}+\left(C_{G,f,k}^{\left(j\right)}\right)^{-\top}\eta_{\theta_{G,f,k}}$
and $h_{1:T,f,k}=\mu_{L,f,k}^{\left(j\right)}+\left(C_{L,f,k}^{\left(j\right)}\right)^{-\top}\eta_{h_{1:T,f,k}}$.
\end{enumerate}
\item Generate Monte Carlo samples for $v_{f,k}$ from the variational distribution
$q_{\lambda_{v_{f,k}}}^{SRN}\left(v_{f,k}\right)$,
\begin{enumerate}
\item Generate $\left(\eta_{v_{f,k}}\right)\sim N\left(0,I_{T}\right)$
and calculating $\xi_{v_{f,k}}=\mu_{\xi_{v_{f,k}}}^{\left(j\right)}+d_{\xi_{v_{f,k}}}^{\left(j\right)}\circ\eta_{v_{f,k}}$.
\item Generate $v_{f,k}=t_{\gamma_{v_{f,k}}}^{-1(j)}\left(\xi_{v_{f,k}}\right)$,
for $t=1,...,T$.
\end{enumerate}
\end{itemize}
\item Generate Monte Carlo samples for the factor loading $\beta$ from
the variational distribution
\begin{enumerate}
\item Generate $z_{\beta}\sim N\left(0,I_{p}\right)$ and $\eta_{\beta}\sim N\left(0,I_{R_{\beta}}\right)$
calculating $\xi_{\beta}=\mu_{\xi_{\beta}}^{\left(j\right)}+B_{\xi_{\beta}}^{\left(j\right)}z_{\beta}+d_{\xi_{\beta}}^{\left(j\right)}\circ\eta_{\beta}$,
and $R_{\beta}$ is the total number of parameters in the factor loading
matrix $\beta$.
\item Generate $\beta_{i}=t_{\gamma_{\beta_{i}}}^{-1(j)}\left(\xi_{\beta_{i}}\right)$,
for $i=1,...,R_{\beta}$.
\end{enumerate}
\item Generate Monte Carlo samples for the $W_{\epsilon,s}$ and $W_{f,k}$
for $s=1,...,S$ and $k=1,...,K$,
\begin{enumerate}
\item Generate $W_{\epsilon,s,t}$ from $IG\left(\frac{v_{\epsilon,s}+1}{2},\frac{v_{\epsilon,s}}{2}+\frac{1}{2}\left(y_{s,t}-\beta_{s,.}f_{t}\right)^{2}\exp\left(-\tau_{\epsilon,s}h_{\epsilon,s,t}-\kappa_{\epsilon,s}\right)\right)\;\textrm{for }s=1,...,S.$
\item Generate $W_{f,k,t}$ from $IG\left(\frac{v_{f,k}+1}{2},\frac{v_{f,k}}{2}+\frac{1}{2}\left(f_{k,t}\right)^{2}\exp\left(-\tau_{f,ks}h_{f,k,t}\right)\right)\;\textrm{for }k=1,...,K.$
\end{enumerate}
\item Generate Monte Carlo samples for the latent factors from its full
conditional distributions
\[
p\left(f_{k,t}|\theta,h_{\epsilon}h_{f},W_{\epsilon},W_{f}\right)=N\left(\mu_{f_{k,t}},\Sigma_{f_{k,t}}\right)\;\textrm{for }k=1,...,K\;\textrm{and}\;\textrm{for }t=1,...,T,
\]
where $\mu_{f_{k,t}}=\Sigma_{f_{k,t}}\beta^{'}\left(\widetilde{V}_{t}^{-1}y_{t}\right)$,
$\Sigma_{f_{k,t}}=\left(\beta\widetilde{V}_{t}^{-1}\beta^{'}+\widetilde{D}_{t}^{-1}\right)^{-1}$.
\end{itemize}

\subsection*{STEP 2}
\begin{itemize}
\item \textbf{For} $s=1:S$, Update the variational parameters $\lambda_{\epsilon,s}$
of the variational distributions $q_{\lambda_{\epsilon,s}}^{TBN}\left(\theta_{\epsilon,s},h_{\epsilon,s,1:T}\right)$
\begin{enumerate}
\item Construct unbiased estimates of gradients $\widehat{\nabla_{\lambda_{\epsilon,s}}\mathcal{L}\left(\lambda_{\epsilon,s}\right)}$.
\item Set adaptive learning rates $a_{j,\epsilon,s}$ using ADAM method.
\item Set $\lambda_{\epsilon,s}^{\left(j+1\right)}=\lambda_{\epsilon,s}^{\left(j\right)}+a_{t,\epsilon,s}\widehat{\nabla_{\lambda_{\epsilon,s}}\mathcal{L}\left(\lambda_{\epsilon,s}\right)}^{\left(j\right)}$,

where $\lambda_{\epsilon,s}=\left\{ \mu_{G,\epsilon,s}^{\top},v\left(C_{G,\epsilon,s}^{*}\right)^{\top},d_{\epsilon,s}^{\top},\textrm{vec}\left(D_{\epsilon,s}\right)^{\top},f_{\epsilon,s}^{*\top},\textrm{vec}\left(F_{\epsilon,s}\right)^{\top}\right\} ^{\top}$
as defined in Section \ref{subsec:Sparse-Cholesky-factorTan2019}.
\end{enumerate}
\item For $k=1:K$,
\begin{itemize}
\item Update the variational parameters $\lambda_{f,k}$ of the variational
distributions $q_{\lambda_{f,k}}^{TBN}\left(\theta_{G,f,k},h_{f,k,1:T}\right)$
\begin{enumerate}
\item Construct unbiased estimates of gradients $\widehat{\nabla_{\lambda_{f,k}}\mathcal{L}\left(\lambda_{f,k}\right)}$.
\item Set adaptive learning rates $a_{j,f,k}$ using ADAM method.
\item Set $\lambda_{f,k}^{\left(j+1\right)}=\lambda_{f,k}^{\left(j\right)}+a_{j,f,k}\widehat{\nabla_{\lambda_{f,k}}\mathcal{L}\left(\lambda_{f,k}\right)}^{\left(j\right)}$,

where $\lambda_{f,k}=\left\{ \mu_{G,f,k}^{\top},v\left(C_{G,f,k}^{*}\right)^{\top},d_{f,k}^{\top},\textrm{vec}\left(D_{f,k}\right)^{\top},f_{f,k}^{*\top},\textrm{vec}\left(F_{f,k}\right)^{\top}\right\} ^{\top}$.
\end{enumerate}
\end{itemize}
\item Update the variational parameters $\lambda_{\beta}$ from the variational
distribution $q_{\lambda_{\beta}}^{SRN}\left(\beta\right)$.
\begin{enumerate}
\item Construct unbiased estimates of gradients $\widehat{\nabla_{\lambda_{\beta}}\mathcal{L}\left(\lambda_{\beta}\right)}$.
\item Set adaptive learning rates $a_{j,\beta}$ using ADAM method.
\item Set $\lambda_{\beta}^{\left(j+1\right)}=\lambda_{\beta}^{\left(j\right)}+a_{j,\beta}\widehat{\nabla_{\lambda_{\beta}}\mathcal{L}\left(\lambda_{\beta}\right)}^{\left(j\right)}$,
where $\lambda_{\beta}=\left(\gamma_{\beta_{1}}^{\top},...,\gamma_{\beta_{R_{\beta}}}^{\top},\mu_{\xi_{\beta}}^{\top},vech(B_{\xi_{\beta}})^{\top},d_{\xi_{\beta}}^{\top}\right)^{\top}$.
\end{enumerate}
\item Update the variational parameters $\lambda_{v_{\epsilon,s}}$ for
$s=1,...,S$,
\begin{enumerate}
\item Construct unbiased estimates of gradients $\widehat{\nabla_{\lambda_{v_{\epsilon,s}}}\mathcal{L}\left(\lambda_{v_{\epsilon,s}}\right)}$.
\item Set adaptive learning rates $a_{j,v_{\epsilon,s}}$ using ADAM method.
\item Set $\lambda_{v_{\epsilon,s}}^{\left(j+1\right)}=\lambda_{v_{\epsilon,s}}^{\left(j\right)}+a_{j,v_{\epsilon,s}}\widehat{\nabla_{\lambda_{v_{\epsilon,s}}}\mathcal{L}\left(\lambda_{v_{\epsilon,s}}\right)}^{\left(j\right)}$,
where $\lambda_{v_{\epsilon,s}}=\left(\gamma_{v_{\epsilon,s}}^{\top},\mu_{\xi_{v_{\epsilon,s}}}^{\top},d_{\xi_{v_{\epsilon,s}}}^{\top}\right)^{\top}$.
\end{enumerate}
\item Update the variational parameters $\lambda_{v_{f,k}}$ for $k=1,...,K$,
\begin{enumerate}
\item Construct unbiased estimates of gradients $\widehat{\nabla_{\lambda_{v_{f,k}}}\mathcal{L}\left(\lambda_{v_{f,k}}\right)}$.
\item Set adaptive learning rates $a_{j,v_{f,k}}$ using ADAM method.
\item Set $\lambda_{v_{f,k}}^{\left(j+1\right)}=\lambda_{v_{f,k}}^{\left(j\right)}+a_{j,v_{f,k}}\widehat{\nabla_{\lambda_{v_{f,k}}}\mathcal{L}\left(\lambda_{v_{f,k}}\right)}^{\left(j\right)}$,
where $\lambda_{v_{f,k}}=\left(\gamma_{v_{f,k}}^{\top},\mu_{\xi_{v_{f,k}}}^{\top},d_{\xi_{v_{f,k}}}^{\top}\right)$.
\end{enumerate}
\end{itemize}

\section{ADAM Learning Rates \label{subsec:Learning-Rate-1}}

The setting of the learning rates in stochastic gradient algorithms is
very challenging, especially for a high dimensional  parameter vector.
 The choice of learning rate affects both the
rate of convergence and the quality of the optimum attained. Learning
rates that are too high can cause instability of the optimisation,
while learning rates that are too low result in a slow convergence
and can lead to a situation where the parameters appear to have converged.
In all our examples, we set the learning rates adaptively using the
ADAM method \citep{Kingma2014a} that gives different step sizes for
each element of the variational parameters $\lambda$. At iteration
$t+1$, the variational parameter $\lambda$ is updated as
\[
\lambda^{\left(t+1\right)}=\lambda^{\left(t\right)}+\triangle^{\left(t\right)}.
\]
Let $g_{t}$ denote the stochastic gradient estimate at iteration
$t$. ADAM computes (biased) first and second moment estimates of
the gradients using exponential moving averages,

\begin{eqnarray*}
m_{t} & = & \tau_{1}m_{t-1}+\left(1-\tau_{1}\right)g_{t},\\
v_{t} & = & \tau_{2}v_{t-1}+\left(1-\tau_{2}\right)g_{t}^{2},
\end{eqnarray*}
where $\tau_{1},\tau_{2}\in\left[0,1\right)$ control the decay rates.
Then, the biased first and second moment estimates can be corrected
by
\begin{eqnarray*}
\widehat{m}_{t} & = & m_{t}/\left(1-\tau_{1}^{t}\right),\\
\widehat{v}_{t} & = & v_{t}/\left(1-\tau_{2}^{t}\right).
\end{eqnarray*}
 Then, the change $\triangle^{\left(t\right)}$ is then computed as
\begin{equation}
\triangle^{\left(t\right)}=\frac{\alpha\widehat{m}_{t}}{\sqrt{\widehat{v}_{t}}+eps}.
\end{equation}
We set $\alpha=0.001$, $\tau_{1}=0.9$, $\tau_{2}=0.99$, and $eps=10^{-8}$
\citep{Kingma2014a}.

\end{document}